%% file: PhD.tex
\documentclass [12pt] {book}

\usepackage{graphicx}
\usepackage{amsmath}
\usepackage{amsfonts}
\usepackage{amssymb}

\def \D {\mbox{D}}

\def \ep {\varepsilon}
\def \cu{{\cal U}}
\def \cq{ Q}
\def \cp{ \Pi}

\def\bea{\begin{eqnarray}}
\def\eea{\end{eqnarray}}
\newcommand{\be}{\begin{equation}}
\newcommand{\ee}{\end{equation}}
\newcommand{\ba}{\begin{eqnarray}}
\newcommand{\ea}{\end{eqnarray}}
\newcommand{\half}{\frac{1}{2}}
\newcommand{\pre}{\frac{1}{4\pi\alpha'}}

\newcommand{\five}{\tilde}
\newcommand{\mb}[1]{\mbox{\boldmath $#1$}}
\begin{document}
\pagestyle{empty}
\input{intestazione.tex}
\pagestyle{headings}
\pagenumbering{roman}
\tableofcontents
\chapter{Foreword}
\footnote{This is a rough and informal view of quantum gravity,
using \cite{R1}.}Just one year after the discovery of general
relativity, Einstein pointed out that quantum effects must lead to
modifications in the theory of general relativity \cite{E1}. The
problems of compatibility between gravity and quantum mechanics were further
considered by Heisenberg \cite{H1}. He realized that since the
gravitational coupling constant is dimensional, a theory which
quantizes gravity will have serious problems. Indeed in the
seventies, t'Hooft and Veltman as well as Deser and Van
Nieuwenhuizen \cite{tH1}, confirmed that the quantum theory of
gravity coupled with matter has non-renormalizable divergences.
This was disappointing for a quantum theory of gravity. Indeed it
showed that the old techniques of quantization could not work for
general relativity. Soon after, Hawking \cite{H2} discovered that
a black hole is not cold, but thanks to quantum effects emits
radiation with temperature
\begin{equation}
   T=\frac{\hbar c^3}{8\pi kGM}.
\end{equation}
Then Unruh \cite{U1} proposed a relation between accelerated
observers, quantum theory, gravity and thermodynamics. This
suggested new profound relations between quantum and classical
worlds. The idea was that a full quantum gravity description must
reproduce these relations.

A connection between classical gravity and quantum field theory
was first seen in the seventies.
In the late sixties indeed, Veneziano \cite{V1}, trying to understand why the theoretically
expected amplitude divergences did not appear when high
energy elastic scattering of baryons produced
particles with higher spins, introduced a duality between the Mandelstam
variables $s$ and $t$ of the process \cite{IZ}. In this
way the divergences in one channel for higher spins, were
exactly cancelled out by the other channel.
The Veneziano theory had some peculiarities
such as the existence of
a massless spin-two particle, which did not
correspond to any renormalizable quantum theory. This particle was
then recognized as the graviton.
Indeed several years later, people realized that the phenomenological
model of Veneziano could be derived from a more fundamental
theory. The idea was to quantize one dimensional objects (strings)
instead of point-like objects (particles). Interpreting the massless spin-two
particle as the graviton, it was
conjectured that string theory leads to a non-divergent theory of
quantum gravity.
As well as producing exciting insights,
the introduction of strings implied also the existence of extra spatial dimensions, which was
at that time considered to be a problem.

In the eighties Polyakov \cite{P1} showed that the extra
dimensions were linked to the necessity of keeping the classical
symmetries for this action at the quantum level.

Later works showed also that string theory provides a consistent
theory at perturbed level. The aim of this theory was to unify all
the forces without introducing phenomenological coupling
constants.

Another important concept was developed by t'Hooft \cite{tH2} and
promoted by Susskind \cite{S1}. They proposed that the information
of a physical state in the interior of a region can be represented
on the region's boundary and is limited by the area of this
boundary. This was motivated to explain why the entropy of a black
hole (which contains all the information of
 this object) depends only on the area of its horizon and not on
 the volume inside the horizon. This principle was called
the ``Holographic principle''. The holography was a tentative way to connect
classical physics with quantum physics in a purely geometrical
way. In 1998 Maldacena \cite{M1} applied this concept in string
theory. Soon after, Witten \cite {W2} clarified the holographic
features of anti-deSitter spacetimes which was the basis for the
Maldacena work. The concept was that a conformal quantum field
theory (CFT) on a boundary of a spacetime with anti-deSitter
background can be described as a classical gravitational theory.
This was later called the AdS/CFT correspondence.

A sociological event was also born with string theory. A dialogue
between General Relativity
and Quantum Physics communities reemerged.

But, as Rovelli explains \cite{R1}, there are still profound
divergences. From the point of view of the General Relativity
community, quantum field theory is problematic. Up to now indeed,
quantum physics with gravity is consistent only for fixed
spacetime backgrounds. This implies that it is inadequate for a
full understanding of a dynamical theory of spacetime. On the
other side, for the Quantum Mechanics community, General
Relativity is only a low energy limit of a much more complex
theory, and thus cannot be taken too seriously as an indication of
the deep structure of Nature.

A bridge between the two approaches must be found. One possible way is to study the
dynamics of gravitational theories inspired by quantum gravity.

If one follows the direction of string theory and the
holographic principle, one must find a mechanism to reduce the dimensions of the
spacetime to four, at least at low energies.

One possibility was introduced in 1999 by Randall and Sundrum
\cite{RS}, where the extra dimension is infinitely large and the
matter is trapped on a four-dimensional submanifold called the
``brane''. The graviton is instead localized on the brane only at
low energies, reproducing the correct Newtonian limit. This is
possible thanks to a negative cosmological constant which away from
the brane induces an anti-deSitter spacetime.

In this thesis I will discuss the astrophysical and cosmological implications
of this mechanism for simple phenomenological set ups.

I use the signature $(-,+,+,+)$ and $(-,+,+,+,+)$ for the
four and five dimensional Lorentzian manifold, the natural
units $c=\hbar=1$ and the definition of the Ricci tensor
$R^\alpha{}_{\mu\alpha\nu}=R_{\mu\nu}$, where $R_{\alpha\mu\nu\beta}$ is the
Riemann tensor.

\mainmatter
\newpage
\chapter{Extra dimensions: motivations}

To give a flavour of how extra dimensions appear in modern
theoretical physics, I will describe simple examples in the
context of String Theory and Holography. Since this chapter must
be seen as only as a motivation for the study of extra
dimensions, I will not directly connect these theories with
the Randall-Sundrum mechanism. I will anyway discuss throughout
the thesis, the influences that these theories have in the
Randall-Sundrum-type models.

\section{String Theory}

Currently the most promising theory for the unification of all the
forces seems to be superstring theory. There are actually five
anomaly-free perturbative string theories which are: type I, type
IIA, type IIB, SO(32) and $E_8\times E_8$ heterotic theories
\cite{GSW,PO}. These are all supersymmetric and require in general
ten dimensions. Another interesting theory is supergravity. In
eleven dimensions it is unique and it has been proved that its
compactification in ten dimensions reproduces the low energy limit
of the type IIA superstring theory. This was the starting point to
conjecture that a more general theory, whose low energy limit is
the eleven-dimensional supergravity \cite{SU}, is actually the
ultimate theory. This concept was also reinforced thanks to the
evidence that the superstring theories are related to each other
by dualities. The ultimate theory was called M-Theory
\cite{MT1,MT2,MT3,MT4}.

String theory contains an infinite tower of massive states,
corresponding to the oscillations of the string. The massless
states are separated from the massive ones by a gap of energy of
order $1/\sqrt{\alpha'}$. $T=1/{2\pi\alpha'}$ is called the string
tension, which is usually taken to be close to the Planck scale.
This is the only arbitrary parameter of the theory. In the limit
of an infinite tension, the massless excitations decouple from
the massive ones and the theory is described by the low energy
limit which in the effective action contains the usual
Einsteinian gravity. The low energy limit, more geometrically, is
dictated by the comparison between the curvature radius of the spacetime and the
string length. This means that the low energy limit is for
$R^{-1/2}\gg \sqrt{\alpha'}$ where $R$ is the Ricci scalar.

\subsection{Classical string action and equation of motion}

\subsubsection{String action in flat background}

In this section I give a taste of how the extra dimensions arise in string theory.
A complete treatment of bosonic and
supersymmetric string theories can be found in \cite{GSW,PO}.

The motion of a free falling relativistic particle is described by
the maximum spacetime length from an event $P_1$ to $P_2$
\cite{Wald}. The spacetime length is described by the integral \be
{\cal L}=-m\int^{P_2}_{P_1} ds, \ee which depends on the path,
where $m$ is the mass of the particle and \be
ds^2=g_{\alpha\beta}dx^\alpha dx^\beta, \ee where
$g_{\alpha\beta}$ is the spacetime metric.

Supposing the particle is massive ($m>0$), we can define the
proper time $\tau$ such that \be \frac{ds}{d\tau}=-1. \ee With
this parameter we can rewrite the total length in terms of the
particle four velocity $u^\alpha=dx^\alpha/d\tau$, as \be {\cal
L}=-m\int^{\tau_2}_{\tau_1}\sqrt{-u^\alpha u_\alpha} d\tau. \ee

Now we introduce a one-dimensional object instead of a point-like
one, i.e. a string. With the same logic we can say that the motion
of a string is described by the extremal area covered from a curve
$S_1$ to $S_2$. So the action for a string is \be {\cal
A}=-\frac{1}{2\pi\alpha'}\int^{S_2}_{S_1} dA, \ee where
$T=1/2\pi\alpha'$ is the tension of the string.

Since the string is a two-dimensional spacetime object, we can
describe its motion as a two-dimensional sheet. This sheet can be
described by two internal coordinates. We use the proper time
$ \tau $ and the proper spatial length $ \sigma $. Of course the
metric which describes this sheet in spacetime is (with signature
$(-,+)$) \be h_{ab}{}^{\mu\nu}=\partial_a x^\mu \partial_b x^\nu,
\ee where the Latin letters $a,b,..$ are the internal coordinates
$(\sigma,\tau)$. This is just the generalization of the point-like
four velocity $u^\alpha=\partial_\tau x^\alpha$. Indeed we have
this simple scheme: \ba
&\mbox{particle}\rightarrow\mbox{string}\cr\nonumber\cr\nonumber
&\sqrt{-u^\alpha
u_\alpha}\rightarrow \sqrt{-\mbox{det}(h_{ab}{}^\alpha{}_\alpha)}
=\sqrt{-\mbox{det}(h_{ab})}.\nonumber\cr\nonumber \ea Then finally
the action for a string is \be\label{ng} {\cal
A}=-\frac{1}{2\pi\alpha'}\int^{\tau_2,\sigma_2}_{\tau_1,\sigma_1}
\sqrt{-\mbox{det}(h_{ab})}d\sigma d\tau. \ee This is called the
Nambu-Goto action. From now on $\mbox{det}(h_{ab})=h$ and the
surface with boundaries $(\sigma_1,\tau_1)$ and
$(\sigma_2,\tau_2)$ is denoted by $M$.

Since we are interested in the quantum theory of strings, we need
to manipulate the action so that it appears in a more ``linear"
form. In this way we can use the standard quantization rules for
non-linear sigma models. To do that we introduce an auxiliary
metric $\gamma^{ab}(\sigma,\tau)$ and rewrite eq. (\ref{ng})
as \be\label{polact} {\cal A}=-\frac{1}{4\pi\alpha'}\int_M d\tau
d\sigma \sqrt{-\gamma}\gamma^{ab}h_{ab}. \ee Then we can see under
which conditions this action is equivalent to the Nambu-Goto one.
If $\gamma_{ab}$ is an auxiliary metric, then the variation of the
action with respect to it must vanish. Varying the action we get
the equation \be h_{ab}=\frac{1}{2}\gamma_{ab}\gamma^{cd}h_{cd},
\ee from which follows
\be
\sqrt{-h}=\sqrt{-\gamma}\gamma_{ab}h^{ab}. \ee

The action (\ref{polact}) is called the Polyakov action. This
action is invariant under the worldsheet ($(\sigma,\tau)$-space)
diffeomorphism
\begin{eqnarray}
&\delta x^\mu=\pounds_{\xi}x^\mu=\xi^a\partial_a x^\mu\ ,\\
&\delta \gamma^{ab}=\pounds_{\xi}\gamma^{ab}=\xi^c\partial_c
\gamma^{ab}- 2\gamma^{c(a}\partial_c\xi^{b)},
\end{eqnarray}
and the two-dimensional Weyl invariance \be \delta
\gamma^{ab}=\omega \gamma^{ab}. \ee
Here $\xi^a$ and $\omega$
are small vector and scalar parameters.
Of course we have also the
spacetime Poincar\'e invariance
\be \delta
x^\mu=\omega^{\mu\nu}x_\nu+b^\mu\ , \ee
where $\omega^{\mu\nu}$ represent a rotation and $b^\mu$ a translation in spacetime.

Now $\gamma^{ab}$ is actually the metric for the gravitational
field on the worldsheet, so we can use it to raise or lower the
worldsheet indices.

In general in the action (\ref{polact}) we could add a term which
describes the field equation for it. In two dimensions (the
worldsheet) the only possibility is to add a cosmological
constant, since the Ricci scalar is a total derivative, and so
therefore does not affect the equations of motion. However this
breaks the Weyl invariance, which is a key point to quantize the
string, and so it is set to zero.

Since the action is invariant under variations of $\gamma^{ab}$,
the energy-momentum tensor associated with it,
\be
T_{ab}=\partial_a x^\mu\partial_b x_\mu-\half
\gamma_{ab}\partial^cx^\mu\partial_cx_\mu\ , \ee
must vanish. In
particular its trace is zero. This is a consequence of the
conformal invariance (Weyl) of the action. We will see that, in
general, quantum mechanically the trace of this tensor does not
vanish. In order to keep this symmetry we will need in general
extra-dimensions.

Thanks to the classical Weyl and Poincar\'e symmetries we can locally choose the
worldsheet metric. We use the Minkowski one \be
\gamma_{ab}=\eta_{ab}=\mbox{diag}(-1,1). \ee Since strings are
one-dimensional objects they can be open or closed. We will see
that in the first case we have some boundary conditions to impose.
The variation of the action with respect to the fields
$x^\mu(\tau,\sigma)$ is \be -4\pi\alpha'\delta{\cal A}=\int_M
d^2\sigma \partial_a x^\mu \partial^a \delta x^\mu, \ee where we
used $\delta\partial_a x^\mu\simeq\partial_a\delta x^\mu$ and
$d\sigma d\tau=d^2\sigma$.

Integrating by parts we get
\be -4\pi\alpha'\delta{\cal A}=\int_M
d^2\sigma \partial _a (\eta^{ab}\partial _b x^\mu \delta
x_\mu)-\int_M d^2\sigma \delta x_\mu \Box x^\mu\ . \ee
If the string
is closed the first integral is zero. So the equations of motion
are \be \label{eqm} \Box
x^\mu=(\partial^2_\tau-\partial^2_\sigma)x^\mu(\tau,\sigma)=0. \ee
For an open string the first integral is \be \int_M d^2\sigma
\partial _a (\eta^{ab}\partial _b x^\mu \delta
x_\mu)=\oint_{\partial M}d\sigma_a\eta^{ab}\partial_b x^\mu\delta
x_\mu. \ee
We can choose $\partial M$ to be a space-like boundary,
so that $d\sigma_a=d\tau\delta^\sigma_a$. Normalizing the length
of the string such that $0\leq \sigma\leq\pi$, in order to have
zero variation we can consider a combination of the following
possibilities for the ends of the string:

{\it Neumann boundary conditions} \ba
\partial_\sigma x^\mu\Big|_\pi=0,\ \ \partial_\sigma x^\mu\Big|_0=0.
\ea

{\it Dirichlet boundary conditions} \ba
x^\mu\Big|_\pi=\mbox{const}\Rightarrow\delta x^\mu\Big|_\pi=0,\ \
x^\mu\Big|_0=\mbox{const}\Rightarrow\delta x^\mu\Big|_0=0. \ea

\subsubsection{General string action}

The string action can be generalized in the following way
\cite{GSW}:
\be\label{coupled} {\cal S}=-\pre \int d^2\sigma
\left[\sqrt{-\gamma}\gamma^{ab}g_{\mu\nu}(x^\alpha)\partial_a
x^\mu\partial_b x^\nu+\epsilon^{ab}B_{\mu\nu} (x^\alpha)\partial_a
x^\mu \partial_b x^\nu\right]. \ee
The coupling functions
$g_{\mu\nu}$ and $B_{\mu\nu}$ can be identified as the background
spacetime graviton and antisymmetric tensor fields in which the
string is propagating. $\epsilon^{ab}$ is the two-dimensional
Levi-Civita symbol. The coupling of these tensors with the string
fields is well justified also at the quantum level. At quantum
level we also get a massless scalar field in the spectrum. Such a
field in general breaks the Weyl invariance. Fradkin and Tseytlin
\cite{F1} have suggested that one should add to the string theory
action the renormalizable but not Weyl invariant term \be
S_{dil}=\frac{1}{4\pi}\int d^2\sigma
\sqrt{\gamma}R^{(2)}\Phi(x^\alpha), \ee where $R^{(2)}$ is the
two-dimensional Ricci scalar and the scalar field $\Phi(x^\alpha)$
is called the dilaton. Now it is true that at a classical level
this term breaks the conformal symmetry. Since string theory is a
quantum theory, we actually can require the weak condition that
the Weyl anomaly is cancelled at least at quantum level.

From now we use the following definition of total string action
\be
I[x,\gamma]={\cal S}+S_{dil}\ .
\ee
In the next
section we show how the quantization of this action leads to the Weyl anomalies.

\subsection{Quantization of the string action}

As an example of how extra-dimensions appear in string theory we first consider
the simplest background in which $\Phi=0$.

A modern concept of quantization is to introduce the partition function via the path
integral
\be
Z=\int {\cal D}\gamma(\sigma){\cal D}x(\sigma)e^{-I[x,\gamma]}\ ,
\ee
where ${\cal D}f$ means the integration over all the possible functions $f$.

Since Weyl symmetry of the classical action we can restrict ourselves in considering
\be
\gamma_{ab}=e^\phi\eta_{ab}\ .
\ee
In this way we can try to make explicit the measure ${\cal D}\gamma(\sigma)$ as the integration
of all the possible reparameterizations of the worldsheet metric. This is possible by introducing
auxiliary fields that basically fix the gauge for any choice of $\phi$ under integration.
These fields, called
Faddeev-Popov ghosts ($b,c$), are anticommuting \cite{GSW}.
Using the fact that the partition function can be rewritten as
\be
Z=\int {\cal D}\phi(\sigma)\int {\cal D}x(\sigma){\cal D}b(\sigma) {\cal D}c(\sigma)
e^{-I[x,b,c]}\ ,
\ee
where now the effective action becomes, in complex coordinates ($ds^2=dzd\overline z$)
\be
I[x,b,c]=I[x,\gamma]-\frac{1}{2\pi\alpha'}\int d^2\sigma \left[b_{zz}\nabla_{\overline z}
c^z+ \mbox{c.c.} \right]\ ,
\ee
the ghost $c^z$($c^{\overline z}$) is a holomorphic (antiholomorphic) vector and the
antighost $b_{zz}$($b_{\overline z\overline z}$) is an holomorphic (antiholomorphic) quadratic
differential.

The effective action $I[x,b,c]$ is classically conformal invariant.
In order to keep this invariance at the quantum level, we have to show under which conditions
the product of the measures ${\cal D}x{\cal D}b {\cal D} c$ and $I[x,b,c]$
are quantum Weyl invariant. Considering the
quantum fluctuations of the string and ghost fields as gaussian, because of their
dependence on the
worldsheet metric, we obtain, under rescaling
$\eta_{ab}\rightarrow e^{\xi} \eta_{ab}$, \cite{Ginsparg2}
\be\label{poly}
{\cal D}_{e^{\xi}\eta}x {\cal D}_{e^\xi\eta}[\mbox{ghost}]=\exp\left(\frac{D-26}{48\pi}S_L(\xi)
\right){\cal D}_{\eta}x{\cal D}_\eta[\mbox{ghost}]\ ,
\ee
where $S_L(\xi)$ is known as the Liouville action
(see \cite{Ginsparg2} for the explicit form) and
$D$ is the spacetime dimension.
Then in order to keep Weyl invariance at the quantum level, the first requirement is that
$D=26$.

Considering now a non vanishing $\Phi$, to see under what conditions $I[x,b,c]$ is quantum scale invariant,
one can calculate the quantum trace of the renormalized energy-momentum tensor of
the string, defined as
\be
\gamma^{ab}\frac{\delta \ln Z}{\delta \gamma_{ab}}\Big|_{\rm ren}=
i\gamma^{ab}\langle T_{ab}\rangle\ .
\ee
The renormalized result is \cite{C1}
\ba 2\pi
\langle T^a{}_a\rangle=\beta^\Phi\sqrt{-\gamma}R^{(2)}+\beta_{\mu\nu}^g\sqrt{\gamma}\gamma^{ab}
\partial_ax^\mu\partial_b
x^\nu+\beta_{\mu\nu}^B \epsilon^{ab}\partial_ax^\mu\partial_b
x^\nu, \ea
where $\beta^\Phi$, $\beta^g$ and $\beta^B$ are local
functionals of the coupling functions $\Phi$, $g_{\mu\nu}$ and
$B_{\mu\nu}$, and in the limit $R^{(2)-1}\sqrt{\alpha'}\ll
1$,
\ba \beta^g_{\mu\nu}=
R_{\mu\nu}+2\nabla_\mu\nabla_\nu\Phi-\frac{1}{4}H_{\mu\rho\sigma}
H^{\rho\sigma}_\nu+O(\alpha')\label{noncri1}\\
\beta^B_{\mu\nu}=\nabla_\lambda H^\lambda_{\mu\nu}-2(\nabla_\lambda\Phi)H^\lambda_{\mu\nu}+
O(\alpha')\\
\beta^\Phi=\frac{D-D_c}{48\pi^2}+\frac{\alpha'}{16\pi^2}\left\{4(\nabla\Phi)^2-4\nabla^2\Phi-
R+\frac{1}{12}H^2\right\}
+O(\alpha'^2).\label{noncri2} \ea
Here
$H_{\mu\nu\lambda}=3\nabla_{[\mu}B_{\nu\lambda]}$, and $D_c=26$ is the number of
critical dimensions. In the beta function for the Weyl
anomaly $\beta^\Phi$, the leading term for bosonic strings was
discovered by Polyakov \cite{P1} in a similar way as we did in finding (\ref{poly}).
The graviton part was found by Friedan et
al. \cite{F2} and the $H$-fields by Witten \cite{W1} and
Curtright and Zachos \cite{C2}. In order to keep the Weyl symmetry at quantum level, we
have to make $\beta^\Phi$, $\beta^g$ and $\beta^B$ vanish.

It is important to note that since the coefficient of
$R^{(2)}\Phi$ is smaller by a factor $\alpha'$ than the other
couplings, its {\it classical} contribution is of the same order
as the one-loop {\it quantum contribution} of the $g_{\mu\nu}$ and
$B_{\mu\nu}$ couplings. This is because $R^{(2)}\Phi$ is scale
non-invariant at the classical level, while the other couplings
only lose scale-invariance at the quantum level. It is very simple
to prove that $\beta^\Phi$ is actually a constant in space time.
Indeed applying the Bianchi identities to $\beta^g$ and $\beta^B$
we get \be \nabla_\mu \beta^\Phi=0. \ee Therefore once we have solved the
equations for $\beta^{G,B}$, $\beta^\Phi$ is determined up to a
constant.

From eqs. (\ref{noncri1}-\ref{noncri2}) we can already argue that
string theory
 does not have to live in critical
dimensions. This kind of String theory is called non-critical
string theory \cite{PO}. In the $\beta^\Phi$ function, even if
the correction to the critical theory with $D=D_c$ is at first
order in $\alpha'$, it is actually possible to solve the equation
consistently to the one-loop correction. Myers \cite{M2} indeed
found that at least for the bosonic string in flat spacetime, where
$D_c=26$, a consistent solution is possible. This is
\ba
B_{\mu\nu}=0\ ,\ \Phi(x^\mu)=V_\mu x^\mu, \ea
where
\ba V_\mu
V^\mu=\frac{26-D}{6\alpha'}. \ea
This solution is compatible with
any dimension of the spacetime. There are not any other
anomaly-free non-critical string theories known.

Introducing supersymmetry in the string
action, which is a natural way to consider matter fields, one can prove
that $D_c=10$
instead of $D_c=26$ in eq. (\ref{noncri2}) \cite{C1}.
The bosonic sector of the supersymmetric string still satisfies
equations (\ref{noncri1}-\ref{noncri2}). These can be described by an
effective action \cite{M3}
\ba S_D=\frac{1}{2\kappa^2_D}\int d^D x
\sqrt{-g}e^{-\Phi}\left[ R+(\nabla\Phi)^2-\frac{1}{12}
H^2-\frac{2(D-10)}{3}\right].
\ea
Since the $\beta^\Phi$ function is at
order $\alpha'$, we can consistently consider the $\alpha'$
corrections to the $\beta^g_{\mu\nu}$ function as well. This
calculation can be done in heterotic string theory, which seems
to contain the gauge group of the Standard Model \cite{G1}. If we
choose a background with $B_{\mu\nu}=0$, the effective action then
becomes \cite{BoDe:85}
\be S_D=\frac{1}{2\kappa^2_D}\int d^Dx
\sqrt{-g}e^{-\Phi}\left[
R+(\nabla\Phi)^2-\frac{2(D-10)}{3}-\frac{\alpha'}{8}
\left\{L_{GB}-(\nabla\Phi)^4 \right\}\right], \ee
where
$L_{GB}=R^{\mu\nu\alpha\beta}R_{\mu\nu\alpha\beta}-4R^{\mu\nu}R_{\mu\nu}+R^2$
is the Gauss-Bonnet term.

Applying a conformal transformation
$g_{\mu\nu}\rightarrow e^{\frac{4}{D-2}\Phi}g_{\mu\nu}$ we can rewrite the
action in a more familiar form \cite {GBE,M3}
\ba
S_D=\frac{1}{2k^2_D}\int d^Dx \sqrt{-g}[
R&-&\frac{4}{D-2}(\nabla\Phi)^2-\frac{2(D-10)}{3}
e^{\frac{4}{D-2}\Phi}+\cr\nonumber
&+&\frac{\alpha'}{8}e^{-\frac{4}{D-2}\Phi}(L_{GB}+4\frac{D-4}{(D-2)^3}(\nabla\Phi)^4)].
\ea
This frame is called the Einstein frame. At this level it is
explicit that string theory contains, as effective low energy
theory, the generalization of the Einstein theory in extra
dimensions, called the
Lovelock theory of gravity \cite{Lov}, when $\Phi$ is constant.

    \section{Holography}
The idea of Holography is to relate quantum physics on a spacetime boundary with
classical geometrical properties of the spacetime.

Initially the aim of this theory was to understand the quantum physics of black holes.
Indeed when a black hole is formed,
classical and quantum physics encounter each other at its boundary, the horizon.
In particular all the quantum degrees of
freedom live holographically on the horizon. This arises from the study of the entropy
of a black hole. We know that the entropy is
the measure of the degrees of freedom of a physical state. Applying covariant quantization and
the analogies between thermodynamics and
black hole physics, Bekenstein and then Hawking \cite{holo1,holo2,holo3,H2} discovered
that the entropy of a black hole
is proportional to its area $A$:
\begin{equation}
S=\frac{A}{4G_N}.
\end{equation}
Quantum mechanically this was a breakthrough. Indeed quantum field theory (QFT)
gives an estimation of entropy
proportional to the volume and not to the area of a physical object.
This seems to say that QFT lives only on the horizon. Consider a spherical
region $\Gamma$ of volume $V$ in an asymptotically flat spacetime.
The boundary $\delta \Gamma$ has area $A$. The maximal entropy is
defined by
\begin{equation}
S_{max}=\ln N_{states}
\end{equation}
where $N_{states}$ is the total number of possible states of $\Gamma$.
Since we are considering gravitational objects, the number of states should
be linked with the degrees of freedom of the spacetime.
Suppose we consider that, at each fixed time, the space can be quantized. This means
that we divide it into elementary cells of width $\alpha$. Each of these cells store
the local information of a physical state. Moreover, assuming that for each cell
we have $m$ possible states, we get approximately
\begin{equation}
N_{states}=m^{V/\alpha^3}\ .
\end{equation}
This implies that the maximum entropy is
\begin{equation}
S_{max}\propto V.
\end{equation}
Now we consider a static star of energy $E$ and radius $R$. A static equilibrium implies that
the energy of the star $E$ must be bounded,
$E<M$. Where $M=R/2$ is the maximum mass one can fit in a static sphere or radius $R$.
The second law of thermodynamics tells us that an increase in energy implies an increment of
entropy. The maximum entropy will
then be reached when the black hole is formed. Indeed after the formation of a black hole,
all the degrees of freedom inside
the horizon are causally disconnected from the exterior, so that they
cannot be taken into account. This means that the maximum entropy is the
black hole entropy
\begin{equation}
S_{max}=\frac{A}{4G_N}.
\end{equation}
If we believe that we can calculate the entropy of a star by simple QFT calculations, we
implicitly break unitarity
of the quantum theory, losing predictability  \cite{predict}. This is because there is a
transition of the number of states
from the collapsing star ($\ln N_{states}\propto V$) to the black hole ($\ln N_{states}
\propto A$). If instead
we accept that the maximum entropy of a
spatial region is proportional to the area of its boundary, rather then its volume, then we
can retain unitarity in the collapse process.
This is how the holographic principle was first formulated. In the extended version this
principle states that
for any Lorentzian
manifold it is possible to find a submanifold (screen) where all the quantum degrees of
freedom are present \cite{Bousso}.
The AdS/CFT correspondence \cite{M1} is a specific and explicit example. Here the screen
is identified as the AdS boundary.
In this holographic correspondence, a quantum conformal field theory on the screen,
can be
described as a boundary effect of a classical geometrical theory of spacetime.
If we believe seriously in this principle,
the evidence
that our world is described by quantum degrees of freedom, makes the study of gravity
in higher dimensions encouraging. In the following
I describe how the relationship between quantum physics and classical
gravity arises in an anti de Sitter manifold (AdS), which is
the geometry in which the Randall-Sundrum mechanism is constructed.

        \subsection{AdS/CFT correspondence}
This test of the holographic principle was developed in particular string theories.
Here it has been shown
that on $D$-dimensional AdS backgrounds there is a correspondence between a classical
perturbed $D$-dimensional super-gravity theory and a
$(D-1)$-dimensional super-conformal field theory.
This duality was then argued to be present also
in deformations of AdS, such as the generalization of the Randall-Sundrum model, where the
quantum theory of the boundary was not necessarily
a conformal field theory (for a very nice introduction to the topic see \cite{padilla}).
In order to have the flavour of this
duality, I give a simple
example of AdS/CFT correspondence. Here I review the result that the zero
point energy of a CFT theory is actually described
by the boundary energy of an anti de Sitter spacetime \cite{BK}.

\subsubsection{The Brown-York tensor}
To define the energy of a manifold boundary we are going to use the definition of the
quasi-local energy given in \cite{BY}. The idea
is very simple and clear. We make an analogy with a classical non-relativistic system.
Suppose this system has an action $S$.
It satisfies the Hamilton-Jacobi equation $H=-\partial S/\partial t$,
where $H$ is the Hamiltonian of the system which describes the
energy and $t$ is the time. We would like now to generalize this concept for a gravitational
 system. Take a $D$ dimensional manifold
$M$ which can be locally described by a product of a $D-1$ dimensional space $\Sigma$ and
a real line interval. The
boundary $\partial \Sigma$ need not be simply connected. The product
of $\partial \Sigma$ with the real line
orthogonal to
$\Sigma$ will be denoted by $B$. By analogy with classical mechanics, the quasi-local
energy associated with the spacelike
hypersurface $\Sigma$, is defined as minus the variation of the action with respect to a unit
increase in proper time separation between
$\partial \Sigma$ and its neighboring $D-2$ surface, as measured orthogonally to $\Sigma$
at $\partial \Sigma$. This basically measures
the variational rate of the action on $B$. In order to define this splitting of spacetime
we will use the ADM \cite{ADM}
decomposition which has a global splitting (at least in the absence of singularities or null
surfaces). Here we can naturally define an
Hamiltonian and therefore an energy.

The spacetime metric is $g_{\mu\nu}$ and $n^\mu$ is the outward pointing spacelike unit
normal to the boundary $B$. The metric and
the extrinsic curvature of $B$ are denoted respectively by $\gamma_{\mu\nu}$ and
$\Theta_{\mu\nu}$.
Now denote by $u^\mu$ the future pointing timelike unit normal to a family of spacelike
hypersurfaces $\Sigma$ that
foliate the spacetime. The metric and the extrinsic curvature of $\Sigma$ are given by the
spacetime tensors $h_{\mu\nu}$ and
$K_{\mu\nu}$, respectively. $h^\mu_\nu$ is also the projection tensor on $\Sigma$. The
spatial coordinates $i,j,...=0,...,D-2$ are adapted
coordinates on $\Sigma$. We then define the momentum $P^{ij}$ conjugate to the spatial
metric $h_{ij}$. The ADM decomposition is
simply
\begin{equation}
ds^2=g_{\mu\nu}dx^\mu dx^\nu=-N^2dt^2+h_{ij}(dx^i+V^idt)(dx^j+V^jdt),
\end{equation}
where $N$ is the lapse function and $V^i$
the shift vector.

For our purposes we use the fact that the foliation $\Sigma$ is orthogonal to $B$, which implies
$(u\cdot n)\Big|_B=0$. Because at this restriction,
the metric at $B$ can be decomposed as
\begin{equation}
\gamma_{\mu\nu}dx^\mu dx^\nu=-N^2dt^2+\sigma_{ab}(dx^a+V^adt)(dx^b+V^bdt),
\end{equation}
where $a,b,...=0,...,D-3$ are adapted coordinates on $\partial \Sigma$.

For general relativity coupled to matter, consider first the action suitable for fixation
of the metric on the boundary \cite{York}
\begin{equation}
S=\frac{1}{2 \kappa}\int_M d^Dx \sqrt{-g} R+\frac{1}{\kappa}\int^{t_2}_{t_1}d^{D-1}x\sqrt{h}K-
\frac{1}{\kappa}\int_Bd^{D-1}x\sqrt{-\gamma}\Theta+S^m\ ,
\end{equation}
where $S^m$ is the matter action, including a possible cosmological term. Now the variation
of this action gives
\begin{eqnarray}
\delta S&=&(\mbox{terms giving the equations of motion})\cr
&+&(\mbox{boundary terms coming from the matter action})\cr &+&\int^{t_2}_{t_1}d^{D-1}
x P^{ij}\delta h_{ij}+\int_Bd^{D-1}x\pi^{ij}\delta\gamma_{ij}.
\end{eqnarray}
Here $\pi^{ij}$ is the conjugate momentum to $\gamma_{ij}$. We assume that the
matter action contains no derivatives of the metric. For the gravitational variables, the
boundary three-metric $\gamma_{ij}$ is
fixed on $B$, and the hypersurface metric $h_{ij}$ is fixed in $t_1$ and $t_2$. Since we are
interested in the variation of the action,
this will have an ambiguity in its definition. The ambiguity in $S$ is taken into account by
subtracting an arbitrary function of the
fixed boundary data. Thus we define the action
\begin{equation}
A=S-S^0,
\end{equation}
where $S^0$ is a functional of $\gamma_{ij}$. The variation in $A$ just differs from the
variation of $S$ by the term:
\begin{equation}\label{S0}
-\delta S^0=-\int_B d^{D-1}x\frac{\delta S^0}{\delta \gamma_{ij}}\delta\gamma_{ij}=-
\int_Bd^{D-1}x\pi_0^{ij}\delta \gamma_{ij},
\end{equation}
where it is clear that $\pi_0^{ij}$ is a function only of $\gamma_{ij}$. We call the
classical action $S_{cl}$ the action evaluated
on the classical solution of $A$. This will be of course a functional of the fixed boundary
data consisting of $\gamma_{ij}$,
$h_{ij}(t_1)$, $h_{ij}(t_2)$ and the matter fields. Then the variation of $S_{cl}$ among the
possible classical solutions give
\begin{eqnarray}
&\delta S_{cl}=(\mbox{terms involving variations in the matter fields})\cr \nonumber
&+\int^{t_2}_{t_1}d^{D-1}
x P^{ij}_{cl}\delta h_{ij}+\int_Bd^{D-1}x(\pi^{ij}_{cl}-\pi_{0}^{ij})\delta\gamma_{ij}.
\end{eqnarray}
The generalization of the Hamilton-Jacobi equation
for the momentum are the equations
\begin{equation}
P^{ij}_{cl}\Big|_{t_2}=\frac{\delta S_{cl}}{\delta h_{ij}(t_2)}.
\end{equation}
The generalization of the energy equation is
\begin{equation}\label{BY}
T^{ij}=\frac{2}{\sqrt{-\gamma}}\frac{\delta S_{cl}}{\delta\gamma_{ij}}\ ,
\end{equation}
which is the quasi-local stress-energy tensor for gravity.

\subsubsection{Holographic zero point energy of AdS}

In general, if the spacetime is not asymptotically flat the Brown-York tensor (\ref{BY})
diverges at infinity. For asymptotically AdS spacetime,
there is a resolution of this difficulty. If we believe there is an
holographic duality between AdS and CFT, the
fact that the Brown-York tensor diverges, it can be interpreted as the standard ultraviolet
divergences of
quantum field theory, and may be classically removed
by adding local counter-terms to the action. These subtractions depend only on the intrinsic
geometry of the boundary. Once
we
renormalize the energy in this way, we should be able via classical and quantum renormalization
schemes to obtain the same
zero point energy of the spacetime. The idea is to renormalize the stress-energy tensor by
adding a finite
series
of boundary invariants to the classical action.
The essential terms are fixed uniquely by requiring finiteness of the stress tensor. For
simplicity we will work it out in three dimensions,
but this is true in higher dimensions as well \cite{BK}.
In particular, global $\mbox{AdS}_5$, with an $S^3\times R$ boundary, has a positive
mass \cite{positive}.
This result is beautifully explained via the proposed
duality with a boundary CFT. The dual super Yang-Mills (SYM) theory on a sphere has a Casimir
 energy that precisely matches this spacetime mass.

The action we are considering now is the three-dimensional Einstein-Hilbert with negative
cosmological constant $\Lambda=-1/l^2$
and boundary terms
\begin{equation}
S=\frac{1}{2 \kappa}\int_M d^3x \sqrt{-g} \left(R-\frac{1}{l^2}\right)+
\frac{1}{\kappa}\int^{t_2}_{t_1}d^2x\sqrt{h}K-
\frac{1}{\kappa}\int_Bd^2x(\sqrt{-\gamma}\Theta)+S_{ct},
\end{equation}
where $S_{ct}$ is the counter-term action added in order to obtain a finite stress tensor.
In order
to preserve the equation
of motion, this extra term must be a boundary term. In particular it lives in $B$.
The Brown-York tensor is then
\begin{equation}
T^{ij}=\frac{1}{\kappa}\left[\Theta^{ij}-\Theta\gamma^{ij}+\frac{2}{\sqrt{-\gamma}}\frac{\delta
S_{ct}}{\delta\gamma_{ij}}\right],
\end{equation}
with everything calculated on the classical solution. Since $S_{ct}$ must only
depend on the boundary geometry, it describes a local
gravitational action in two dimensions. In particular, we wish to cancel out the divergences
coming from the negative energy from the
cosmological constant. Therefore the most general form is
\begin{equation}
S_{ct}=-\int_B\frac{1}{l}\sqrt{-\gamma}d^2 x\ \Rightarrow\ T^{ij}=\frac{1}{\kappa}\left[\Theta^{ij}-
\Theta\gamma^{ij}-\frac{1}{l}\gamma^{ij}\right].
\end{equation}
Now consider $\mbox{AdS}_3$ spacetime in light-cone coordinates
\begin{equation}\label{ads}
ds^2=\frac{l^2}{r^2}dr^2-r^2dx^+dx^-.
\end{equation}
In this case $T^{ij}=0$. What we would like to prove is that a dual CFT is living on the
surface $ds^2=-r^2dx^+dx^-$ with $r$
eventually taken to infinity (AdS boundary). If this is true, asymptotically conformal
perturbations of AdS
correspond to excitations of CFT fields on its boundary.
In particular, on the CFT side even if classically $T^i_i=0$, because quantum excitation
of the vacuum, $\gamma^{ij}<T_{ij}>\neq 0$. The scale of
this energy is linked to the renormalization scale. Connecting it with the
gravitational scale $l$ of AdS, we identify the Brown-York tensor on the boundary
of AdS as corresponding to the quantum CFT energy-momentum tensor on flat background.
In \cite{Brohe} Brown and Henneaux proved that a perturbation of the AdS metric with
asymptotic
AdS behaviour must have the following expansion
\begin{eqnarray}
\delta g_{+-}=O(1),\ \ \delta g_{++}=O(1),\ \ \delta g_{--}=O(1)\\
\delta g_{rr}=O\left(\frac{1}{r^4}\right),\ \ \delta g_{+r}=O\left(\frac{1}{r^3}\right),\ \
\delta g_{-r}=
O\left(\frac{1}{r^3}\right).
\end{eqnarray}
We rewrite this expansion using the following diffeomorphism
\begin{eqnarray}
x^+\rightarrow x^+-\xi^+-\frac{l^2}{2r^2}\partial^2_-\xi^-\ ,\label{diff1}\\
x^-\rightarrow x^--\xi^--\frac{l^2}{2r^2}\partial^2_+\xi^+\ ,\\\
r\rightarrow r+\frac{r}{2}(\partial_+\xi^++\partial_-\xi^-)\ ,\label{diff2}
\end{eqnarray}
where $\xi^\pm=\xi^\pm(x^\pm)$, which yields at
first order
\begin{equation}
ds^2=\frac{l^2}{r^2}dr^2-r^2dx^+dx^--\frac{l^2}{2}\partial^3_+\xi^+ (dx^+)^2-
\frac{l^2}{2}\partial^3_- \xi^-(dx^-)^2.
\end{equation}
Then we obtain the following non-zero components for the Brown-York tensor
\begin{equation}\label{grav}
\delta_{\xi^{\pm}}T_{\pm\pm}=-\frac{l}{2k}\partial_\pm^3\xi^\pm.
\end{equation}
This therefore represents the zero point energy of the boundary of AdS
due to perturbations of this spacetime that preserve the AdS symmetries on the boundary.
Now we study the
CFT side. The boundary of AdS is just a Minkowski spacetime with metric
\begin{equation}
ds^2=-r^2dx^+dx^-.
\end{equation}
A conformal theory is such that under conformal transformation of the type
$g_{\alpha\beta}\rightarrow \Omega \tilde g_{\alpha\beta}$, it is
classically invariant; then its energy momentum tensor satisfies $T^{\alpha\beta}=\tilde
T^{\alpha\beta}$. In particular this means that
classically $T^\alpha_\alpha=0$. If we now introduce a conformal transformation under
the diffeomeorphism (\ref{diff1}-\ref{diff2}), when
$r\rightarrow \infty$ we get quantum mechanically (starting again from a zero energy
momentum tensor)
\footnote{It
is beyond the scope of this thesis to prove this result; a very good introduction can be found
in \cite{ginsparg}.}
\begin{equation}\label{cft}
\delta_{\xi^\pm} T_{\pm\pm}=-\frac{c}{24\pi}\partial_\pm^3\xi^\pm.
\end{equation}
This term comes from the commutation rules.
In particular since this does not depend on the particular
energy-momentum tensor used, it is called the zero point energy.
The constant $c$ is called the
central charge and encode the renormalization scale of
the theory. Surprisingly eqs. (\ref{grav}) and (\ref{cft}) coincide if we measure
the scale of the renormalization with the AdS
scale, or $c=12\pi l/\kappa$. This means that the quantum vacuum energy, due
to the excitation of the fields in the CFT side, is nothing else than the residual
energy due to classical gravitational perturbations of AdS that preserve the AdS boundary symmetries.

\newpage

\chapter{A dynamical alternative to compactification}

The possibility that spacetime has more than four dimensions is an
old idea. Originally the extra dimensional spaces were
thought to be compactified. This kind of model is usually
called Kaluza-Klein compactification (see for example \cite{KK}).
Here, to have a correct Newtonian and Standard Model limit, the
size of the extra dimensions must be less than the Electroweak
scale ($ \sim 1\ \mbox{TeV}^{-1}$). These theories were introduced to
incorporate geometrically scalar and vector fields in the
projected four-dimensional gravity, thus giving a geometrical interpretation of electromagnetism.

Later, extra dimensions arose very naturally in string theory and associated phenomenology.

In this thesis I will discuss only models with an infinitely large extra
dimension. In particular, I am interested in models where
gravity is dynamically localized at low energies \cite{RS}.
Another type of localization is also possible \cite{DGP,GRS}, in which
gravity has the correct Newtonian limit
at short distances. This arises from introducing an induced
gravity on the four-dimensional submanifold. In this way, at short
distances, the induced effects dominate, generating an effective
four-dimensional Einstein gravity consistent with Newtonian gravity
experiments. At large distances, by contrast, the
extra-dimensional gravity becomes more and more important. This
modifies, at large distances, the effective four-dimensional
gravity.

The fact that particles can be trapped gravitationally in a submanifold,
was argued already by Visser \cite{Visser}. He realized that
a particle living in a five-dimensional spacetime with warped time component,
\begin{equation}\label{w1}
ds^2=-e^{-2\Phi(\xi)}dt^2+d{\vec x}\cdot d\vec x+d\xi^2,
\end{equation}
($\xi$ is the extra-dimensional coordinate) is exponentially
trapped by the scalar field $\Phi(\xi)>0$. In 1999 Randall and
Sundrum \cite{RS} used a similar mechanism to localize gravity. In
order to do that, the whole four-dimensional metric must be warped.
In this way a curved background supports a ``bound state'' of the
higher-dimensional graviton with respect to the co-dimensions.
So although space is indeed infinite in extent, the
graviton is confined to a small region within this space.

Suppose for simplicity we have only one extra dimension
\footnote{The mechanism is general even with more than one co-dimension.}.
Moreover suppose the five-dimensional gravity is governed by the Einstein-Hilbert action.
We then consider a domain wall with
vacuum energy $\lambda$ where, as a macroscopic approximation, all the matter fields live.
The action is then
\begin{equation}\label{act}
S=\frac{1}{\tilde\kappa^2}\int d^5x\sqrt{-\tilde
g}\{\tilde R-2\tilde\Lambda\}+\int d^4x\sqrt{-g}\{{\cal L}_m-2\lambda\},
\end{equation}
where $\tilde \kappa$ and $\tilde g_{AB}$ are respectively the five
dimensional Planck mass and metric, $\tilde\Lambda$ and $\tilde R$ are the five
dimensional negative cosmological constant and Ricci scalar, ${\cal L}_m$ and
$g_{\mu\nu}$ are the four-dimensional Lagrangian for matter and
the four-dimensional metric.  The three-dimensional submanifold is
called the ``brane'' and the total spacetime will be
called the ``bulk''. We use the notation $A,B,...=0,...,4$ for
the bulk coordinates and $\mu,\nu,...=0,...,3$ for the brane
coordinates. The role of the five-dimensional cosmological
constant is very important. Physically, we can understand that
such a term will induce a positive ``pressure''. This localizes the
gravitons produced in the four-dimensional brane on the brane
itself. Indeed, as we shall see, the warp factor
is produced by the cosmological constant.

In order to simplify the system we introduce
$Z_2$-symmetry in the co-dimension. This can be also motivated from the orbifold structure
appearing in M-theory to
incorporate heterotic string theory. In this way M-theory contains the
gauge groups of the standard model of particle physics \cite{HW}.

We start with an empty brane, ${\cal L}_m=0$. We would like to find a warped
solution which has, as a slice, a Minkowski
spacetime. In the next section we will introduce the
general formalism in the presence of matter, then we will study small perturbations
around Minkowski, obtaining the correct Newtonian limit.
We start with the metric:
\begin{equation}\label{ansatz}
ds^2=e^{-2\sigma(y)}\gamma_{\mu\nu}(x^\alpha)dx^\mu dx^\nu+dy^2,
\end{equation}
where $y$ is the extra-dimensional coordinate and
$\gamma_{\mu\nu}(x^\alpha)$ is the four-dimensional metric. In order to
solve the variational problem $\delta S=0$, we have two ways. The first one is
to solve the Einstein equations together with junction
conditions between the brane and the bulk. Since the brane is a singular object, with this
approach we need to involve the theory of
distributions. We will discuss this technique in the next section and study it extensively
in the last chapter and Appendix \ref{Appjc}.
Here we follow another way. Instead of introducing a
thin shell (brane), we consider a spacetime with a boundary at
$y=0$. This means that we will allow the four-dimensional
metric to vary only on the boundary, whereas we will keep it fixed in the
bulk as Minkowski slices, in such a way that the global spacetime is AdS.
If we would like to
implement the $Z_2$ symmetry, we can just make a copy of the
same spacetime on the two sides of the boundary. In this way we
don't have to deal with distributions. Now if we substitute the
ansatz (\ref{ansatz}) into the action (\ref{act}), we are left
with a variational problem for the scalar field $\sigma(y)$ in the
bulk and a variational problem for the metric $\gamma_{\mu\nu}$ on
the brane. Then we use the constraint
$\gamma_{\mu\nu}=\eta_{\mu\nu}$. The
Lagrangian for gravity will be \cite{Wald}\footnote{I use the notation $\partial_y f= f'$.}
\begin{equation}
\sqrt{-\tilde g}\tilde R=12e^{-4\sigma}(\sigma')^2+2(K\sqrt{-g})'+e^{-2\sigma}{\cal R}\ ,
\end{equation}
where ${\cal R}$ is the Ricci scalar associated with $\gamma_{\mu\nu}$ and $K={1\over 2}
g^{\mu\nu}\pounds_n g_{\mu\nu}$,
is the trace of the extrinsic curvature orthogonal to the brane. Using this notation we get
\begin{equation}
\int d^5x\sqrt{-\tilde g}\tilde R=12\int d^5x \{e^{-4\sigma}(\sigma')^2+e^{-2\sigma}
{\cal R}\}+2\int d^4x K \sqrt{-g},
\end{equation}
where the boundary is at $y=0$. The last term is called the Gibbons-Hawking term \cite{GW}.

Since the degrees of freedom of the four-dimensional Ricci scalar and the degrees of freedom
of the scalar field $\sigma$ are now
completely separate, we can set ${\cal R}=0$ before the variation.
Considering only the first integral, since $\delta \sigma=0$ on the boundary we have
\begin{equation}
12\delta\int d^5x e^{-4\sigma}(\sigma')^2=-12\int d^5x \{2\sigma''-(4\sigma')^2\}e^{-4\sigma}
\delta\sigma.
\end{equation}
The variation of the five dimensional cosmological constant is
\begin{equation}
-2\delta\int d^5x\sqrt{-\tilde g}\tilde\Lambda=12\int d^5x \frac{2}{3}\tilde\Lambda
e^{-4\sigma}\delta\sigma.
\end{equation}
Combining the two variations we get the equation
\begin{equation}
\sigma''-2(\sigma')^2=\frac{\tilde\Lambda}{3}.
\end{equation}
This equation has two separate branches, one is the trivial
\begin{equation}
\sigma'=\pm\sqrt{-\frac{\tilde\Lambda}{6}}\ ,
\end{equation}
and the other is
\begin{equation}
\sigma'=\pm\sqrt{1+\tanh \left(2\sqrt{-\frac{\tilde\Lambda}{6}}y+C\right)}.
\end{equation}
To implement the $Z_2$-symmetry we have to cut and paste the solution from one side to the
other of the boundary. The second solution does not have parity. Moreover, the Einstein tensor
describing this system contains another constraint equation
hidden in the integration process we are using here.
One can show that the second solution does not satisfy this additional constraint (see the $\tilde G_{yy}$ equation in \ref{second}
with $e^{-\sigma(y)}=a(y)$, and considering the static case).
This implies that we can only use the first solution
\begin{equation}\label{jcRS}
\sigma'=\sqrt{-\frac{\tilde\Lambda}{6}}\mbox{sgn}(y).
\end{equation}
Here we introduced the sign function $\mbox{sgn}(y)$, in order to describe the full spacetime.
In particular we choose the positive
branch of the square root. This is because we want a decreasing warp factor.
Now we have to make the variations of the boundary action with
respect to $g_{\mu\nu}$ vanish, keeping $\partial_y g_{\mu\nu}$ fixed. We have
the combination
\begin{equation}
\frac{1}{\tilde\kappa^2}\int
d^4x\sqrt{-g}\left\{ K_{\mu\nu}-g_{\mu\nu}K-\frac{\tilde\kappa^2\lambda}{2}
g_{\mu\nu}\right\} \delta g_{\mu\nu},
\end{equation}
using the constraint $\gamma_{\mu\nu}=\eta_{\mu\nu}$, this boundary action vanishes for
\begin{equation}\label{jcRS2}
\sigma'(0)=\frac{\tilde\kappa^2\lambda}{6}.
\end{equation}
Combining (\ref{jcRS}) with (\ref{jcRS2}) we get the fine tuning
\begin{equation}\label{min}
\frac{\tilde \kappa^4\lambda^2}{6}+\tilde\Lambda=0.
\end{equation}
We can now write down the global geometry. This is an AdS spacetime in the bulk and
a Minkowski
spacetime on the brane:
\begin{equation}\label{Minkmetric}
ds^2=e^{-2\sqrt{-\tilde\Lambda/6}\mid y\mid}\eta_{\mu\nu}dx^\mu dx^\nu+dy^2.
\end{equation}
In the next paragraph we show how
generalize this model in the presence of matter.
Moreover we will see that if the fine tuning (\ref{min}) does not hold, an
effective four-dimensional cosmological constant appears.

\section{The Randall-Sundrum braneworld scenario}
The mechanism provided by Randall-Sundrum can be generalized to include
matter. We will follow
\cite{SMS}. In this work a geometrical approach is used. We suppose that
the bulk spacetime is governed by the Einstein
equations and then we impose the junction conditions.

The normal to the brane is $n^A$. The induced metric on
the brane is $g_{AB}= \tilde g_{AB}-n_A n_B$. The Gauss equation
which relates the $D-$dimensional Riemann tensor to the
$(D-1)$-dimensional one is\footnote{I will use the notation with
tilde for five-dimensional objects and without for
four-dimensional ones.}
\begin{equation}\label{gauss}
R^A{}_{BCD}=\five R^E{}_{FGH}\ g^{A}_{~E}\ g^{F}_{~B}\ g^{G}_{~C}\ g^{H}_{~D}\ +K^A{}_C K_{BD}
-K^A{}_D K_{BC}
\end{equation}
where
\begin{equation}
K_{AB}=\frac{1}{2}\pounds_n \tilde g_{AB}
\end{equation}
is the extrinsic curvature of the brane. Then we have the Codacci equations
\begin{equation}\label{codacci}
g^{B}_{~C}\nabla_B K^{C}{}_A- g^{B}_{~A}\nabla_B K=\five R_{CD}n^C g^{D}_{~A},
\end{equation}
where we use the notation $K^A{}_A=K$. From the Gauss equation we
can build up the four-dimensional Einstein tensor
\ba
G_{AB}&=&\five G_{CD}\ g^C_{~A}\ g^D_{~B}\ +\five R_{CD}\ n^C\ n^D\ g_{AB}+KK_{AB}\cr
&-&K^C{}_A\ K_{BC}-
\frac{1}{2}g_{AB}(K^2-K^{AB}\ K_{AB})-\tilde
E_{AB}\ ,
\ea
where we introduced
\begin{equation}
\tilde E_{AB}\equiv\five R^C{}_{DFG}\ n_C\ n^F\ g^{D}_{~A}\ g^{G}_{~B}.
\end{equation}
Now we require that the bulk spacetime is governed by the five-dimensional Einstein equations
\begin{equation}\label{einstein5}
\tilde G_{AB}=\tilde\kappa^2 \tilde T_{AB}\ ,
\end{equation}
where $\tilde T_{AB}$ is the five dimensional energy-momentum tensor. We
then use the decomposition of the Riemann tensor
into Weyl curvature, Ricci tensor and scalar curvature,
\begin{equation}\label{decRie}
\five R_{ABCD}=\frac{2}{3}\left(\tilde g_{A[C}\five R_{D]B}-\tilde g_{B[C}
\five R_{D]A}\right)-\frac{1}{6}\tilde g_{A[C}\tilde g_{D]B}\five R+
\five C_{ABCD}\ ,
\end{equation}
where $\tilde C_{ABCD}$ is the Weyl tensor or the traceless part
of the Riemann tensor. Finally the four-dimensional Einstein
tensor in adapted coordinates to the brane is
\ba
G_{\mu\nu}
&=& {2 \tilde \kappa^2 \over 3}\left[T_{\mu\nu}
+\left(T_{\rho\sigma}n^\rho n^\sigma-{1 \over 4}T^\rho_{~\rho}\right)
 g_{\mu\nu} \right]\cr
&+& KK_{\mu\nu}
-K^{~\sigma}_{\mu}K_{\nu\sigma} -{1 \over 2}g_{\mu\nu}
  \left(K^2-K^{\alpha\beta}K_{\alpha\beta}\right) - \tilde {\cal E}_{\mu\nu},
\label{4dEinstein}
\ea
where
\begin{equation}
\tilde {\cal E}_{AB} \equiv \tilde C^C_{~DEF}n_C n^E\
g_A^{~D}\ g_B^{~F},
\end{equation}
and it is traceless. From the Codacci equations (\ref{codacci}) and the five
dimensional Einstein equations (\ref{einstein5}) we get
\begin{equation}\label{quasicon}
g^{~B}_C\nabla_B K^{~C}_A- g^{~B}_A\nabla_B K=\five T_{CD}n^C g^{~D}_A.
\end{equation}
Up to this point we have not introduced a braneworld. A braneworld is
a four-dimensional hypersurface which can be described by the
equation $y=0$ where $y$ is the extra-dimensional
coordinate. Close to the brane we can use Gaussian normal
coordinates such that the metric has the form
\begin{equation}
ds^2=g_{\mu\nu}dx^\mu dx^\nu+dy^2.
\end{equation}
The energy momentum tensor is
\begin{equation}
\tilde T_{AB}=-\frac{\tilde\Lambda}{\tilde \kappa^2} g_{AB}+\delta(y)\left(-\lambda g_{AB}+
T_{AB}\right)\ ,
\end{equation}
where we assume that, at macroscopic level, the matter ($T_{AB}$), is defined only on the brane.
This is
mathematically achieved by introducing a
Dirac distribution $\delta(y)$. $\lambda$ is the vacuum energy of the brane which
coincides with its tension for $T_{AB}=0$\footnote{The tension of a hypersurface is defined as its
extrinsic curvature.}. In particular
we also suppose that there is not a energy-momentum exchange between bulk and brane, or
\begin{equation}
T_{AB}\ n^A=0.
\end{equation}
As I will show in Appendix \ref{Appjc}, reasonable junction conditions require that the singular
behaviour of the metric is encoded in the singular behaviour of the Lie
derivative of the extrinsic curvature along the normal. From that one can get
Israel's junction conditions \cite{israel},
\begin{eqnarray} \label{jcsms}
\left[g_{\mu\nu}\right]
&=&0\,,
\nonumber\\
\left[K_{\mu\nu}\right]&=& -\tilde\kappa^2
\Bigl(T_{\mu\nu}-\frac{1}{3}g_{\mu\nu}(T-\lambda)\Bigr),
\end{eqnarray}
where $[X]:=\lim_{y \to +0}X - \lim_{y \to -0}X=X^+-X^-$.
Substituting then the junction conditions (\ref{jcsms}) into (\ref{4dEinstein}), we get the
effective four-dimensional gravitational
equations
\begin{eqnarray}
G_{\mu\nu}=-\Lambda g_{\mu\nu}
+ 8 \pi G_N T_{\mu\nu}+\frac{48\pi G_N}{\lambda}\,S_{\mu\nu}
-{\cal E}_{\mu\nu}\,, \label{eq:effective}
\end{eqnarray}
where
\begin{eqnarray}
{\cal E}_{\mu\nu}&=&{1\over 2}\left[\tilde {\cal E}_{\mu\nu}\right]\,,\\
\Lambda &=&\frac{1}{2}\left(\tilde\Lambda
+\frac{1}{6}\tilde\kappa^4\,\lambda^2\right)\,,
\label{Lamda4}\\
G_N&=&{\tilde \kappa^4\,\lambda\over48 \pi}\,,
\label{GNdef}\\
S_{\mu\nu}&=&
-\frac{1}{4} T_{\mu\alpha}T_\nu^{~\alpha}
+\frac{1}{12}T T_{\mu\nu}
+\frac{1}{8}g_{\mu\nu}T_{\alpha\beta}T^{\alpha\beta}-\frac{1}{24}
g_{\mu\nu}T^2\ .
\label{pidef}
\end{eqnarray}
From the effective equations (\ref{eq:effective}), the localization of gravity is clearly
obtainable in the limit
$\lambda\rightarrow\infty$, $\tilde\kappa\rightarrow 0$, such that $G_N$ remains finite,
and ${\cal E}_{\mu\nu}\rightarrow 0$. Indeed
in this limit we recover Einsteinian gravity. The fact that $G_N>0$ constrains
$\lambda$ to be positive \cite{CGKT}.

We now study the matter conservation
equations. Since $T_{AB}n^A=0$ from (\ref{quasicon}) and using the
junction conditions
\begin{equation}
\nabla^\mu T_{\mu\nu}=0.
\end{equation}
This implies that matter follows the standard conservation equations of general relativity.
Thanks to the
four-dimensional Bianchi identities we get differential equations for the projection of the
Weyl tensor, ${\cal E}_{\mu\nu}$. These effective gravitational
equations do not in general close the system.
Indeed the full determination of the projected Weyl tensor requires in general the study of the
five-dimensional gravitational problem.
As we will see, we can sometimes
get around that by introducing particular symmetries for the metric. Then this formalism
becomes a very
powerful tool.

\section{1+3 formalism in the braneworld}
In the last section we commented that the four-dimensional effective equations are not in
general closed. This is because the tensor ${\cal E}_
{\mu\nu}$ is not in general completely constrained by the projected equations. However,
often four-dimensional symmetries constrain sufficiently
the five-dimensional geometry, and families of solutions can be found.
To understand better this concept it is very useful to
use the formalism developed by Maartens (see for example \cite{maacov}) which I explain in the
following. In particular since we will be
interested only in perfect fluids, we can drastically simplify all the equations, using the
perfect fluid energy-momentum
tensor
\begin{equation}\label{decT}
T_{\mu\nu}=\rho u_\mu u_\nu+ p h_{\mu\nu},
\end{equation}
where $u^\mu$ is the unit four-velocity of the matter ($u^\mu u_\mu=-1$) and $h_{\mu\nu}$
is the space-like metric that projects
orthogonal to $u^\mu$ ($h_{\mu\nu}=g_{\mu\nu}+u_\mu u_\nu$). $\rho$
and $p$ are respectively the energy density
and the pressure of the perfect fluid.
In the same way as (\ref{decT}), we decompose the local correction $S_{\mu\nu}$ and the
non-local one ${\cal E}_{\mu\nu}$. ${\cal E}_{\mu\nu}$
gives a non-local contribution in the sense that it codifies all the bulk geometrical
back-reaction on the brane.
The decomposition of the matter-correction is
\be
S_{\mu\nu}={{1\over12}}\rho\left[\rho u_\mu u_\nu +\left(\rho+2
p\right)h_{\mu\nu}\right]\,.
 \ee
Using the fact that ${\cal E}_{\mu\nu}$ is traceless we can decompose it as
\footnote{ $\kappa^2=8\pi G_N$ .}
 \be
-{1\over \kappa^2} {\cal E}_{\mu\nu} = \cu\left(u_\mu u_\nu+{ {1\over3}}
h_{\mu\nu}\right)+ {\cq_\mu} u_{\nu} + {\cq_\nu} u_{\mu}+\cp_{\mu\nu}\, ,
 \ee
where we introduced an effective ``dark" radiative
energy-momentum on the brane, with energy density $\cal U$,
pressure ${\cal U}/3$, momentum density $Q_\mu$ and anisotropic
stress ${\Pi}_{\mu\nu}$.

The brane-world corrections can conveniently be consolidated into
an effective total energy density, pressure, momentum density and
anisotropic stress \cite{m}:
\begin{eqnarray}
\rho^{\rm eff} &=& \rho\left(1 +\frac{\rho}{2\lambda} +
\frac{\cal U}{\rho} \right)\,,\label{reff} \\ \label{peff}
p^{\rm eff } &=& p  + \frac{\rho}{2\lambda}
(2p+\rho)+\frac{\cal U}{3}\;, \\ q^{\rm eff }_\mu &=& Q_\mu\;, \\
\label{e:pressure2} \pi^{\rm eff }_{\mu\nu} &=& \Pi_{\mu\nu}\;.
\end{eqnarray}
Note that nonlocal bulk effects can contribute to effective
imperfect fluid terms even when the matter on the brane has
perfect fluid form: there is in general an effective momentum
density and anisotropic stress induced on the brane by
the 5D graviton.

The effective total equation of state and sound speed follow from
eqs.~(\ref{reff}) and (\ref{peff}) as
 \bea
w^{\rm eff} &\equiv & {p^{\rm eff}\over\rho^{\rm eff}} =
{w+(1+2w)\rho/2\lambda+ \cu/3\rho \over 1+\rho/2\lambda +\cu/\rho}
\,,\label{vh1}\\ c_{\rm eff}^2 &\equiv & {\dot{p}^{\rm
eff}\over\dot{\rho}^{\rm eff}}= \left[c_{\rm s}^2+{\rho+p \over
\rho+\lambda} +{4\cu\over 9(\rho+p)(1+\rho/\lambda)}\right]
\left[1+ {4\cu\over 3(\rho+p)(1+\rho/\lambda)}\right]^{-1}
\, \label{vh2}
 \eea
where $w=p/\rho$ and $c_{\rm s}^2=\dot p/\dot\rho$. We also used the notation
$\dot f=d f/d\tau$ where $\tau$ is the proper
time of the perfect fluid.

\subsection{Conservation equations}
${\cal E}_{\mu\nu}$, the projection of the bulk Weyl tensor on the
brane, encodes corrections from the 5D graviton effects (often called Kaluza-Klein or KK
modes). From
the brane-observer viewpoint, the energy-momentum corrections in
$S_{\mu\nu}$ are local, whereas the KK corrections in ${\cal
E}_{\mu\nu}$ are nonlocal, since they incorporate 5D gravity wave
modes. These nonlocal corrections cannot be determined purely from
data on the brane. In the perturbative analysis
which leads
to the corrections in the gravitational potential,
the KK modes that generate this correction are
responsible for a nonzero ${\cal E}_{\mu\nu}$; this term is what
carries the modification to the weak-field field equations. An equivalent picture is that
these modes arise as a geometrical bulk
back-reaction to the variations of
the matter fields on the brane. We can see how the matter can source these modes as follows.

The standard conservation equations
\begin{equation}
\nabla^\mu T_{\mu\nu}=0
\end{equation}
together with the four-dimensional Bianchi identities applied to the effective
four-dimensional gravitational equations
(\ref{eq:effective}) lead to the following non-local equations
\begin{equation}\label{non-local}
\nabla^\mu {\cal E}_{\mu\nu}=\frac{6 \kappa^2}{\lambda}\nabla^\mu S_{\mu\nu}.
\end{equation}
It is clear that the projection of the Weyl tensor, which encodes the curvature of
the co-dimension, can be sourced (as
a back-reaction) by the variation of the matter fields.

It is useful to rewrite the local and non-local conservation equation in a $1+3$ formalism.
Introducing the covariant projected Levi-Civita
tensor $\ep_{abc}$, the spatial covariant
derivative $D_a$ (where $D_a S^{b...}{}_{...c}=h^e{}_a h^b{}_f...h^g{}_c \nabla_e S^{f...}
{}_{...g}$),
the volume expansion $\Theta=\nabla^\alpha u_\alpha$, the
proper time derivative $\dot S^{a...}{}_{...b}=u^\alpha
\nabla_\alpha S^{a...}{}_{...b}$, the acceleration $A_a=\dot u_a$, the shear
$\sigma_{ab}=D_{(a} u_{b)}-\Theta/3 h_{ab}$ and the
vorticity $\omega_a=-{1 \over 2}\mbox{curl}\
u_a$, we get the local conservation equations
\begin{eqnarray}
&&\dot{\rho}+\Theta(\rho+p)=0\,,\label{pc1}\\ && \D_ a
p+(\rho+p)A_ a =0\,,\label{pc2}
\end{eqnarray}
and the nonlocal equations
\begin{eqnarray}
&& \dot{\cu}+{{4\over3}}\Theta{\cu}+\D^ a\cq_ a+2A^ a\cq_
a+\sigma^{ ab }\cp_{ ab }=0\,, \label{pc1'}\\&& \dot\cq_{
a}+{{4\over3}}\Theta\cq_ a +{{1\over3}}\D_
a{\cu}+{{4\over3}}{\cu}A_ a +\D^ b\cp_{ ab }+A^ b\cp_{ ab
}+\sigma_{ a}{}^b\cq_b-\ep_{a}{}^{bc}\omega_b\cq_c
\nonumber\\&&~~{} =-{(\rho+p)\over\lambda} \D_a\rho\,.\label{pc2'}
\end{eqnarray}
The non-local equations do not contain evolution equations for the anisotropic part of the
projected Weyl tensor $\Pi_{ab}$.
This makes the system of ``brane'' equations not closed. However if we
ask for spatially homogeneous and isotropic solutions, the system becomes closed.

\section{Newtonian limit: localization of gravity}\label{secNewton}

Equations (\ref{eq:effective})
show that in the limit $\lambda\rightarrow \infty$ and in the case of a
bulk with ${\cal E}_{\mu\nu}=0$,
we recover
General Relativity. In this section we discuss the correction to the Newtonian
potential due to a point-like particle. In order to do that we have to study the full five
dimensional
equations because such a perturbation is due to purely
five dimensional effects. In doing that we follow \cite{DD}.

We start by considering the unperturbed Randall-Sundrum braneworld.
The solution for the metric in Gaussian normal
coordinates is given by eq. (\ref{Minkmetric}), i.e.
\be
ds^2=dy^2+e^{-2\mid y\mid/l}\eta_{\mu\nu}dx^\mu dx^\nu\ ,
\ee
where ${l}$ is the curvature scale of the AdS spacetime, ${l}=\sqrt{-6/\tilde \Lambda}$.
In order to solve the perturbations of this metric it is
better to use a conformally Minkowskian coordinate system.
Consider only one side of the spacetime, say $y\geq 0$. In this
side, we can use the transformation of coordinates
\be
e^{y/l}dy=dz,
\ee
and then the metric transforms to
\be
ds^2=\left(\frac{l}{z}\right)^2\eta_{AB}dx^A dx^B.
\ee
In these coordinates the brane is at $z={l}$.

We now perturb the Minkowskian metric. We also expect that,
due to these perturbations, the position of the brane will change, generating a ``bending''.
The metric is
\be\label{metan}
ds^2=\left(\frac{l}{z}\right)^2\left(\eta_{AB}+\gamma_{AB}\right)dx^A dx^B,
\ee
where $\gamma_{AB}$ is a perturbation and the position of the brane is at
\be
z={l}+\xi,
\ee
where $\xi\ll {l}$ is a function describing the bending. We have now the freedom
to choose the transversal gauge
$\gamma_{Az}=0$, and in addition we can use the traceless gauge
$\gamma^\mu{}_\mu=\partial_\rho \gamma^\rho{}_\mu=0$.
In particular we expect that the bending function will depend only on the four-dimensional
brane coordinates, since the bending is due
to a four-dimensional perturbation, therefore we require $\partial_z \xi=0$. Under these
conditions, the unit normal to the brane
is
\be
n^A=-\frac{z}{l}\left(\delta^A_z-\delta^A_\mu\partial^\mu \xi\right).
\ee
From this we can calculate the extrinsic curvature to first order in perturbations
\footnote{$\partial_{\mu\nu}=\partial_\mu
\partial_\nu$.}
\be K^A{}_B=-\frac{1}{2} \tilde g ^{AC}\pounds_n g_{CB}\simeq \delta^\nu{}_B
\delta^A{}_\mu\left(\frac{1}{l}\eta^\mu{}_\nu-\partial^\mu{}_\nu\xi
-\frac{1}{2}\partial_z \gamma^\mu{}_\nu\right). \ee From the
junction conditions (\ref{jcsms}) and using the fine tuning
(\ref{min}) we get on the brane $\Sigma$
\be\label{pluba}
\frac{\tilde\kappa^2}{2}\left(T_{\mu\nu}-{1 \over 3}
g_{\mu\nu}T\right)=\left(\partial_{\mu\nu} \xi-\frac{1}{2}\partial_z
\gamma_{\mu\nu}\right)\Big|_\Sigma. \ee
Taking now the trace and using the
traceless condition for the perturbation we get
\be
-\frac{\tilde\kappa^2}{6} T=\Box_4 \xi\ . \ee
Using a pointlike
particle with mass $M$ \be T_{00}=M\delta(\vec r),\ \ \ \
T_{0i}=T_{ij}=0, \ee we obtain \be\label{xi}
\xi=-\frac{\tilde\kappa^2}{24\pi}\frac{M}{r}. \ee Applying this to
(\ref{pluba}) we get the boundary conditions \be \label{boco}
\partial_z\gamma_{00}\Big|_\Sigma=-{2\tilde\kappa^2 M\over 3}\delta (\vec r)\quad,\quad
\partial_z\gamma_{0i}\Big|_\Sigma=0\quad,\quad
\partial_z\gamma_{ij}\Big|_\Sigma=-{\tilde\kappa^2  M\over 3}\delta (\vec r)\delta_{ij}-
{\tilde\kappa^2 M\over
12\pi}\partial_{ij}{1\over r}\,.
\ee
Now the procedure is to solve the bulk equations and then impose the boundary conditions
(\ref{boco}).
The bulk equations are nothing else than the linearization
of the Einstein equations
\be
\tilde G_{AB}=-\tilde \Lambda \tilde g_{AB},
\ee
that, using the metric ansatz (\ref{metan}), give \cite{5}
\be
\Box_4 \gamma_{\mu\nu}+\partial_z^2\gamma_{\mu\nu}-\frac{3}{z}\partial_z\gamma_{\mu\nu}=0.
\ee
Finally  the general solution of these equations in the static case is a
superposition of Fourier modes~:
\be
\gamma_{\mu\nu}(x^\mu,z)=\int\! {d^3\vec
k\over(2\pi)^{{3\over2}}} e^{i \vec k\,.\vec r}\hat\gamma_{\mu\nu}(\vec k, z)
\ee
with
\be
\hat\gamma_{\mu\nu}(\vec k, z)=z^2
\left[e_{\mu\nu}^{(1)}(\vec k)H_2^{(1)}(ikz)+e_{\mu\nu}^{(2)}(\vec k)H_2^{(2)}(ikz)\right]
\ee
where (because of the traceless conditions) the polarization tensors $e_{\mu\nu}^{(1,2)}
(\vec k)$ are transverse and traceless,
and  where
$H_2^{(1,2)}(ikw)$ are the Hankel functions of first and second kind of order $2$.
The junction conditions (\ref{boco}) therefore determine a combination of the polarization
tensors such that
\begin{eqnarray}
e_{00}^{(1)}(\vec k)H_1^{(1)}(ik{l})+e_{00}^{(2)}(\vec
k)H_1^{(2)}(ik{l})=-{2\tilde\kappa^2 M\over 3}
{1\over(2\pi)^{3\over2}}{1\over ik{l}^2}\ ,\cr e_{0i}^{(1)}(\vec
k)H_1^{(1)}(ik{l})+e_{0i}^{(2)}(\vec k)H_1^{(2)}(ik{l})=0\ ,\cr e_{ij}^{(1)}(\vec k)H_1^{(1)}
(ik{l})+e_{ij}^{(2)}(\vec k)H_1^{(2)}(ik{l})=-{\tilde\kappa^2
M\over 3} {1\over(2\pi)^{3\over2}}{1\over ik{l}^2}\left(\delta_{ij}-{k_ik_j\over4\pi k^2}
\right)\,.
\end{eqnarray}
We are now interested in the solution which converges in the limit
$z\rightarrow\infty$. This is possible by choosing
$e_{\mu\nu}^{(2)}=0$. Then we have the following complete solution
\be \gamma_{\mu\nu}(\vec
r,z)=\int\!{d^3k\over(2\pi)^{3\over2}}\,e^{i\vec k\,.\vec
r}\hat\gamma_{\mu\nu}(\vec k, z)\quad
,\quad\hat\gamma_{\mu\nu}(\vec k,z)= {\tilde\kappa^2 M\over3{l}(2\pi)^{3\over2}} \,z^2\,
{K_2(kz)\over k{l}\,K_1(k{l})}\,c_{\mu\nu}\ , \ee with $c_{00}=2$, $c_{0i}=0$ and
$c_{ij}=\delta_{ij}-k_ik_j/4\pi k^2$, and where $K_\nu(z)$ is the
modified Bessel function defined as
$K_\nu(z)=i{\pi\over2}e^{i\nu{\pi\over2}}H_\nu^{(1)}(iz)$. Near
the brane this metric reduces to, setting $\epsilon={z\over{l}}-1$,
\be \label{gamma}
\hat\gamma_{\mu\nu}(\vec k,\epsilon)=
{\tilde\kappa^2 M{l}\over3(2\pi)^{3\over2}}
\left\{{K_2(k{l})\over k{l}\,K_1(k{l})}-\epsilon-{k{l}\over4}\epsilon^2\left[{K_2(k{l})
\over K_1(k{l})}-3{K_0(k{l})\over K_1(k{l})}\right]+{\cal O}(\epsilon^3)\right\}c_{\mu\nu}
. \ee
The
appearance of Dirac distributions in the expansion of
$\gamma_{\mu\nu}(\vec r,z)$ does not however necessarily mean that
$\gamma_{\mu\nu}(\vec r,z)$ is singular at $\vec r=0$ as the sum
may be regular. We now see what the perturbed metric
on the brane is. We know that the brane is at $z={l}+\xi$.
Therefore if we want to expand the metric around $z={l}$ we
have
\be ds^2\Big|_{\Sigma}=\left[\frac{l}{{l}+\xi}\left(\eta_{AB}+\gamma_{AB}\right)dx^A
dx^B\right]\Big|_{\Sigma}\simeq\left(\eta_{AB}+h_{AB}\right) dx^A dx^B, \ee where \be
h_{AB}=\gamma_{AB}\Big|_{\Sigma}-2\frac{\xi}{l}\eta_{AB}. \ee
Therefore
the actual perturbed four-dimensional metric is \be \label{actual}
g_{\mu\nu}=\eta_{\mu\nu}+h_{\mu\nu}. \ee The Newtonian potential
is then $h_{00}$. If we now Fourier transform $h_{00}$ considering
(\ref{xi}) and (\ref{gamma}), we get
\be \hat h_{00}(\vec
k)=\hat{\tilde h}_{00}(\vec
k)=\hat\gamma_{00}\Big|_\Sigma+2{\hat\xi\over{l}}=
{\tilde\kappa^2 M\over k^2{l}(2\pi)^{3\over2}}\left[1-{2k{l}\over3}{K_0(k{l})\over
K_1(k{l})}\right]\,. \ee
Taking the inverse-Fourier transform
and integrating over angles, we obtain, setting $\alpha=r/{l}$,
\be h_{00}(\vec r)={\tilde\kappa^2 M\over4\pi{l}}{1\over
r}\left(1+{4\pi\over3}{\cal K}_\alpha\right)\quad\hbox{with}\quad
{\cal K}_\alpha=\lim_{\epsilon\to0}\int_0^{+\infty}\!
du\,\sin(u\alpha)\,{K_0(u)\over K_1(u)}e^{-\epsilon u}\,. \ee
We have a short and long distance limit,
$\lim_{\alpha\to0}{\cal K}_\alpha=\alpha^{-1}={l}/r$,
$\lim_{\alpha\to\infty}{\cal K}_\alpha=\pi/2\alpha^2=\pi ({l}/r)^2/2$. We hence recover
that at short distances the
correction to Newton's law is ${l}/r$, whereas at
distances large compared with the AdS curvature scale ${l}$,
the correction is reduced by another
${l}/r$ factor, in agreement with \cite{RS,bunchNew}
\be\label{weakexp} \lim_{r/{l}\to\infty}h_{00}(\vec
r)={2G_NM\over r}\left[1+{2\over3}\left({{l}\over
r}\right)^2\right]\,, \ee where, as before, we identify the
Newtonian constant as $8\pi G_N=\tilde\kappa^2/l$. This
expansion can be taken as a ``quality-test'' that solutions of the
four-dimensional effective field equations have to pass to lead to
a regular bulk. Indeed, as we will discuss in the next chapter,
stellar solutions which do not have the weak regime expansion
(\ref{weakexp}), probably lead to a non-regular Cauchy horizon for
the bulk \cite{static}.

\subsection{Quantum correction to the Newtonian potential and holographic interpretation}

\footnote{To show the importance of studying braneworld scenarios from the holographic
point of view, I will
occasionally insert sections on the quantum side of the correspondence (e.g. AdS/CFT
and its deformations). The main idea of this thesis
is to study the classical astrophysics and cosmology of braneworlds. Therefore these
sections aim to give the reader only a
flavour of the
quantum counterpart of the correspondence.} Following
\cite{DL} we show that the four-dimensional quantum gravity correction to the
Newtonian law
corresponds to the classical correction (\ref{weakexp}) found in the context of the
braneworld scenario. This is a test of the
deformed AdS/CFT
correspondence at perturbative level.

We start with the linearized Einstein equations in four dimensions, using the metric
(\ref{actual}),
where $h_{\mu\nu}$ is a perturbation of the Minkowskian metric, the harmonic gauge
$\partial_\mu g^{\mu\nu}=0$ implies
\be
\partial_\mu(h^{\mu\nu}-{1\over 2} \eta^{\mu\nu}h)=0.
\ee
Defining $\tilde h_{\mu\nu}=h^{\mu\nu}-{1\over 2} \eta^{\mu\nu}h$
we get the
linearized Einstein equations
\be
\Box \tilde h_{\mu\nu}=-16\pi G_N T_{\mu\nu},
\ee
so that in Fourier space
\be
\tilde h_{\mu\nu}(p)=16\pi G_N {1\over p^2}T_{\mu\nu}(p).
\ee
For quantum corrections, the total perturbation will be of form
\begin{equation}
\tilde h_{\mu\nu}=h^c_{\mu\nu}+h^q_{\mu\nu}\ ,
\end{equation}
where the superscript $c$ means ``classical'' and the superscript $q$ means ``quantum".
The quantum spin-two tensor $h^q_{\mu\nu}$
must vanish with the vanishing of the
classical perturbations. We have
then two other
ingredients, the graviton propagator and the self-energy of the gravitons.
These two objects must be represented by tensors of rank 4. This is because they have
to be applied to the classical perturbation of
the metric and they have to reproduce a rank 2 tensor (the quantum perturbation of the metric).

We start with the propagator, calling it
$D^{\alpha\beta\gamma\delta}$. Taking two points $x$ and
$x'$ at the same proper time, the propagator describes the short-distance quantum effects
of the products $h_{\alpha\beta}(x)h_{\gamma\delta}(x')$. This
implies that it should be proportional to $\delta(x-x')$. Another
ingredient is that it must depend on the unperturbed metric. In
particular since it must be local, it cannot depend on the derivatives
of the background metric. So in general the propagator will depend
only on the products $\eta^{\alpha\beta}\eta^{\gamma\delta}$.
Bearing in mind that $D_{\mu\nu\alpha\beta}\sim
h_{\mu\nu}h_{\alpha\beta}$, by symmetry we can already guess the
form of this propagator as
\be D^{\mu\nu\alpha\beta}={1 \over
2}\delta(x-x')\left(\eta^{\mu\alpha}\eta^{\nu\beta}+\eta^{\mu\alpha}\eta^{\nu\alpha}+
\lambda\eta^{\mu\nu}\eta^{\alpha\beta}\right),
\ee
where $\lambda$ is a constant that must be determined.

For the Hamiltonian constraint to be non-singular, the propagator in Fourier
space becomes \cite{BdW}
\be D^{\mu\nu\alpha\beta}(p^2)=\frac{1}{2
p^2}\left(\eta^{\mu\alpha}\eta^{\mu\beta}+\eta^{\mu\alpha}\eta^{\nu\alpha}-
\eta^{\mu\nu}\eta^{\alpha\beta}\right).
\ee
Up to now we have
\be
h_q^{\mu\nu}=D^{\mu\nu\alpha\beta}\Pi_{\alpha\beta\gamma\delta}h_c^{\gamma\delta},
\ee
where $\Pi_{\alpha\beta\gamma\delta}$ is the graviton self
energy. Now in Fourier space the self energy is a function of the
momentum $p_\alpha$ of the graviton and of the unperturbed metric
(we still are at the first order in quantum corrections). In
particular since the tensor is a rank four tensor and
the metric is dimensionless, the self energy must have all the
possible combinations of $p_\alpha$ products of order four. This is
because in general we will have at least a term $p_\alpha p_\beta
p_\gamma p_\delta$. If we would like to preserve the classical
isometries at quantum level, we can also use the Slanov-Ward
gravitational identity \cite{SWi}
\be p_\mu p_\nu
D^{\mu\nu\alpha\beta}\Pi_{\alpha\beta\gamma\delta}D^{\gamma\delta\rho\sigma}=0.
\ee
This at first linearized level determines the self-energy up to
two functions $\Pi_1(p)$ and $\Pi_2(p)$. Combining the classical
and the one-loop quantum results at the linearized level, we get
\ba
h_{\mu\nu}&=&16 \pi \frac{G_N}{p^2}\left[T_{\mu\nu}-{1\over
2}\eta_{\mu\nu}T(p)\right]\cr\nonumber &-&16\pi
G_N\left[2\Pi_2(p)T_{\mu\nu}(p)+\Pi_1(p)\eta_{\mu\nu}T(p)\right].
\ea
The actual form of the $\Pi_i$'s depend on the theory we
consider. However for any massless theory in four dimensions,
after cancelling the infinities with the appropriate counterterms,
the finite remainder must have the form \cite{CLR} \be\label{self}
\Pi_i(p)=32\pi G_N\left(a_i\ln \frac{p^2}{\mu^2}+b_i\right), \ee
where $\mu$ is an arbitrary subtraction mass linked with the
renormalization energy. Using now a point source
$T_{00}=M\delta(\vec r)$, we obtain the perturbed time component
of the metric
\be g_{00}=-(1-\frac{2 G_N M}{r}-\frac{2\alpha G_N^2
M}{r^3}), \ee
where $\alpha=4\cdot 32\pi(a_1+a_2)$. Explicit
calculations of the self-energy (\ref{self}) for different spins
give \cite{DL,245Duff}
\be a_i(s=1)=4a_i(s=1/2)=12a_i
(s=0)=\frac{1}{120(4\pi)^2}(-2,3). \ee
Considering the number of
particle species of spin $s$ going around the loop $N_s$ and
considering that for a single CFT these numbers can be rewritten
in terms of the dimensionality of the gauge group of the CFT ($N$)
\be (N_1,N_{1/2},N_0)=(N^2,4N^2,6N^2), \ee
we have that the
one-loop correction to the Newtonian potential is
\be
V(r)=\frac{G_N M}{r}\left(1+\frac{2N^2 G_N}{3\pi r^2}\right). \ee
Using the AdS/CFT relation \cite{M1} $N^2=\pi{l}^3/2\tilde\kappa^2$ we get exactly
the result (\ref{weakexp}),
which was found considering a five dimensional classical
braneworld scenario.

\newpage
\chapter{Stars in the braneworld}

The first step in considering infinitely large extra dimensions is
the discovery that gravity can be localized at low energies \cite{RS}.
This basically gives the strongest constraint for such theories.
The second step is therefore to see how deviations from general relativity may
explain the nature of our Universe.

The natural scenario,
where deviations from general relativity occur, is cosmology.
Indeed cosmological implications of these braneworld models have been
extensively investigated (see e.g. the review~\cite{maacov} for
further references). But this is not all. Significant deviations from Einstein's theory
in fact occur also in astrophysics. Indeed
very compact objects and gravitational collapse to black holes, can leave
traces of the extra dimensions.
For example, when an horizon
forms, even if the high-energy
effects eventually become disconnected
from the outside region on the brane, they could leave a
signature on the brane \cite{collapse}.

In addition to local high-energy effects,
there are also nonlocal corrections arising from the imprint on
the brane of Weyl curvature in the bulk, i.e. from 5-dimensional
graviton stresses. These nonlocal Weyl stresses arise on the brane
whenever there is inhomogeneity in the density; the inhomogeneity
on the brane generates Weyl curvature in the bulk which
`backreacts' on the brane. Note that we can have these nonlocal Weyl
stresses even if the density is homogeneous \cite{static}.

The high-energy (local) and bulk graviton stress (nonlocal)
effects combine to significantly alter the matching problem on the
brane, compared with the general relativistic case. For spherical
compact or collapsing objects (uncharged and non-radiating), matching in general
relativity shows that the asymptotically flat exterior spacetime
is Schwarzschild. High-energy corrections to the pressure,
together with Weyl stresses from bulk gravitons, mean that on the
brane, matching no longer leads to a Schwarzschild exterior in
general \cite{static, collapse}. These stresses also mean that the matching conditions do
not have unique solution on the brane \cite{static}; knowledge of the
5-dimensional Weyl tensor is needed as a minimum condition for
uniqueness.

\section{The static star case}

\footnote{I will base this section essentially on \cite{static}.}In
this section we consider the simplest case of a static spherical
star with uniform density. We find an exact interior solution,
thus generalizing the Schwarzschild interior solution of general
relativity. We show that the general relativity compactness limit
given by $GM/R<{4\over9}$ is reduced by high-energy 5-dimensional
gravity effects. The existence of neutron stars allows us to put a
lower bound on the brane tension, which is stronger than the bound
from big bang nucleosynthesis, but weaker than the bound from
experiments probing Newton's law on sub-millimetre scales. We also
give two different exact exterior solutions, both of which satisfy
the braneworld matching conditions and field equations and are
asymptotically Schwarzschild, but neither of which is the
Schwarzschild exterior. One of these solutions is the
Reissner-N\"ordstrom-type solution found in~\cite{dmpr}, in which
there is no electric charge, but instead a Weyl `charge' arising
from bulk graviton tidal effects. The other is a new solution.
Both of these exterior solutions carry the imprint of bulk
graviton stresses, and each has an horizon on the brane which is
larger than the Schwarzschild horizon.

Both of our solutions (i.e. the full solution, interior plus exterior) are consistent
braneworld solutions, but we do not know the bulk solutions of
which they are boundaries. In fact, no exact 5-dimensional
solution for astrophysical brane black holes is known, and the
uniform star case is even more complicated.

We have seen in the section (\ref{secNewton}) that the Newtonian potential on the brane is
modified as follow
\begin{equation}\label{pert}
\Phi={GM\over r}\left(1+{2l^2\over3r^2}\right)\,,
\end{equation}
where $l$ is the curvature scale of AdS. This result assumes that the bulk
perturbations are bounded in conformally Minkowski coordinates,
and that the bulk is nearly AdS. It is not clear whether there
is a covariant way of uniquely characterizing these perturbative
results~\cite{DD}, and therefore it remains unclear what the
implications of the perturbative results are for very dense stars
on the brane. However, it seems reasonable to conjecture that the
bulk should be asymptotically AdS, and that its Cauchy horizon
should be regular. Then perturbative results suggest that on the
brane, the weak-field potential should behave as in
eq.~(\ref{pert}). In fact, perturbative analysis also constrains
the weak-field behaviour of other metric components on the
brane~\cite{bunchNew}, as well as the nonlocal stresses on the
brane induced by the bulk Weyl tensor~\cite{ssm}.
This is also supported at non-linear level if one assumes a
bounded bulk in conformally Minkowskian coordinates.
In this case indeed, it is possible to integrate numerically simple models of
homogeneous and isotropic stars \cite{wiseman}.

\subsection{Field equations and matching conditions}\label{sectionField}

We have already discussed that the local and nonlocal extra-dimensional modifications to
Einstein's equations on the brane may be consolidated into an
effective total energy-momentum tensor:
\begin{equation}
G_{\mu\nu}=\kappa^2 T^{\rm eff}_{\mu\nu}\,, \label{6'}
\end{equation}
where $\kappa^2=8\pi G_N$ and the bulk cosmological constant is
chosen so that the brane cosmological constant vanishes. The
effective total energy density, pressure, anisotropic stress and
energy flux for a fluid are given by eqs. (\ref{reff}-\ref{e:pressure2}).

From big bang nucleosynthesis constraints, $\lambda \gtrsim
1$~MeV$^4$, but a much stronger bound arises from null results of
sub-millimetre tests of Newton's law\footnote{From the definition (\ref{GNdef}) and the fine-tuning (\ref{min}) we get $\lambda=3/4\pi (l^2 G_N)^{-1}$.
Table-top tests of Newton's laws currently find no deviation down to about $0.2$ mm. This implies from (\ref{pert}) that $l\lesssim 0.2$ mm, or
$\lambda \gtrsim
10$~TeV$^4$.}: $\lambda \gtrsim
10$~TeV$^4$.

The local effects of the bulk, arising from the brane extrinsic
curvature, are encoded in the quadratic terms, $\sim
(T_{\mu\nu})^2/\lambda$, which are significant at high energies,
$\rho\gtrsim\lambda$. The nonlocal bulk effects, arising from the
bulk Weyl tensor, are carried by nonlocal energy density ${\cal
U}$, nonlocal energy flux ${Q}_\mu$ and nonlocal anisotropic
stress ${\Pi}_{\mu\nu}$. Five-dimensional graviton stresses are
imprinted on the brane via these nonlocal Weyl terms.

Static spherical symmetry implies ${Q}_\mu=0$ and
\begin{equation}\label{p}
{\Pi}_{\mu\nu}={\Pi}(r_\mu r_\nu-{\textstyle{1\over3}}
h_{\mu\nu})\,,
\end{equation}
where $r_\mu$ is a unit radial vector. For static spherical symmetry, the
conservation equations (\ref{pc1}-\ref{pc2'}) reduce to
\begin{eqnarray}
&&\D_\mu p+(\rho+p)A_\mu=0\,,\label{c1}\\ && {\textstyle{1\over3}}
\D_\mu{\cal U}+ {\textstyle{4\over3}}{\cal U}A_\mu+\D^\nu {\Pi}_{\mu\nu}=
-\frac{(\rho+p)}{\lambda} \D_\mu\rho\, .
\label{c2}
\end{eqnarray}
In static coordinates the metric is
\begin{eqnarray}\label{m}
 ds^2=-A^2(r)dt^2+B^2(r)dr^2+r^2d\Omega^2\,,
\end{eqnarray}
and eqs.~(\ref{6'})--(\ref{m}) imply
\begin{eqnarray}
&&{1\over r^2}-{1\over B^2}\left({1\over r^2}-{2\over r}{B'\over
B}\right) = 8\pi G_N\rho^{\rm eff}\,,\label{f1}\\ &&-{1\over
r^2}+{1\over B^2}\left({1\over r^2}+{2\over r}{A'\over A}\right) =
8\pi G_N\left(p^{\rm eff}+ {2\over 3}{\Pi}
\right)\,,\label{f2}\\&&p'+{A'\over A}(\rho+p)=0\,,\label{f3}\\
&&{\cal U}'+4{A'\over A} {\cal U}+ 2{\Pi}'+2{A'\over A}{\Pi}+
{6\over r}{\Pi}=-3\frac{(\rho+p)}{\lambda}\rho'\,. \label{f4}
\end{eqnarray}

The exterior is characterized by
\begin{equation}
\rho=0=p\,,~~{\cal U}={\cal U}^+\,,~{\Pi}={\Pi}^+\,,
\end{equation}
so that in general $\rho^{\rm eff}$ and $p^{\rm eff}$ are nonzero
in the exterior: there are in general Weyl stresses in the
exterior, induced by bulk graviton effects. These stresses are
radiative, since their energy-momentum tensor is traceless. The system of equations
for the exterior is not closed until a further condition is given
on ${\cal U}^+$, ${\Pi}^+$ (e.g., we could impose ${\Pi}^+=0$ to close the system).
In other words, from a brane
observer's perspective, there are many possible static spherical
exterior metrics, including the simplest case of Schwarzschild
(${\cal U}^+=0={\Pi}^+$).

The interior has nonzero $\rho$ and $p$; in general, ${\cal U}^-$
and ${\Pi}^-$ are also nonzero, since by eq.~(\ref{f4}), {\em
density gradients are a source for Weyl stresses in the interior}.
For a uniform density star, we can have ${\cal U}^-=0={\Pi}^-$,
but nonzero ${\cal U}^-$ and/ or ${\Pi}^-$ are possible,
subject to eq.~(\ref{f4}) with zero right-hand side.

From eq.~(\ref{f1}) we obtain
\begin{eqnarray}
B^2(r)=\left[1-\frac{2Gm(r)}{r}\right]^{-1}\,,
\end{eqnarray}
where the mass function is
\begin{eqnarray}
m(r)=4\pi\int^r_a\rho^{\rm eff}(r')r'^2dr'\,,
\end{eqnarray}
and $a=0$ for the interior solution, while $a=R$ for the exterior
solution.

The Israel matching conditions at the stellar surface
$\Sigma$ give~\cite{israel}
\begin{eqnarray}
[G_{\mu\nu}r^\nu]_{\Sigma}=0\,,
\end{eqnarray}
where $[f]_\Sigma\equiv f(R^+)-f(R^-)$. By the brane field
equation~(\ref{6'}), this implies $[T^{\rm
eff}_{\mu\nu}r^\nu]_\Sigma=0$, which leads to
\begin{eqnarray}  \label{m2}
\left[p^{\rm eff}+ {2\over 3}{\Pi}\right]_\Sigma=0\,.
\end{eqnarray}
Even if the physical pressure vanishes at the surface, the
effective pressure is nonzero there, so that in general a radial
stress is needed in the exterior to balance this effective
pressure.

The general relativity limit of eq.~(\ref{m2}) implies
\begin{equation}\label{p0}
p(R)=0\,.
\end{equation}
This can also be obtained from a slightly different
point of view. Consider the conservation equation (\ref{f3})
\be
p'+(\rho+p)\frac{A'}{A}=0.
\ee
The junction conditions (see Appendix \ref{Appjc}) require that on the stellar
surface the metric and
its first derivative along the orthogonal direction to the surface must be continuous and
non-singular.
This implies that $A'/A$ is a continuous and non-singular function across the boundary of
the star. The model we
are going to consider is such that
\be
\rho=\tilde\rho\ \theta(r-R)\ ,\ p=\tilde p\ \theta(r-R)\ ,
\ee
where the functions with tilde are continuous and non-singular functions and
\ba
\theta(r-R)=\begin{cases} 1\ \mbox{for}\ r<R\cr 0\ \mbox{for}\ r>R\end{cases}\ ,
\ea
is the Heaviside function. Now we consider the first derivative of the pressure
\be
p'=\tilde p'\ \theta(r-R)+\tilde p\ \delta(r-R)\ ,
\ee
where $\delta(r-R)$ is the Dirac distribution. In order to balance this distribution
in eq. (\ref{f3}), we have two possibilities. The first one is that $p(R)=0$ and the
second is that the metric becomes singular\footnote{Even if we relax the standard junction
conditions we are left with the non trivial problem of defining the product of distributions
$\delta(r-R)\theta(r-R)$.}. If we consider the junction conditions, requiring $A'/A$
to be continuous and non-singular, the second choice becomes unavailable. Then the only
physically sensible model is with $p(R)=0$. We have only used the
conservation equations (\ref{f3}) and the geometrical junction conditions, valid for
both general relativity and braneworld effective gravity.
This implies that unlike general relativity, where (\ref{p0}) and (\ref{m2})
correspond to the same
constraint, they become separate conditions in the braneworld.

Using $p(R)=0$, the constraint (\ref{m2}) becomes
\begin{equation}\label{m3}
3\frac{\rho^2(R)}{\lambda}+{\cal U}^-(R) +2 {\Pi}^-(R)= {\cal U}^+(R)
+2 {\Pi}^+(R)\,.
\end{equation}
Equation~(\ref{m3}) gives the matching condition for any static
spherical star with vanishing pressure at the surface. If there
are no Weyl stresses in the interior, i.e. ${\cal U}^-=0= {\Pi}^-$,
and if the energy density is non-vanishing at the surface,
$\rho(R)\neq0$, then there must be Weyl stresses in the exterior,
i.e. the exterior cannot be Schwarzschild. Equivalently, {\em if
the exterior is Schwarzschild and the energy density is nonzero at
the surface, then the interior must have nonlocal Weyl stresses}.

\subsection {Braneworld generalization of exact uniform-density solution}

Here we are interested in the most simple model of a compact star (such as
for example a neutron star \cite{gravitation}). In this model the density is considered
constant inside the star and zero outside and the geometry is isotropic.
This of course implies that $\rho'=0$ everywhere. Moreover, consistently with the
isotropy of the star, we set ${\Pi}^-=0$.

Equation~(\ref{f3}) integrates in the interior for
$\rho=$\,const to give
\begin{equation}
A^-(r)={\alpha\over \rho+p(r)}\,,
\end{equation}
where $\alpha$ is a constant.
Eq. (\ref{f4}) reads
\be
{\cal U}'+4\frac{A'}{A}{\cal U}=0\ ,
\ee
with solution
\begin{equation}\label{u}
{\cal U}^-(r)={\beta\over \left[A^-(r)\right]^4}\,,
\end{equation}
where $\beta$ is a constant. The matching condition in
eq.~(\ref{m3}) then reduces for a {\em uniform} star to
\begin{equation}\label{m4}
3\frac{\rho^2}{\lambda}+{\beta\over\alpha^4}\rho^4= {\cal U}^+(R) +2
{\Pi}^+(R)\,.
\end{equation}
It follows that in general the exterior of a uniform star cannot be
Schwarzschild.

Combining eqs. (\ref{f2},\ref{f3}) we obtain the generalization of the Tolman-Oppenheimer-
Volkoff
equation of hydrostatic equilibrium for the braneworld
\be \label{hydro}
\frac{dp}{dr}=-(p+\rho)\frac{G_Nm(r)+4\pi r^3 p^{\rm eff}}{r(r-2G_Nm(r))},
\ee
where the interior mass function is
\begin{eqnarray}\label{mass}
m^-(r)=M\left[1+\frac{3M}{8\pi\lambda R^3}\right]\left({r\over R}
\right)^3+4\pi\frac{\beta}{\alpha^4}\int^r_0 r'{}^2{(p(r')+\rho)^4} dr'\,,
\end{eqnarray}
with $M=4\pi R^3\rho/3$, and the effective pressure is
\be
p^{\rm eff}=p+\frac{\rho}{2\lambda}(2p+\rho)+\frac{\beta}{3\alpha^4}(p+\rho)^4.
\ee
Eq. (\ref{hydro}) can be analytically solved if and only if
$\beta=0$. Indeed in this case it is possible to separate the variables $p$ and $r$.

Here we are interested in typical astrophysical stars. Their density
$\rho\sim 10^{-3}{}\mbox{GeV}^4$ is much smaller then the tension of the brane
$\lambda\sim 10\mbox{~TeV}^4$. Non-local corrections are produced
by back-reaction of the five-dimensional gravitational field on the brane. Therefore we
naively expect
that their order of magnitude is much smaller than the local effects.
This is also supported by numerical models \cite{wiseman}.
Since we expect ${\cal U}^-\ll \rho^2/\lambda$, we assume in the following $\beta=0$
and
then we can analytically solve (\ref{hydro}).

With uniform density and ${\cal U}^-=0= {\Pi}^-$, we have the
case of purely local (high-energy) modifications to the general
relativity uniform-density solution, i.e. to the Schwarzschild
interior solution~\cite{exact}.

We can now calculate the pressure. Considering that
\begin{eqnarray}
B^-(r)={1\over\Delta(r)}\,,
\end{eqnarray}
the pressure is given by
\begin{eqnarray} \label{3}
{p(r)\over\rho}=\frac{[{\Delta(r)}-{\Delta(R)}] (1+\rho/\lambda)}
{[3{\Delta(R)}-{\Delta(r)}]+[3{\Delta(R)}-2{\Delta(r)}]\rho/
\lambda}\,,
\end{eqnarray}
where
\begin{eqnarray}
\Delta(r)=\left[1-{2G_NM\over r}\left({r\over R}\right)^3\left\{ 1+
{\rho\over 2\lambda }\right\}\right]^{1/2}\,.
\end{eqnarray}
In the general relativity limit, $\lambda^{-1}\to 0$, we regain
the known exact solution~\cite{exact}. The high-energy corrections
considerably complicate the exact solution.

\begin{figure}
\begin{center}
\includegraphics[width=6cm,height=7cm,angle=-90]{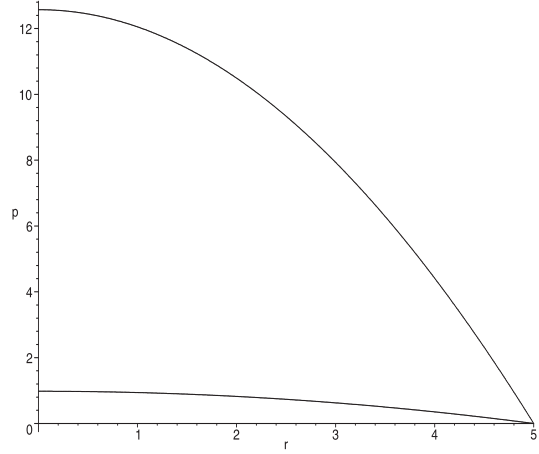}
\end{center}
\caption{ \label{g1} Qualitative comparison of the pressure $p(r)$,
in general relativity (upper curve), and in a braneworld model
with $\lambda= 5\times 10^{8}$~MeV$^4$ (lower curve).}
\end{figure}

Since $\Delta(R)$ must be real, we find {\em an astrophysical
lower limit on $\lambda$, independent of the Newton-law and
cosmological limits:}
\begin{equation}\label{al}
{\lambda}\geq \left(\frac{G_NM}{R-2G_NM}\right)\rho~~\mbox{for all
uniform stars}\,.
\end{equation}
In particular, since $\lambda,\rho>0$, $\Delta(R)^2>0$ implies $R>2GM$, so that the Schwarzschild
radius is still a limiting radius, as in general relativity.
Taking a typical neutron star (assuming uniform density) with
$\rho\sim 10^{9}$~MeV$^4$ and $M\sim 4\times 10^{57}$~GeV, we find
\begin{eqnarray}\label{al'}
\lambda > 5\times 10^{8}~\mbox{MeV}^4\,.
\end{eqnarray}
This is the astrophysical limit, below which stable neutron stars
could not exist on the brane. It is much stronger than the
cosmological nucleosynthesis constraint, but much weaker than the
Newton-law lower bound. Thus stable neutron stars are easily
compatible with braneworld high-energy corrections, and the
deviations from general relativity are very small. If we used the
lower bound in eq.~(\ref{al'}) allowed by the stellar limit, then
the corrections to general relativistic stellar models would be
significant, as illustrated in fig. (\ref{g1}).

We can also obtain an upper limit on compactness from the
requirement that $p(r)$ must be finite. Since $p(r)$ is a
decreasing function, this is equivalent to the condition that
$p(0)$ is finite and positive, which gives the condition
\begin{eqnarray}\label{comp}
\frac{G_NM}{R} \leq {4\over 9}\left[{1+7\rho/4\lambda +
5\rho^2/8\lambda^2
\over(1+\rho/\lambda)^2(1+\rho/2\lambda)}\right]\,.
\end{eqnarray}
It follows that high-energy braneworld corrections {\em reduce}
the compactness limit of the star. For the stellar bound on
$\lambda$ given by eq.~(\ref{al'}), the reduction would be
significant, but for the Newton-law bound, the correction to the
general relativity limit of $4\over9$ is very small. The lowest
order correction is given by
\begin{equation}\label{comp2}
{G_NM\over R}\leq  \frac{4}{9}\left[1-\frac{3\rho}{4\lambda}
+O\left({\rho^2\over\lambda^2}\right)\right]\,.
\end{equation}
For $\lambda\sim10$~TeV$^4$, the minimum allowed by
sub-millimetre experiments, and $\rho\sim 10^{9}$~MeV$^4$, the
fractional correction is $\sim 10^{-16}$.

\subsection{Two possible non-Schwarzschild exterior solutions}

The system of equations satisfied by the exterior spacetime on the
brane is not closed. Essentially, we have two independent unknowns
${\cal U}^+$ and ${\Pi}^+$ satisfying one equation, i.e.
eq.~(\ref{f4}) with zero right-hand side. Even requiring that the
exterior must be asymptotically Schwarzschild does not lead to a
unique solution. Further investigation of the 5-dimensional
solution is needed in order to determine what the further
constraints are. We are able to find two exterior solutions for a
uniform-density star (with
${\cal U}^-=0$) that are consistent with all equations
and matching conditions on the brane, and that are asymptotically
Schwarzschild.

The first is the Reissner-N\"ordstrom-like solution given
in~\cite{dmpr}, in which a tidal Weyl charge plays a role similar
to that of electric charge in the general relativity
Reissner-N\"ordstrom solution. We stress that there is {\em no}
electric charge in this model: nonlocal Weyl effects from the 5th
dimension lead to an energy-momentum tensor on the brane that has
the same form as that for an electric field, but without any
electric field being present. The formal similarity is not
complete, since the tidal Weyl charge gives a {\em positive}
contribution to the gravitational potential, unlike the negative
contribution of an electric charge in the general relativistic
Reissner-N\"ordstrom solution.

The braneworld solution is~\cite{dmpr}
\begin{eqnarray}
&&\left(A^+\right)^2=\left(B^+\right)^{-2}=1-\frac{2G_N{\cal
M}}{r}+\frac{q}{r^2}\,,\label{rn1} \\&& {\cal U}^+=-\frac{{\Pi}^+}{2} = \frac{4}{3}\pi G_N
q\lambda \,{1\over r^4}\,,\label{rn2}
\end{eqnarray}
where the matching conditions imply
\begin{eqnarray}
q&=&-3G_NMR\,{\rho\over\lambda}\,,\\ {\cal M}&=& M \left(
1-\frac{\rho}{\lambda }\right)\,, \\ \alpha&=& \rho\Delta (R)\,.
\end{eqnarray}
Note that the Weyl energy density in the exterior is {\em
negative}, so that 5-dimensional graviton effects lead to a
strengthening of the gravitational potential (this is discussed
further in~\cite{dmpr,ssm}). Since ${\cal M}>0$ is required for
asymptotic Schwarzschild behaviour, we have a slightly stronger
condition on the brane tension:
\begin{equation}
\lambda>\rho \,.
\end{equation}
However, this still gives a weak lower limit, $\lambda>
10^{9}~\mbox{MeV}^4$. In this solution the horizon is at
\begin{equation}
r_{\rm h}=G_N{\cal M}\left[1+\left\{1+\left({3R\over
2G_NM}-2\right){\rho\over \lambda} +{\rho^2\over
\lambda^2}\right\}^{1/2}\right].
\end{equation}
Expanding this exact expression shows that the horizon is slightly
beyond the general relativistic Schwarzschild horizon:
\begin{equation}
r_{\rm h}=2G_NM\left[1+{3(R-2G_NM)\over
4G_NM}\,{\rho\over\lambda}\right] +O\left({\rho^2\over
\lambda^2}\right)
>2GM\,.
\end{equation}
The exterior curvature invariant ${\cal R}^2=
R_{\mu\nu}R^{\mu\nu}$ is given by
\begin{eqnarray}\label{r2}
{\cal R}= 8\pi G_N\left({\rho\over \lambda}\right)^2\left({R\over
r}\right)^4\,.
\end{eqnarray}
Note that for the Schwarzschild exterior, ${\cal R}=0$.

\begin{figure}
\begin{center}
\includegraphics[width=6cm,height=7cm,angle=-90]{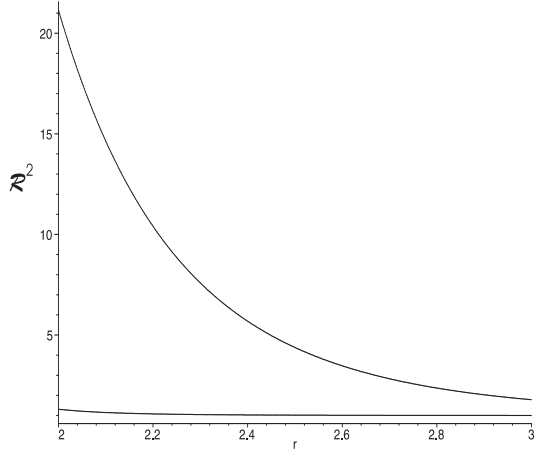}
\end{center}
\caption{ \label{g3} Qualitative behaviour of the curvature
invariant ${\cal R}^2$: the upper curve is the
Reissner-N\"ordstrom-like solution given by eqs.~(\ref{rn1}) and
(\ref{rn2}); the lower curve is the new solution given by
eqs.~(\ref{n1})--(\ref{n4}) ($\lambda= 5\times 10^{8}$~MeV$^4$).}
\end{figure}

The second exterior is a new solution. Like the above solution, it
satisfies the braneworld field equations in the exterior, and the
matching conditions at the surface of the uniform-density star. It
is given by
\begin{eqnarray}
\left(A^+\right)^2&=&1-\frac{2G_N{\cal N}}{r}\,, \label{n1} \\
\left(B^+\right)^{-2}&=&\left(A^+\right)^2
\left[1+\frac{C}{\lambda(r-{\textstyle{3\over2}} G_N{\cal
N})}\right]\,,\label{n2} \\ {\cal U}^+&=& {2\pi G_N^2{\cal
N}C\over(1-3G_N{\cal N}/2r)^2}\,{1\over r^4}\,,\label{n3}\\ {\Pi}^+&=& \left({2\over3}-
{r \over G_N{\cal N}}\right){\cal U}^+\,.
\label{n4}
\end{eqnarray}
From the matching conditions:
\begin{eqnarray}
{\cal N} &=& M\left[\frac{1+2\rho/\lambda}{1+3G_NM\rho/
R\lambda)}\right]\,,\\ C &=& 3G_NM\rho\left[ \frac{1-3G_NM/2R}{ 1
+3G_NM\rho/R\lambda}\right]\,,\\ \alpha &=&{\rho\Delta(R)\over (1+3
G_NM\rho/R\lambda)^{1/2}}\,.
\end{eqnarray}
The horizon in this new solution is at
\begin{equation}
r_{\rm h}=2G_N{\cal N}\,,
\end{equation}
which leads to
\begin{equation}
r_{\rm h}=2G_NM\left[1+\left({2R-3G_NM\over 2R}\right){\rho\over
\lambda}\right]+O\left({\rho^2\over\lambda^2}\right)>2G_NM\,.
\end{equation}
The curvature invariant is
\begin{eqnarray}
{\cal R}&=&\sqrt{{\textstyle{3\over2}}}RC\left({4\pi R\over
3M}\right)^2{(1-8G_N{\cal N}/3r+2G_N^2{\cal N}^2 /r^2)^{1/2}\over
1-3G_N{\cal N}/2r}\times\nonumber\\ &&~~{}\times \left({\rho\over
\lambda}\right)^2\left({R\over r}\right)^3 \,.
\end{eqnarray}
Comparing with eq.~(\ref{r2}), it is clear that these two
solutions are different. The difference in their curvature
invariants is illustrated in fig. (\ref{g3}).

\subsection{Interior solution with Weyl contribution}

We now consider the case where the contribution of the projected Weyl tensor
is important in the interior or when $\beta\neq 0$.

Following \cite{m} we can calculate the tidal acceleration on the brane measured by
a co-moving observer with four velocity $u^A$.
Its modulus is
\ba
\tilde A=-\lim_{y\rightarrow 0}n_A\tilde R^A_{BCD}u^B n^C u^D\ .
\ea
Using the decomposition of the Riemann tensor (\ref{decRie}) and recalling that
$T_{AB}n^A=0$, we have
\ba
\tilde A=\kappa^2{\cal U}+\frac{\tilde \Lambda}{6}.
\ea
Now if ${\cal U}<0$ the localization of the gravitational field near the brane is enhanced
reinforcing the Newtonian potential.
In this case therefore a negative Weyl energy contributes to binding the star to
the brane. Since this effect is independent of the brane tension, it makes
the star more stable than
in the general relativistic case. We see it in a very special limiting case when
the exterior is Schwarzschild. In this case
(${\cal U}^+=0={\Pi}^+$), eqs.~(\ref{u}) and (\ref{m4})
imply that the interior must have dark radiation
density:
\begin{eqnarray}
{\cal U}^-(r)=-\left(\frac{4\pi G_N}{\rho}\right)^2[\rho+p(r)]^4\,.
\end{eqnarray}
It follows that the mass function in eq.~(\ref{mass}) becomes
\begin{eqnarray}
m^-(r)&=& M\left(1+\frac{\rho}{2\lambda
}\right) \left({r\over R} \right)^3
-\frac{6\pi}{\lambda\rho^2}\int^r_0 [\rho+p(r')]^4 r'^2
dr'\,,
\end{eqnarray}
which is {\em reduced} by the negative Weyl energy density,
relative to the solution in the previous section and to the
general relativity solution. The effective pressure is given by
\begin{equation}\label{pschw}
p^{\rm eff}=p-{\rho\over2\lambda}(2+6w+4w^2+w^3)\,,
\end{equation}
where $w=p/\rho$. Thus $p^{\rm eff}<p$, so that 5-dimensional
high-energy effects reduce the pressure in comparison with general
relativity. This means, as we already anticipated, that this star is more
stable than the general relativistic one.

\subsection{Unique exterior solution: a conjecture}

We found that the Schwarzschild solution is no longer the unique asymptotically
flat vacuum exterior; in general, the exterior carries an imprint
of nonlocal bulk graviton stresses. The exterior is not uniquely
determined by matching conditions on the brane, since the
5-dimensional metric is involved via the nonlocal Weyl stresses.
We demonstrated this explicitly by giving two exact exterior
solutions, both asymptotically Schwarzschild. Each exterior which
satisfies the matching conditions leads to different bulk metrics.
Without any exact or
approximate 5-dimensional solutions to guide us, we do not know
how the properties of the bulk metric, and in particular its
global properties, will influence the exterior solution on the
brane.

Guided by perturbative analysis of the static weak field
limit~\cite{RS,bunchNew,ssm,DD}, we make the following conjecture:
{\em if the bulk for a static stellar solution on the brane is
asymptotically AdS and has regular Cauchy horizon, then the
exterior vacuum which satisfies the matching conditions on the
brane is uniquely determined, and agrees with the perturbative
weak-field results at lowest order.} An immediate implication of
this conjecture is that the exterior is not Schwarzschild, since
perturbative analysis shows that there are nonzero Weyl stresses
in the exterior~\cite{ssm} (these stresses are the manifestation
on the brane of the massive Kaluza-Klein bulk graviton modes). In
addition, the two exterior solutions that we present would be
ruled out by the conjecture, since both of them violate the
perturbative result for the weak-field potential,
eq.~(\ref{pert}).

The static problem is already complicated, so that analysis of
dynamical collapse on the brane will be very difficult. However,
the dynamical problem could give rise to more striking features.
Energy densities well above the brane tension could be reached
before horizon formation, so that high-energy corrections could be
significant. We expect that these corrections, together with the
nonlocal bulk graviton stress effects, will leave a non-static,
but transient, signature in the exterior of collapsing matter.
This what the next section considers.

\section{Gravitational collapse}

\footnote{This section is based on \cite{collapse}.}The study of gravitational collapse
in general relativity is
fundamental to understanding the behaviour of the theory at high
energies. The Oppenheimer-Snyder model still provides a
paradigmatic example that serves as a good qualitative guide to
the general collapse problem in general relativity. It can be solved analytically,
as it simply assumes a collapsing homogeneous dust cloud of finite
mass and radius, described by a Robertson-Walker metric and
surrounded by a vacuum exterior. In general relativity, this exterior is
necessarily static and given by the Schwarzschild
solution~\cite{s}. In other theories of gravity that differ from
general relativity at high energies, it is natural to look for similar examples.
In this section we analyze an Oppenheimer-Snyder-like collapse in the
braneworld scenario, in order to shed light on some fundamental differences
between collapse in general relativity and on the brane.

Braneworld gravitational collapse is complicated by a number of
factors. The confinement of matter to the brane, while the
gravitational field can access the extra dimension, is at the root
of the difficulties relative to Einstein's theory, and this is
compounded by the gravitational interaction between the brane and
the bulk. Matching conditions on the brane are more complicated to
implement, and one also has to impose regularity and
asymptotic conditions on the bulk, and it is not obvious what
these should be.

In general relativity, the Oppenheimer-Snyder model of collapsing dust has a Robertson-Walker
interior matched to a Schwarzschild exterior. We show that even
this simplest case is much more complicated on the brane. However,
it does have a striking new property, which may be part of the
generic collapse problem on the brane. The exterior is not
Schwarzschild, and nor could we expect it to be, as discussed
in the previous section, but the exterior is not even {\em static}, as shown by our
no-go theorem. The reason for this lies in the nature of the
braneworld modifications to general relativity.

The dynamical equations for the projected Weyl tensor (\ref{non-local})
are the complete set of
equations on the brane. They are not closed, since ${\cal
E}_{\mu\nu}$ contains 5D degrees of freedom that cannot be
determined on the brane. However, using only the 4D
projected equations, we prove a no-go theorem valid for the full
5D problem: {\em given the standard matching conditions on the
brane, the exterior of a collapsing homogeneous dust cloud cannot be static}.
We are not able to determine the non-static exterior metric, but
we expect on general physical grounds that the non-static
behaviour will be transient, so that the exterior tends to a
static form.

The collapsing region in general contains dust and also energy
density on the brane from KK stresses in the bulk (``dark radiation"). We show that in
the extreme case where there is no matter but only collapsing
homogeneous KK energy density, there is a unique exterior which is
static for physically reasonable values of the parameters. Since
there is no matter on the brane to generate KK stresses, the KK
energy density on the brane must arise from bulk Weyl curvature.
In this case, the bulk could be pathological. The collapsing KK
energy density can either bounce or form a black hole with a 5D
gravitational potential, and the exterior is of the Weyl-charged
de\,Sitter type \cite{dmpr}, but with
no mass.

\subsection{Gravitational collapse: a no-go theorem}\label{star}

A spherically symmetric collapse region has a Robertson-Walker metric
\begin{equation}\label{1}
ds^2=-d\tau^2+a(\tau)^2(1+ {\textstyle {1\over4}}
kr^2)^{-2}\left[dr^2+r^2d\Omega^2\right]\, ,
\end{equation}
where $\tau$ is the proper time of the perfect fluid. This implies that the
four velocity is $u^\alpha=\delta^\alpha_\tau$, so that
\be
T^\alpha_\beta=\mbox{diag}\left[-\rho(\tau),p(\tau),p(\tau),p(\tau)\right]\ .
\ee
Moreover the symmetries of the geometry force the four-dimensional Einstein tensor
to have the following properties
\ba
G^r_r=G^\theta_\theta=G^\phi_\phi\ ,
\ea
and to have the off-diagonal terms identically zero. This implies
\ba
Q_\mu=0=\Pi_{\mu\nu}\ .
\ea
Whether or not this model has a regular bulk solution is not possible
to determine here. However if the metric (\ref{1}) applies throughout the brane (i.e. there is
no collapse and no exterior region) we can fully integrate the five dimensional equations
\cite{bdl} obtaining a regular bulk. We will see this in the last chapter dedicated to cosmology.

In this case the only non-local non-zero equation left is
\be
\dot{\cal U}+\frac{4}{3}\Theta{\cal U}=0\ ,
\ee
where $\Theta=3\ \dot a/a$. Then
\be
{\cal U}=\frac{C}{\lambda a^4}\ ,
\ee
where $C$ is an integration constant.

Then from the $G^0_0$ components of (\ref{eq:effective}) we get the modified Friedmann equation
\begin{eqnarray} \label{evol}
\frac{\dot{a}^2}{a^2}= {\textstyle{8\over3}}\pi G_N\rho\left(1 +
\frac{\rho}{2\lambda}\right) + \frac{C}{\lambda a^4} -\frac{k}{a^2}+
{\textstyle{1\over3}} {\Lambda}\, .
\end{eqnarray}
The $\rho^2$ term,
which is significant for $\rho\gtrsim \lambda$, is the high-energy
correction term, following from $S_{\mu\nu}$. Standard Friedmann
evolution is regained in the limit $\lambda^{-1} \to 0$.

At this point we would like to use this geometry as the interior of a collapsing region and try to
match it with a static exterior. Since we are in comoving coordinates, the boundary of the
collapsing region is described by the implicit function
\be
\Phi=r-r_0=0\ ,
\ee
so that its normal unit vector is $n_\alpha=a(1+kr^2/4)^{-1}\delta_\alpha^r$.
From the junction conditions (\ref{Appjc2})
\ba\label{jc}
\left[G^r_r\right]=\left[p^{\rm eff}\right]=0\ ,
\ea
where again $\left[f\right]:=f(r_0^+)-f(r_0^-)$, and $r_0^\pm$ is respectively the limit
from the exterior and the interior to the boundary. Using the same discussion of par.
(\ref{sectionField}) we can argue also that
\be
\left[p\right]=0\ ,
\ee
so that for a vacuum exterior $p^\pm=0$. Therefore the collapsing region must be dust.

In this case in the interior
\be
p^{\rm eff}=\frac{\rho^2}{2\lambda}+\frac{\cal U}{3}\ ,
\ee
implying that the exterior can not be static.

In the following we find a ``measure" of the non-static behaviour of the exterior solution.
This is encoded in a Ricci anomaly. As we will see in the final section this anomaly can be
interpreted holographically as the Weyl anomaly due to the Hawking evaporation for a collapsing
body.

The conservation equations (\ref{pc1}) for a dust cloud gives
\ba
\rho=\rho_0\left(\frac{a_0}{a}\right)^3 ,
\ea
where $a_0$ is
the epoch when the cloud started to collapse. The proper radius
from the centre of the cloud is $R(\tau)=r a(\tau)/(1+ {\textstyle
{1\over4}} kr^2)$. We call the collapsing boundary surface $\Sigma$,
which has as a proper radius $R_\Sigma(\tau)= r_0 a(\tau)/(1+
{\textstyle {1\over4}} kr_0^2)$.

We can rewrite the modified Friedmann equation on the interior
side of $\Sigma$ as
\begin{equation}\label{geo1}
\dot{R}^2= \frac{2G_NM}{R}+ 3\frac{G_NM^2}{4\pi\lambda R^4}+ \frac{Q}{\lambda
R^2}+ E+ \frac{\Lambda}{3} R^2\,,
\end{equation}
where the ``physical mass" $M$ (total energy per proper star
volume) and the total ``tidal charge" $Q$ are
\begin{equation}\label{mq}
M={\textstyle{4\over3}}\pi a_0^3 r_0^3\frac{\rho_0}{(1+ {\textstyle
{1\over4}}kr_0^2)^3},\ \ Q=C\frac{r_0^4}{(1+ {\textstyle
{1\over4}}kr_0^2)^4},
\end{equation}
and the ``energy" per unit mass is given by
\begin{equation}\label{en}
E=-\frac{kr_0^2}{(1+ {\textstyle {1\over4}}kr_0^2)^2}>-1 \,.
\end{equation}

Now we assume that the exterior is static, and satisfies the
standard 4D junction conditions. Then we check whether this
exterior is physical consistent by imposing the modified Einstein
equations (\ref{eq:effective}) for vacuum, i.e.\ for $T_{\mu\nu} =0 =
S_{\mu\nu}$. The standard 4D Israel matching conditions, which we
assume hold on the brane, require that the metric and the
extrinsic curvature of $\Sigma$ be continuous. The extrinsic
curvature is continuous if the metric is continuous and if $\dot
R$ is continuous~\cite{s}. We therefore need to match the metrics
and $\dot R$ across $\Sigma$.

The most general static spherical metric that could match the
interior metric on $\Sigma$ is
\begin{eqnarray}
ds^2= -F(R)^2A(R)dt^2 +\frac{dR^2}{A(R)} +R^2d\Omega^2,\label{s1}
\end{eqnarray}
where
\be \label{A}
A(R) =1-2G_Nm(R)/R.
\ee

We need two conditions to determine the functions $F(R)$ and
$m(R)$. Now $\Sigma$ is a freely falling surface in both metrics.
Therefore the first condition comes from the comparison of the geodesic
equations of the two metrics. The geodesic equation for a radial trajectory seen
from the exterior is calculated as follows. For a time-like geodesic we have
\be\label{geodesic}
\frac{ds^2}{d\tau^2}=-1=-F(R)^2A(R)\dot t^2+\frac{\dot R^2}{A}\ .
\ee
Now since we are requiring that the exterior metric is static, it exist a
time-like Killing vector $\xi^\alpha$ such that
\ba
\pounds_\xi g_{\alpha\beta}&=\xi_{\alpha}{}_;{}_\beta+\xi_{\beta}{}_;{}_\alpha=0\ ,\
\xi^\alpha \xi_\alpha<0\nonumber
\ea
where $\pounds$ denotes the Lie derivative. If $u^\alpha$ is a geodesic vector,
$u^\alpha\nabla_\alpha u_{\beta}=0$, we have that
\be\label{tildeE}
\tilde E^{1/2}=-\xi^\alpha u_{\alpha}=-F(R)^2A(R)\dot t\ ,
\ee
is a constant along the geodesic motion, or $u^\alpha\nabla_\alpha \tilde E=0$.
Using (\ref{geodesic}) and (\ref{tildeE}), the radial geodesic
equation for the exterior metric gives
$\dot{R}^2=-A(R)+\tilde{E}/F(R)^2$.
Comparing this with eq.~(\ref{geo1}) gives one condition. The
second condition is easier to derive if we change to null
coordinates\footnote{These coordinates, denoted by $(v,r,\theta,\phi)$ are
such that the light-cones are described by the equation $v=\mbox{const}$.}. The
exterior static metric, with
$dv=dt+dR/[F(1-2Gm/R)]$, becomes
\begin{eqnarray}
ds^2&=& -F^2Adv^2 +2FdvdR+R^2d\Omega^2\,.\label{s1'}
\end{eqnarray}
The interior Robertson-Walker metric takes the form~\cite{tesi}
\begin{eqnarray}
ds^2&=& -\tau_{,v}^2\frac{1-(k+\dot{a}^2)R^2/a^2}{1-kR^2/a^2} dv^2
+2 \tau_{,v} \frac{dvdR}{\sqrt{1-kR^2/a^2}}
+R^2d\Omega^2\,,\label{s1''}
\end{eqnarray}
where
\be
d\tau=\tau_{,v}dv+\frac{1+{1\over4}kr^2}{r\dot
a-1+{1\over4}kr^2} dR\ .
\ee
Comparing eqs.~(\ref{s1'}) and (\ref{s1''})
on $\Sigma$ gives the second condition. The two conditions
together imply that $F$ is a constant, which we can take as
$F(R)=1$ without loss of generality (choosing $\tilde E=E+1$), and
that
\begin{equation}\label{s3}
m(R)= M+\frac{3M^2}{8\pi\lambda R^3}+ \frac{Q}{2G_N\lambda R}+ \frac{\Lambda
R^3}{6G_N}\,.
\end{equation}
In the limit $\lambda^{-1}\to 0$, we recover the 4D general relativity
Schwarzschild-de\,Sitter solution. However, we note that the above
form of $m(R)$ violates the weak-field perturbative limit in
eq.~(\ref{pert}), and this is a symptom of the problem with a
static exterior. Equations~(\ref{s1}) and (\ref{s3}) imply that
the brane Ricci scalar is
\begin{eqnarray}\label{m1}
R^\mu{}_\mu=4\Lambda+\frac{9G_N M^2}{2\pi\lambda R^6}\,.
\end{eqnarray}
However, the modified Einstein equations (\ref{eq:effective}) for a vacuum exterior imply that
\ba
R_{\mu\nu}&=&\Lambda g_{\mu\nu}- {\cal E}_{\mu\nu}\\
R^\mu{}_\mu &=& 4\Lambda\,. \label{vac}
\ea
Comparing with eq.~(\ref{m1}), we see that a static exterior is
only possible if $M/\lambda=0\,$. We can therefore interpret $R^\mu{}_\mu$
as a kind of ``potential energy" that must be released from the star during the collapse, due
only to braneworld effects. We emphasize
that this no-go result does not require any assumptions on the
nature of the bulk spacetime.

In summary, we have explored the consequences for
gravitational collapse of braneworld gravity effects, using the
simplest possible model, i.e.\ an Oppenheimer-Snyder-like collapse on a
generalized Randall-Sundrum type brane. Even in this simplest
case, extra-dimensional gravity introduces new features. Using
only the projected 4D equations, we have shown, independent of the
nature of the bulk, that the exterior vacuum on the brane is
necessarily {\em non-static}. This contrasts strongly with general relativity,
where the exterior is a static Schwarzschild spacetime. Although
we have not found the exterior metric, we know that its non-static
nature arises from (a)~5D bulk graviton stresses, which transmit
effects nonlocally from the interior to the exterior, and (b)~the
non-vanishing of the effective pressure at the boundary, which
means that dynamical information on the interior side can be
conveyed outside. Our results suggest that gravitational collapse
on the brane may leave a signature in the exterior, dependent upon
the dynamics of collapse, so that astrophysical black holes on the
brane may in principle have KK hair.

We expect that the non-static exterior will be transient and partially {\em
non-radiative}, as follows from a perturbative study of non-static
compact objects, showing that the Weyl term ${\cal E}_{\mu\nu}$ in
the far-field region falls off much more rapidly than a radiative
term~\cite{ssm}. It is reasonable to assume that the exterior
metric will become static at late times and tend to Schwarzschild,
at least  at large distances.

\subsection{Gravitational collapse of pure Weyl energy}

The one case that escapes the no-go theorem is $M=0$.  In general relativity,
$M=0$ would lead to vacuum throughout the spacetime, but in the
braneworld, there is the tidal KK stress on the brane, i.e.\ the
$Q$-term in eq.~(\ref{geo1}).  The possibility of black holes
forming from KK energy density was suggested in~\cite{dmpr}. The
dynamics of a Friedmann universe (i.e.\ without exterior),
containing no matter but only KK energy density (``dark
radiation") has been considered in~\cite{bcg}. In that case, there
is a black hole in the Schwarzschild-AdS bulk, which sources
the KK energy density. Growth in the KK energy density corresponds
to the black hole and brane moving closer together; a singularity
on the brane can arise if the black hole meets the brane. Here we
investigate the collapse of a bound region of homogeneous KK
energy density within an inhomogeneous exterior. It is not clear
whether the bulk black hole model may be modified to describe this
case, and we do not know what the bulk metric is. However, we know
that there must be 5D Weyl curvature in the bulk, and that the
bulk could be pathological, with a more severe singularity than
Schwarzschild-AdS. Even though such a bulk would be unphysical
(as in the case of the Schwarzschild black string \cite{blackstring}), it is
interesting to explore the properties of a brane with collapsing
KK energy density, since this idealized toy model may lead to
important physical insights into more realistic collapse with
matter and KK energy density.

The exterior is static and unique, and given by the Weyl-charged
de\,Sitter metric
\begin{equation}\label{wd}
ds^2=-Adt^2+\frac{dR^2}{A}+R^2d\Omega^2\,,~M=0\,,
\end{equation}
if $A>0$ ($A$ is given by (\ref{A}) together with the definition (\ref{s3})). For $Q=0$ it is de\,Sitter, with horizon
$H^{-1}=\sqrt{3/\Lambda}$. For $\Lambda=0$ it is the special case
$M=0$ of the solutions given in~\cite{dmpr}, and the length scale
$H_Q^{-1}=\sqrt{|Q|/\lambda}$ is an horizon when $Q>0$; for $Q<0$,
there is no horizon. As we show below, the interplay between these
scales determines the characteristics of collapse.

For $\Lambda=0$, the exterior gravitational potential is
\begin{equation}
\Phi=\frac{Q}{2\lambda R^2}\,,
\end{equation}
which has the form of a purely 5D potential when $Q>0$. When
$Q<0$, the gravitational force is repulsive. We thus take $Q>0$ as
the physically more interesting case, corresponding to {\em
positive} KK energy density in the interior. However we note the
remarkable feature that $Q>0$ also implies {\em negative} KK
energy density in the exterior:
\begin{equation}
-{\cal E}_{\mu\nu}u^\mu u^\nu=\left\{ \begin{array}{ll}
+3Q/(\lambda R_\Sigma^4)\,, & R<R_\Sigma\,, \\ 
-Q/(\lambda R^4)\,, & R> R_\Sigma\,. \end{array}\right.
\end{equation}
Negativity of the exterior KK energy density in the general case
with matter has been previously noted~\cite{ssm,dmpr}.

The boundary surface between the KK ``cloud" and the exterior has
equation of motion $\dot{R}^2=E-V(R)$, where $V=A-1$.

For $\Lambda=0$, the cases are:\\
$\bf Q>0$:\, The cloud collapses
for all $E$, with horizon at $R_{\rm
h}=H_Q^{-1}=\sqrt{Q/\lambda}$. For $E<0$, given that $E>-1$, the
collapse can at most  start from rest at
\be
R_{\rm
max}=\sqrt{\frac{Q}{\lambda|E|}}>R_{\rm h}\ .
\ee
{$\bf Q<0$}:\, It follows that if
$E>0$, there is no horizon, and the cloud bounces at
\be
R_{\rm
min}=\sqrt{|Q|/(\lambda E)}\ .
\ee

\begin{figure}[t]
\begin{center}
\includegraphics[width=8cm,height=4cm,angle=0]{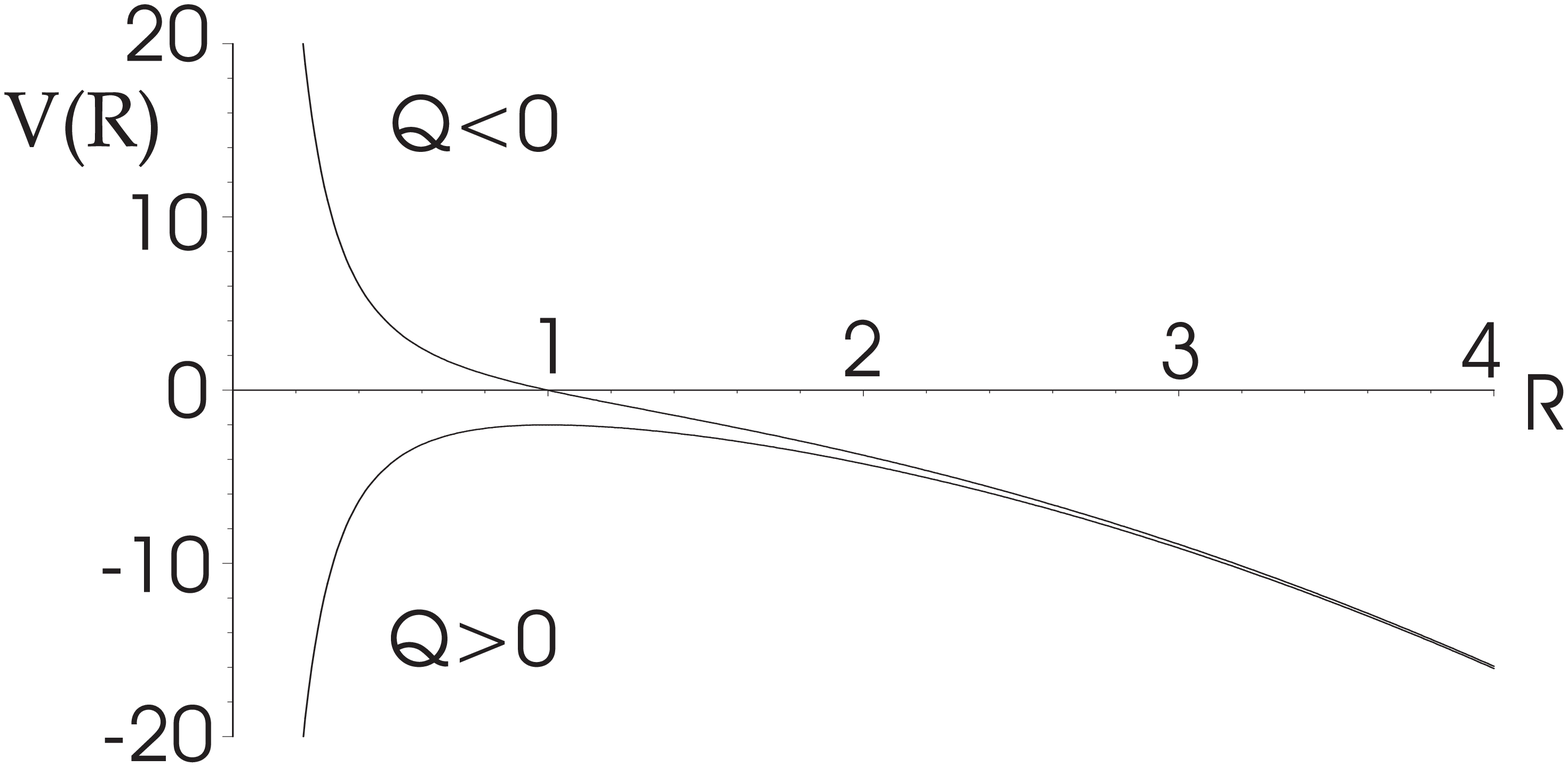}
\end{center}
\caption{ \label{fig3} The potential $V(R)$ for $\Lambda> 0$, with
$R$ given in units of $R_{\rm c}$ and $V$ given in units of
$V_{\rm c}$.}
\end{figure}

For $\Lambda> 0$, the potential is given by
\be
V/ V_{\rm c}=-\left(\frac{R}
{R_{\rm c}}\right)^2\left[1+\epsilon(R_{\rm c}/ R)^4\right]\,,
\ee
where
$V_{\rm c}={H/ H_Q}\,,$ $R_{\rm c}=1/\sqrt{HH_Q}\,,$ and
$\epsilon={\rm sgn}\,Q$ (see fig. (\ref{fig3})). The horizons are given by
\begin{equation}
\left(R_{\rm h}^\pm\right)^2= \frac{R_{\rm c}^2}{2V_{\rm c}} \left[1\pm
\sqrt{1-4\epsilon V_{\rm c}^2}\right]\,.\label{hor}
\end{equation}
If $\epsilon>0$ there may be two horizons; then $R_{\rm h}^-$ is
the black hole horizon and $R_{\rm h}^+$  is a modified de\,Sitter
horizon. When they coincide the exterior is no longer static, but
there is a black hole horizon. If $\epsilon<0$ there is always one
de\,Sitter-like horizon, $R_{\rm h}^+$.
{$\bf Q>0$}:\, The potential has a
maximum $-2V_{\rm c}$ at $R_{\rm c}$. If $E>-2V_{\rm c}$ the cloud
collapses to a singularity. For $V_{\rm c}>{1\over2}$, i.e.\
$Q>3\lambda/4\Lambda$, there is no horizon, and a naked
singularity forms. For for $V_{\rm c}={1\over2}$ there is one
black hole horizon $R_{h}^{-}=R_{h}^{+}=H^{-1}/\sqrt{2}$. If
$E\leq -2V_{\rm c}$, then eq.~(\ref{en}) implies $V_{\rm
c}<{1\over2}$, so there are always two horizons in this case.
Either the cloud collapses from infinity down to $R_{\rm min}$ and
bounces, with $R_{\rm min}<R_{\rm h}^+$ always, or it can at most
start from rest at $R_{\rm max} (>R_{\rm h}^-)$, and collapse to
a black hole, where ($\epsilon=1$)
\begin{eqnarray}
R^2_{\stackrel{\scriptstyle \rm min}{\rm max}} =\frac{R_{\rm c}^2}{2V_{\rm c}}
\left[-E \pm \sqrt{E^2- 4\epsilon V_{\rm c}^2}\right]
\,.\label{mini}
\end{eqnarray}
{$\bf Q<0$}:\, The potential is
monotonically decreasing, and there is always an horizon, $R_{\rm
h}^+$. For all $E$, the cloud collapses to $R_{\rm min} (<R_{\rm
h}^+)$, and then bounces, where $R_{\rm min}$ is given by
eq.~(\ref{mini}) with $\epsilon=-1$.

In summary we have analyzed the idealized collapse of homogeneous KK energy
density whose exterior is static and has purely 5D gravitational
potential. The collapse can either come to a halt and bounce, or
form a black hole or a naked singularity, depending on the
parameter values. This may be seen as a limiting idealization of a
more general spherically symmetric but inhomogeneous case. The
case that includes matter may be relevant to the formation of
primordial black holes in which nonlinear KK energy density could
play an important role.

\subsection{Holographic limit for $\lambda$ via Hawking process}

Recently it has been suggested that Hawking evaporation of a four dimensional black hole
may be described holographically by the five-dimensional classical black hole metric \cite{hawholo}.
Here we see how this correspondence works in the case of an Oppenheimer-Snyder collapse.
It seems that the Hawking effect does not have memory of
the collapsing process \cite{bida}, and we expect that the conclusions
we obtain are general.

A Schwarzschild black hole produces quantum
mechanically a trace anomaly for the vacuum energy-momentum tensor. This is due
to the gravitational energy via curvature of spacetime which excites the vacuum.
We discuss the analogy between the
quantum stress-tensor anomaly and the Ricci anomaly found for a dust cloud
gravitational collapse, (\ref{m1}).

In order to simplify the problem we will calculate explicitly the conformal anomaly only
for a massless scalar field in a curved background. We will then comment on the general
vacuum excitations and give a measure for the tension of the brane using the correspondence.

\subsubsection{Green function for a massless scalar field}

Here we will basically follow \cite{bida}.

A general local Lagrangian for a scalar field $\Phi$ coupled to gravity is
\be\label{lagmassless}
{\cal L}=\frac{1}{2}\sqrt{-g}\left[\partial_\mu \Phi\partial^\mu \Phi-(m^2-\xi R)\Phi^2\right]\ ,
\ee
where $\xi$ is an a dimensionless constant, $R$ is the Ricci scalar and $m$ is the mass
of the field. The classical equations follow as
\be\label{evolscalar}
\left[\Box + m^2-\xi R\right]\Phi=0\ .
\ee
Since in the following we will consider vacuum excitations,
we make the Lagrangian invariant under conformal transformation of the metric. The only non-trivial possibility is for
$m=0$ and $\xi=1/6$ in four dimensions. Now as usual we define the momentum
\be
\pi=\frac{\partial {\cal L}}{\partial(\partial_t\Phi)}\ ,
\ee
and the quantum commutation rules are
\ba
\left[\Phi(t,x),\Phi(t,x')\right]&=&0\ ,\cr
\left[\pi(t,x),\pi(t,x')\right]&=&0\ ,\cr
\left[\Phi(t,x),\pi(t,x')\right]&=&i\delta^3(x-x')/\sqrt{-g}\ .\nonumber
\ea
If we define the Green function as the two-point correlation function
\be
iG_F(x-x')=\langle 0|T(\Phi(x)\Phi(x'))|0\rangle\ ,
\ee
where $T$ is the temporal order operator and $|0\rangle$ is the vacuum, we have
\be\label{scalarGreen}
\left[\Box -\frac{1}{6} R\right]G_F(x-x')=-\frac{\delta^4(x-x')}{\sqrt{-g}}\ .
\ee
In order to define a meaningful concept of a particle in curved space, we have to work only
with local quantities. This means that we consider, as proper particles, only the high
frequencies in the Fourier space of $\Phi$. Therefore we will be interested in having
a solution for (\ref{scalarGreen}) only
in the limit $x\rightarrow x'$.

Introducing the Riemann coordinates $y^\mu=x^\mu-x'^\mu$, we have \cite{kre,pet}
\ba
g_{\mu\nu}&=\eta_{\mu\nu}+\frac{1}{3}R_{\mu\alpha\nu\beta}y^\alpha y^\beta-
\frac{1}{6}R_{\mu\alpha\nu\beta}{}_{;\gamma}y^\alpha y^\beta y^\gamma\cr
&+\left[\frac{1}{20}R_{\mu\alpha\nu\beta}{}_{;\gamma\delta}+\frac{2}{45}R_{\alpha\mu\beta\lambda}
R^\lambda{}_{\gamma\nu\delta}\right]y^\alpha y^\beta y^\gamma y^\delta+...\ ,\nonumber
\ea
where all the coefficients are evaluated at $y^\alpha=0$.

We can now solve (\ref{scalarGreen}) around $y^\alpha=0$
in Fourier space and apply an anti-Fourier-transform, to obtain
\be
G_F(x)=\frac{-i}{(4\pi)^2\sqrt{-g}}\int^\infty_0i ds (is)^{-2}F(x,s)\ ,
\ee
where
\be
F(x,s)=1+a(x)(is)^2\ ,
\ee
and
\be
a(x)=\frac{1}{180}\left[R_{\alpha\beta\gamma\delta}R^{\alpha\beta\gamma\delta}-R^{\alpha\beta}
R_{\alpha\beta}+\Box R\right]\ .
\ee

\subsubsection{Trace anomaly and the Hawking effect}\label{holocollapse}

In semiclassical gravity the gravitational field is treated classically and the matter
fields are
treated quantum mechanically. In this case what actually sources the gravitational field is the
expectation value of the energy-momentum tensor,
\be\label{R}
R_{\mu\nu}-{1\over 2}R g_{\mu\nu}+\Lambda_R g_{\mu\nu}=8\pi G_R \langle T_{\mu\nu}\rangle\ ,
\ee
where the subscript $R$ indicates that the cosmological and Newton constants must be
renormalized. It is possible to prove that for a conformal theory we have
$G_R=G_N$, whereas $\Lambda_R$ must be experimentally evaluated \cite{bida}
\footnote{In general the left hand side can have additional higher curvature and derivative
terms
due to the renormalization process. It is possible to find exactly the expression of these
corrections up to multiplicative constants that must be evaluated experimentally \cite{bida}.
We do not really need in the following the equation (\ref{R}), but we wish to comment that
such corrections are not divergent-free, and therefore allow the energy-momentum tensor to be
non-conserved. Since we believe in the Einstein principle that the stress tensor is conserved
by gravitational identities (e.g. Bianchi) and not by additional constraints introduced by hand,
we set these constants zero.}. Now
the classical energy-momentum tensor
is connected to the variation of the action $S_m$ of matter,
\be
\frac{2}{\sqrt{-g}}\frac{\delta S_m}{\delta g^{\mu\nu}}=T_{\mu\nu}\ .
\ee
Given the introduction of quantum degrees of freedom, the semiclassical energy-momentum tensor
differs from the classical one.
This implies that we should find an effective action to encode the
quantum degrees freedom as well. Calling it $W$ we can define
\be
\frac{2}{\sqrt{-g}}\frac{\delta W}{\delta g^{\mu\nu}}=\langle T_{\mu\nu}\rangle\ .
\ee
From path integral formalism, the $n$-point correlation function can be obtained by
the generating functional
\be
Z[J]=\int{\cal D}[\Phi]\exp\left[iS_m[\Phi]+i\int J(x)\Phi(x)\sqrt{-g}d^4x\right]\ ,
\ee
where ${\cal D}[\Phi]$ indicates the integration over all the functions $\Phi$ and
\be
\left(\frac{\delta^n Z}{\delta J(x^1)...\delta J(x^n)}\right)\Big|_{J(x^i)=0}=\langle \mbox{out}
,0|T(\Phi(x^1)...\Phi(x^n))|0,\mbox{in}\rangle\ .
\ee
Now we have \cite{schwinger}
\be\label{schwinger}
\delta Z[0]=i\int {\cal D}[\Phi]\delta S_m e^{iS_m[\Phi]}=i\langle \mbox{out},0|
\delta S_m|0,\mbox{in}\rangle\ ,
\ee
and therefore we obtain
\be
\frac{2}{\sqrt{-g}}\frac{\delta Z[0]}{\delta g^{\mu\nu}}=i\langle \mbox{out},0|
T_{\mu\nu}|0,\mbox{in}\rangle\ .
\ee
Then
\be
\frac{2}{\sqrt{-g}}\frac{\delta\ln Z[0]}{\delta g^{\mu\nu}}=i\frac{\langle \mbox{out},0|
T_{\mu\nu}|0,\mbox{in}\rangle}{\langle \mbox{out},0|0,\mbox{in}\rangle}=
i\langle T_{\mu\nu}\rangle\ .
\ee
This means that we can identify $Z[0]=e^{iW}$.

Integrating by parts the action for a massless scalar field with Lagrangian (\ref{lagmassless}),
we get\footnote{The boundary term is taken to be zero.}
\be
S_m=-\frac{1}{2}\int d^4 x\sqrt{-g}\Phi \left(\Box -\frac{1}{6}R\right)\Phi=
-\frac{1}{2}\int d^4 x d^4 y\sqrt{-g}\Phi(x) K_{xy}\Phi(y)\ ,
\ee
where
\be
K_{xy}=\left(\Box-\frac{1}{6}R\right)\delta^4(x-y)\sqrt{-g}\ ,
\ee
so that
\be
K_{xy}^{-1}=-G_F(x,y)\ .
\ee
If we change now the variable in (\ref{schwinger}), using
\be
\Phi'(x)=\int d^4y K_{xy}^{1/2}\Phi(y)\ ,
\ee
we obtain
\be
Z[0]\propto (\mbox{det} K_{xy}^{1/2})^{-1}=\sqrt{\mbox{det}(-G_F)}\ .
\ee
Then
\be
W=-i\ln Z[0]=-\frac{1}{2}i\mbox{tr}[\ln(-G_F)]\ .
\ee
We are here interested in the trace of the energy-momentum tensor, which is
\be
\frac{2}{\sqrt{-g}}g^{\mu\nu}\frac{\delta W}{\delta g^{\mu\nu}}=g^{\mu\nu}\langle T_{\mu\nu}\rangle
=\langle T\rangle\ .
\ee
After laborious calculations \cite{duff} and using renormalization techniques,
one obtains the following trace anomaly for the energy-momentum
tensor
\be\label{traceanomaly}
\langle T\rangle=\frac{1}{2880\pi^2}\left[R_{\alpha\beta\gamma\delta}R^{\alpha\beta\gamma\delta}
-R_{\alpha\beta}R^{\alpha\beta}+\Box R\right]\ .
\ee
We now specialize this anomaly to the Schwarzschild background \cite{chrifu}, obtaining
\be
\langle T\rangle =\frac{G_N^2}{60\pi^2}\frac{M^2}{R^6}\ .
\ee
These calculations are made having in mind that the vacuum is only described by a scalar field,
but for more general fields one can find \cite{chridu}
\be\label{anomalyTq}
\langle T\rangle =-u\frac{G_N^2}{60\pi^2}\frac{M^2}{R^6}\ .
\ee
where
\be
u=12N_1-N_0-14N_{1/2}\ ,
\ee
and $N_{0,1/2,1}$ are respectively the number of species of spins $0,1/2,1$ of the theory considered.

Now we go back to the gravitational collapse in the braneworld. We can interpret the anomaly
(\ref{m1}) of the Ricci scalar as due to a non-zero energy momentum tensor for the exterior
which is extracting energy (via evaporation) from the collapsing object\footnote{This
in fact is what happens quantum mechanically \cite{bida}.}. Indeed
if we are not too close to the body and we consider only short periods in the evaporation
process, as in standard
semiclassical calculations, we can neglect the
back-reaction on the metric and consider this process as nearly time-independent \cite{DeWitt}.

The braneworld calculation tells us that in order to release the potential energy
of eq. (\ref{m1}), we must have an effective non-zero energy momentum tensor with trace
\be
T^{\rm eff}=-\frac{9}{16\pi^2\lambda}\frac{M^2}{R^6}\ .
\ee
But this is the same quantum anomaly as (\ref{anomalyTq})
if we identify
\be
\lambda=\frac{135}{4uG_N^2}\ .
\ee
If we take this analogy seriously, then we have an indirect measure for the tension of the brane
\be
\lambda\sim 10 M_p^4 u^{-1}\ ,
\ee
where $M_p\sim 10^{19}\ \mbox{GeV}$ is the Planck mass. Moreover we obtain
information on the theory which describes the Hawking process. Indeed since $\lambda$ is
positive we should have
\be\label{emp}
N_1>\frac{N_0}{12}+\frac{7}{6}N_{1/2}\ .
\ee
As pointed out by \cite{Emparan}, this is incompatible with a SYM theory with ${\cal N}=4$ at
large $N$, which is the quantum counterpart of the AdS/CFT correspondence \cite{M1}.
This means that this process could be a new test of the holographic principle.

\newpage
\chapter{Cosmology in a generalized braneworld}

\footnote{This chapter is based on \cite{barcelo, GS1, GS2, DG}.}In recent decades,
developments in cosmology have been strongly
influenced by high-energy physics.  A remarkable example of this
is the inflationary scenario and all its variants.
This influence has been growing and becoming more and more important.

It is a general belief that Einstein gravity is a
low-energy limit of a quantum theory of gravity which is still unknown.
Among promising candidates we have string theory, which suggests that
in order to have a ghost-free action, quadratic curvature corrections to
the Einstein-Hilbert action must be proportional to the Gauss-Bonnet
term~\cite{BoDe:85}. An example has been given already in the first chapter.
This term also plays a fundamental role in Chern-Simons
gravitational theories~\cite{Cha:89}.
However, although being a string-motivated scenario, the RS model and
its generalizations~\cite{SMS} do not include these terms.
From a geometric point of view, the combination of the
Einstein-Hilbert and Gauss-Bonnet term constitutes, for 5D spacetimes,
the most general Lagrangian producing second-order field
equations~\cite{Lov} (see also~\cite{lovelock}).

These facts provide a strong motivation for the study of braneworld
theories including a Gauss-Bonnet term.
Recent investigations of this issue have shown~\cite{MeOl:00}
that the metric for a vacuum 3-brane (domain wall) is, up to a redefinition
of constants, the warp-factor metric of the RS scenarios. This is because AdS is conformally
Minkowskian and the Gauss-Bonnet term is topological on the boundary. The existence
of a KK zero-mode localized on the
3-brane producing Newtonian gravity at low energies, has also been demonstrated
~\cite{local} (see
also~\cite{Ne:01a,ChNeWe:01}). This can be simply proved by considering that higher
curvature terms in the action can produce only
higher order corrections to the Newtonian gravity.
Properties of black hole solutions in
AdS spacetimes have been studied in~\cite{bhads,Ca:01}.  The cosmological
consequences of these scenarios are less well understood.  This issue has
been studied in~\cite{AbMo:01}. However, only simple ans\"atze for
the 5D metric (written in Gaussian coordinates as
in~\cite{bdl})  were considered, e.g. the separability of the
metric components in the time and extra-dimension coordinates.
One can see that this assumption is too strong even in
RS cosmological scenarios~\cite{bdl}, where it leads to a
very restrictive class of cosmological models, not representative of
the true dynamics.
Other results with higher-curvature terms in braneworld scenarios
are considered in~\cite{others}.

In this chapter we study the equations governing the dynamics of
FRW cosmological models in braneworld
theories with a Gauss-Bonnet (GB) term (Lanczos gravity \cite{Lanczos}). In doing that
we study the cosmological behaviour of shells
(or branes) that are thin but still of a finite
thickness $T$. In this way we want to shed some light on how
the zero thickness is attained in the presence of GB interactions.
This limit has been studied for Einstein gravity in \cite{Mounaix}.
Thick shells in the context of GB interactions
have been already studied in \cite{Corradini} and \cite{Giovannini}, but
with a focus on different aspects than those here.
The conclusion of our analysis here is twofold.
On the one side, our results show that there is a generalized Friedmann equation
 \cite{Charmousis} that can be found by using
a completely general procedure, in which  the energy density of the brane
in the thin-limit is related to the averaged density.  However,
considering specific geometric configurations, one can find another form for
the Friedmann equation, such as in \cite{GS1},
with a procedure in which
the energy density of the brane in the thin-limit comes from
the value of the boundary density in the thick-brane model.
We also argue that the information lost when treating a
real thin shell
as infinitely thin, is in a sense larger in Lanczos
gravity than in the analogous situation in standard General Relativity.

Let us explain further this last point. From a physical point of view,
in the process of passing from the notion of function to that of
distribution, one loses information. Many different series of functions
define the same limiting distribution. For example, the series
\begin{equation}
f_T(y)=\left\{
\begin{array}[c]{l}
0 ~~~{\rm for}~~~ |y| > T/2\,, \\
\mbox{} \\
{1 \over T}  ~~~{\rm for}~~~ |y| < T/2\,;
\end{array}\right.
~~~~~~~~
g_T(y)=\left\{
\begin{array}[c]{l}
0 ~~~{\rm for}~~~ |y| > T/2\,, \\
\mbox{} \\
{12 y^2 \over T^3} ~~~{\rm for}~~~ |y| < T/2\,,
\end{array}\right.
\end{equation}
define the same limiting Dirac delta distribution. The distribution
only takes into account the total conserved area delimited by the
series of functions. The gravitational field equations relate geometry
with matter content.  If we take the matter content to have some
distributional character, the geometry will acquire also a
distributional character. When analyzing the thin-limit of branes in
Einstein gravity, by constructing series or families of solutions
parametrized by their thickness, we observe that the divergent parts
of the series of functions that describe the density-of-matter profile
transfer directly to the same kind of divergent parts in the
description of the associated geometry. Very simple density profiles
[like the above function $f_T(y)$] are associated with very simple
geometric profiles, and vice-versa.  However, when considering
Lanczos gravity this does not happen. The divergent parts
of the series describing the matter density and the geometry are
inequivalent. A simple density profile does not correspond to a very
simple geometric profile and vice-versa; on the contrary, we observe
that they have some sort of complementary behaviour. This result
leads us to argue that the distributional description of the
cosmological evolution of a brane in Lanczos gravity
is hiding important aspects of the microphysics, not present when
dealing with pure Einstein gravity. Also, we find that for simple models
of the geometry, one can make compatible the two seemingly distinct
generalized Friedmann equations found in the literature.
We will then extend these concepts for a domain wall formed by a scalar field.

\section{Static thick shells in Einstein and Lanczos gravity}
\label{S:static}

\subsection{Einstein gravity}

To fix ideas and notation let us first describe the simple case
in which we have a static thick brane in an AdS
bulk. We take an ansatz for the metric of the form
\begin{eqnarray}
ds^2=  e^{-2A(y)} \eta_{\mu\nu}dx^\mu dx^\nu +dy^2\,,
\label{brane}
\end{eqnarray}
where $\eta_{\mu\nu}$ is the four-dimensional Minkowski metric.
Comparing with the formulas given in Appendix~\ref{appa} this means
taking $a(t,y)=n(t,y)=\exp(-2A(y))$, and $b(t,y)=1$.
The energy-momentum tensor has the form
\begin{eqnarray}
\tilde\kappa^2\, \tilde T_{AB} = \rho u_{A}u_{B} + p_Lh_{AB} + p_Tn_{A}n_{B}\,,
\label{tAB}
\end{eqnarray}
where $u_A=(-e^{2A},\mb{0},0)$ and $n_A=(0,\mb{0},1)$. Here, $\rho$, $p_L$
and $p_T$ represent respectively the energy density, the longitudinal pressure
and the transverse pressure, and are taken to depend only on $y$.
The Einstein equations $\tilde G_{AB}=-\tilde\Lambda \tilde g_{AB}+\tilde\kappa^2 \tilde T_{AB}$
with a negative cosmological constant,
$\tilde\Lambda\equiv -6/l^2$,
result in the following independent equations for the metric function
$A(y)$:
\begin{eqnarray}
&&3A''-6A'^2=\rho-{6 \over l^2}\,, \label{eins1}  \\
&&6A'^2=p_T+{6 \over l^2}\,, \label{eins2} \\
&&p_L=-\rho\,.
\label{E-equations}
\end{eqnarray}
For convenience, we will hide the $\tilde\kappa^2$ dependence inside the matter magnitudes,
$\rho=\tilde\kappa^2\rho_{\rm true}$, etc.
We also consider a $Z_2$-symmetric geometry around $y=0$.
The brane extends in thickness from $y=-T/2$ to $y=+T/2$. Outside
this region $\rho=p_T=0$, so we have a purely AdS spacetime:
$A(y)=-y /l+b$ for $y\in (-\infty,-T/2)$ and $A(y)=y /l+b$ for
$y\in (T/2,+\infty)$. The junction conditions at $y=-T/2,+T/2$ [see Appendix~\ref{appa}]
tell us that
\begin{eqnarray}
&& A(-T^-/2)=A(-T^+/2),~~~~ A(T^-/2)=A(T^+/2) \,, \\
&& A'(-T^-/2)=A'(-T^+/2),~~~~ A'(T^-/2)=A'(T^+/2)\,.
\label{E-junctions}
\end{eqnarray}
From this and using (\ref{eins2}), we deduce that the
transversal pressure is zero at the brane boundaries
$p_T(-T/2)=p_T(T/2)=0$. Since we are imposing
$Z_2$-symmetry with $y=0$ as {\em fixed point}, hereafter we
will only specify the value of the different functions in the interval
$(-T/2,0)$.

The function $A'$ is odd and therefore interpolates
from $A'(-T/2)=-1/l$ to $A'(0)=0$. If in addition
we impose that the null-energy condition $\rho+p_T=3A'' \geq 0$ be satisfied
everywhere inside the brane, then $p_T$ has to be a negative and monotonically
decreasing function from $p_T(-T/2)=0$ to $p_T(0)=-6/l^2$.
This condition will turn out to be fundamental in defining
a thin-shell limit.

By isolating $A''$ from equations (\ref{eins1}) and (\ref{eins2})
we can relate the
total bending of the geometry on passing through the brane
with its total $\rho+p_T$
\begin{eqnarray}
{6 \over l} =3A'\Big|_{-T/2}^{T/2}=\int_{-T/2}^{T/2}  (\rho+p_T)\; dy.
\end{eqnarray}
At this stage of generality,
one can create different one-parameter families of thick-brane versions
of the Randall-Sundrum thin brane geometry, by parameterizing each
member of a given family by its thickness $T$. The only requirement needed
to do this is that the value of the previous integral must be kept fixed independently
of the thickness of the particular thick-brane geometry.   Thus, each particular
family can be seen as a regularization of Dirac's delta distribution.

We can realize that, provided the condition $\rho+p_T \geq 0$ is satisfied,
there exists a constant $C$, independent of the thickness $T$, such that
$p_T<C$, that is, the profile for $p_T$ is bounded and will not
diverge in the thin-shell limit.
Therefore, in the limit in which the thickness of the branes goes to
zero, $T\rightarrow 0$, the integral of $p_T$ goes to zero with the
thickness. (Strictly speaking, the thin-shell limit is reached when
$T/l\rightarrow 0$ but throughout this paper we are going to
maintain $l$ as a finite constant.)  Instead, the profile of $\rho$ has to
develop arbitrarily large values in order to fulfil
\begin{eqnarray}
{6 \over l} =\lim_{T\rightarrow 0} \int_{-T/2}^{T/2}  \rho\;dy.
\label{average}
\end{eqnarray}

In the thin-shell limit, we can think of the Einstein equations as
providing a relation between the characteristics of the density
profile and the shape of the internal geometry. A very complicated
density profile will be associated with very complicated function
$A(y)$.  Physically we can argue that when a shell becomes very thin,
one does not care about its internal structure and, therefore, one
tries to describe it in the most simple terms. But what exactly is
the meaning of simple? Here we will adopt two different definitions of
simple: The first is to consider that the internal density is
distributed homogeneously throughout the shell when the shell becomes
very thin; the second is to consider that the profile for $A'$ is
such that it interpolates from $A'(-T/2)=-1/l$ to $A'(0)=0$ through
a straight line, or what is the same, that the internal profile of
$A''$ is constant. Again, we require this for very thin shells. This
geometric prescription is equivalent to asking for a constant internal
scalar curvature, since $R=8A''-20A'^2$, and for every thin shell the term
$A'^2$ is negligible relative to the constant $A''$
term. Hereafter, we will use interchangeably the terms {\em straight
interpolation} or {\em constant curvature} for these models. In
building arbitrarily thin braneworld models, one needs the profiles
for the internal density $\rho$ and the internal $A''$ to acquire
arbitrarily high values (they will become distributions in the limit of
strictly zero thickness). In the first of the two simple models
described, the simplicity applies to the divergent parts of the
matter content; in the second, the simplicity applies to the divergent parts of the geometry.
From the physical point of view advocated in the introduction,
simple profiles are those that do not involve losing information in the
process of taking the limit of strictly zero thickness.

We analyze each case independently.

\subsubsection{Constant density profile}

We first define for convenience $z\equiv y/T$ as a scale invariant
coordinate inside the brane. Then,
mathematically, the idea that the density profile, which we will assume
to be analytic inside the brane for simplicity, becomes constant
in the thin-shell limit, can be expressed as follows:
\begin{eqnarray}
\rho(z)=\sum_n \beta_n(T) \; z^{2n},
\label{constant-density-est}
\end{eqnarray}
where
\begin{eqnarray}
\lim_{T\rightarrow 0} T ~ \beta_n(T)\rightarrow 0,
~~~~\forall n\neq 0;~~~~
\lim_{T\rightarrow 0} T ~ \beta_0(T)\rightarrow \rho_b = {\rm constant}\,.
\label{condi}
\end{eqnarray}
For these density profiles, the Einstein equations in the
thin-shell limit tell us that
\begin{eqnarray}
3A''=\beta_0(T)-{6 \over l^2}+6A'^2\,.
\end{eqnarray}
From here we get the profile for $A'$:
\begin{eqnarray}
A'=\sqrt{{\beta_0(T) \over 6}-{1 \over l^2}}
\tan \left(2\sqrt{{\beta_0(T) \over 6}-{1 \over l^2}}\; y \right)\, .
\label{apri}
\end{eqnarray}
Notice that this expression only makes sense for $\beta_0(T)>6/l^2$,
but this is just the regime we are interested in. We have to impose now
the boundary condition $A'(T/2)=1/l$ on the previous expression (\ref{apri}),
\begin{eqnarray}
{1 \over l}=\sqrt{{\beta_0(T) \over 6}-{1 \over l^2}}
\tan \left(\sqrt{{\beta_0(T) \over 6}-{1 \over l^2}}\; T \right).
\end{eqnarray}
In this manner, we have implicitly determined the form of the function
$\beta_0(T)$. In the limit
in which $T\rightarrow 0$ with $T\beta_0(T)\rightarrow \rho_b$,
we find the relation
\begin{eqnarray}
{6 \over l}=\rho_b.
\end{eqnarray}
This condition is just what we expected from the average
condition (\ref{average}).

\subsubsection{Straight interpolation}

In this case, the mathematical idea that in thin-shell limit
the profile for $A'$ corresponds to a straight interpolation,
can be formulated as
\begin{eqnarray}
A''(z)=\sum_n \gamma_n(T) \; z^{2n},
\label{straight-interpolation}
\end{eqnarray}
with
\begin{eqnarray}
\lim_{T\rightarrow 0} T ~ \gamma_n(T)\rightarrow 0,
~~~~\forall n\neq 0;~~~~
\lim_{T\rightarrow 0} T ~ \gamma_0(T)\rightarrow {2 \over l}\ .
\label{straight-interpolation-cond}
\end{eqnarray}
For these geometries, we find that the associated profiles for
$p_T$ and $\rho$ in the thin-shell limit have the form,
\begin{eqnarray}
&&p_T=-{6 \over l^2}\left(1 - 4 z^2 \right) + \omega_1(T,z) \,, \\
&&\rho={6 \over l T}+{6 \over l^2}\left(1 - 4 z^2 \right) +
\omega_2(T,z)\,,
\end{eqnarray}
where here and throughout this chapter, $\omega_n(T,z)$ denotes functions
that vanish in the limit $T\rightarrow 0$.
Now, from this density profile we can see that
\begin{eqnarray}
\lim_{T\rightarrow 0} \int_{-T/2}^{T/2}  \rho\;dy = {6 \over l},
\end{eqnarray}
as we expected.  Moreover, we can see that the boundary value of the
density satisfies $T\rho\Big|_{T/2} \rightarrow 6/l$ in the thin
shell limit, which is the same condition satisfied by the averaged
density, $T \langle \rho \rangle \rightarrow 6/l$.

An additional interesting observation for what follows is the following.
The set of profiles that yield constant density in the thin-shell limit
(\ref{constant-density-est}) and straight interpolation for the
geometric profile (\ref{straight-interpolation}) coincide. Therefore,
in the thin-shell limit one can assume at the same time a constant
internal structure for the density and a straight-interpolation for
the geometry.

\subsection{Lanczos gravity}\label{EinsteinGB}

Let us move now to the analysis of the same ideas in the presence
of the Gauss-Bonnet term. The field equations are now \cite{Lanczos}
\begin{eqnarray}
\tilde G_{AB}+\alpha \tilde H_{AB}=-\tilde \Lambda \tilde g_{AB}+\tilde \kappa^2 \tilde T_{AB}
\,, \label{fieldeq}
\end{eqnarray}
where $\tilde H_{AB}$ is the Lanczos tensor \cite{Lanczos}:
\begin{eqnarray}
\tilde H_{AB}=2\tilde R_{ACDE} \tilde R_{B}^{\; CDE}-4\tilde R_{ACBD}\tilde R^{CD}
-4\tilde R_{AC} \tilde R_{B}^{\; C}+2\tilde R\tilde R_{AB}-\frac{1}{2}\tilde g_{AB}L_{GB}
\,, \label{habterm}
\end{eqnarray}
where $\sqrt{-\tilde g}L_{GB}$ is the Gauss-Bonnet Lagrangian density,
\begin{eqnarray}
L_{GB} = \tilde R^2 -4\tilde R^{AB}\tilde R_{AB}+\tilde R^{ABCD}\tilde R_{ABCD} \,.
\end{eqnarray}
For the ansatz (\ref{brane}) we obtain (see Appendix~\ref{appa})
\begin{eqnarray}
&&3A''(1 - 4\alpha A'^2)-6A'^2(1 - 2\alpha A'^2)=\rho-{6 \over l^2},
\label{EGB-equations-rho} \\
&&6A'^2(1 - 2\alpha A'^2)=p_T+{6 \over l^2}, \label{EGB-equations-pt} \\
&&p_L=-\rho.
\label{EGB-equations-pl}
\end{eqnarray}
The junction conditions for the geometry are the same as before, eq.
(\ref{E-junctions}), implying again the vanishing of the transversal
pressure at the boundaries, $p_T=0$.

In the outside region the solution is a pure AdS spacetime
but with a modified length scale
\begin{eqnarray}\label{modlength}
{1 \over \tilde l} \equiv \sqrt{ {1 \over 4\alpha}
\left(1-\sqrt{1- {8\alpha \over l^2}}\right) }.
\end{eqnarray}
Now, isolating $A''$ from~(\ref{EGB-equations-rho}) and~(\ref{EGB-equations-pt}),
we can relate the total bending of the geometry on passing through the brane
with the integral of $\rho+p_T$ :
\begin{eqnarray}
{6 \over \tilde l}\left(1- {4 \over 3}{\alpha \over {\tilde l}^2}\right)
=(3A'-4 \alpha A'^3)\Big|_{-T/2}^{T/2}=\int_{-T/2}^{T/2}  (\rho+p_T)\; dy.
\end{eqnarray}
Again, if the condition $\rho+p_T \geq 0$ is fulfilled throughout the brane
we have that in the thin shell limit,
\begin{eqnarray}
{6 \over \tilde l}\left(1- {4 \over 3}{\alpha \over {\tilde l}^2}\right)
=\lim_{T \rightarrow 0} \int_{-T/2}^{T/2}  \rho\; dy.
\label{gb-average-condition}
\end{eqnarray}
At this point we can pursue this analysis in the two simple
cases of constant density profile and straight interpolation.
%

\subsubsection{Constant density profile}

Following the same steps as before for a constant density profile
(\ref{constant-density-est})-(\ref{condi}), the equation that one
has to solve in the thin-shell limit is
\begin{eqnarray}
3A''(1-4\alpha A'^2)=\beta_0(T)-{6 \over l^2}+6A'^2(1-2\alpha A'^2)\,.
\end{eqnarray}
Introducing the notation $B\equiv A'$ we reduce this equation to the following
integral
\begin{eqnarray}
y={1 \over 4\alpha } \int_0^B {(4\alpha B^2-1) \; dB
\over B^4 - {1 \over 2\alpha} B^2
- {1 \over 2\alpha} \left({\beta_0(T) \over 6}-{1 \over l^2}\right)}\,.
\end{eqnarray}
The result of performing the integration is
\begin{eqnarray}
y={1 \over 2}
\left[
{1 \over \sqrt{-R_{-}}}\tan^{-1}\left({B \over \sqrt{-R_{-}}} \right)
-{1 \over \sqrt{R_{+}}}\tanh^{-1}\left({B \over \sqrt{R_{+}}}\right)
\right],
\end{eqnarray}
where
\begin{eqnarray}
R_{\pm}={1 \over 4\alpha }
\left[ 1\pm \sqrt{1+ 8\alpha \left({\beta_0(T) \over 6}-{1 \over l^2}\right) }
\right]\,.
\end{eqnarray}
Again, by imposing the boundary condition
\begin{eqnarray}
{T \over 2}={1 \over 2}
\left[
{1 \over \sqrt{-R_{-}}}\tan^{-1}\left({1 \over \tilde l \sqrt{-R_{-}}} \right)
-
{1 \over \sqrt{R_{+}}}\tanh^{-1}\left({1 \over \tilde l \sqrt{R_{+}}}\right)
\right],
\end{eqnarray}
we find the appropriate form for $\beta_0(T)$. With a lengthy but
straightforward calculation, we can check that in the limit $T\rightarrow 0$,
$\beta_0(T)\rightarrow \infty$, we have
\begin{eqnarray}
T\beta_0(T)\rightarrow
{6 \over \tilde l}\left(1- {4 \over 3}{\alpha \over {\tilde l}^2}\right),
\end{eqnarray}
in agreement with condition (\ref{gb-average-condition}).

\begin{figure}[htb]
\vbox{
\hfil
\scalebox{0.5}{\includegraphics{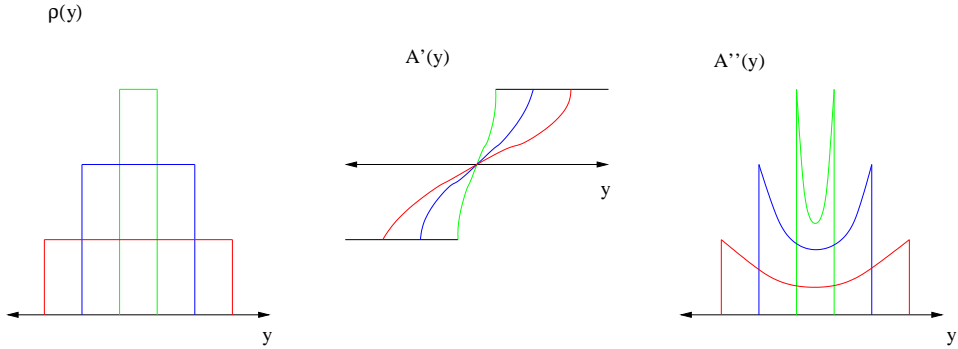}}
\hfil
}
\bigskip
\caption{\label{F:fig1} Family of constant density profiles with decreasing
thickness and associated geometric profile for $A'$ and $A''$.}
\end{figure}

Using this same asymptotic expansion, we can see that, in the
thin-shell limit, the profile for $A'(y)$ satisfies
\begin{eqnarray}
A'(y)-{4 \alpha\over 3}A'(y)^3={1 \over 3} \beta_0(T)\; y.
\end{eqnarray}
Recursively, one can create a Taylor expansion for $A'(y)$.
The first two terms are given by
\begin{eqnarray}
A'(y)&=&{1 \over 3} \beta_0(T)\; y+
{4 \alpha \over 81} \beta_0(T)^3\; y^3+ {\cal O}(y^5)\cr&=&
{1 \over 3} T \;\beta_0(T)\; z+
{4 \alpha \over 81}T^3 \; \beta_0(T)^3\; z^3 + {\cal O}(z^5).
\end{eqnarray}
By differentiating this expression we find
\begin{eqnarray}
A''(y)={1 \over 3} \beta_0(T)+{4 \alpha \over 27}T^2 \; \beta_0(T)^3\; z^2
+{\cal O}(z^4).
\end{eqnarray}
Now, contrary to what happens in Einstein theory, this profile
does not correspond to the set considered in the straight interpolation before
(see
fig. (\ref{F:fig1})).
By looking at (\ref{straight-interpolation}) we can identify
\begin{eqnarray}
\gamma_0(T)\equiv{1 \over 3} \beta_0(T), ~~~~
\gamma_1(T)\equiv{4 \alpha \over 27} T^2 \; \beta_0(T)^3.
\end{eqnarray}
Then, we can see that
\begin{eqnarray}
\lim_{T \rightarrow 0} T\gamma_0(T)=
{2 \over \tilde l}\left(1- {4 \over 3}{\alpha \over {\tilde l}^2}\right)
\neq {2 \over \tilde l}, ~~~~
\lim_{T \rightarrow 0} T\gamma_1(T)=
{32 \alpha \over \tilde l^3}
\left(1- {4 \over 3}{\alpha \over {\tilde l}^2}\right)^3 \neq 0 .
\end{eqnarray}
The coefficients $\gamma_n$ do not satisfy the conditions in
(\ref{straight-interpolation-cond}). Therefore, the scalar
curvature does not have a constant profile as does the energy density.

\subsubsection{Straight interpolation}

As in the Einstein case, the straight interpolation profile for
$A''$ corresponds to
\begin{eqnarray}
A''(z)=\sum_n \gamma_n(T) \; z^{2n},
\label{straight-interpolation2}
\end{eqnarray}
with
\begin{eqnarray}
\lim_{T\rightarrow 0} T ~ \gamma_n(T)\rightarrow 0,
~~~~\forall n\neq 0;~~~~
\lim_{T\rightarrow 0} T ~ \gamma_0(T)\rightarrow {2 \over \tilde l}.
\label{straight-interpolation2-cond}
\end{eqnarray}
From here we can deduce the associated profiles for $p_T$ and $\rho$
by substituting in (\ref{EGB-equations-rho}) and (\ref{EGB-equations-pt}).

\begin{figure}[htb]\label{straightfig}
\vbox{
\hfil
\scalebox{0.5}{\includegraphics{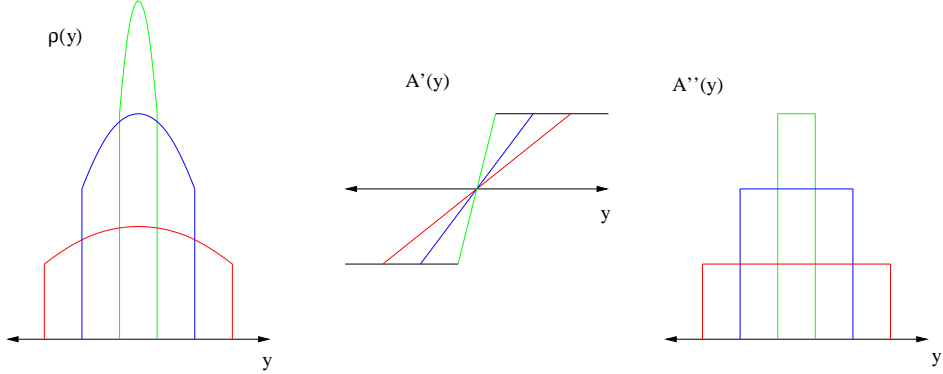}}
\hfil
}
\bigskip
\caption{\label{F:fig2} Plot of the straight interpolation profile
for the geometric factor $A'$ and its associated density profile.}
\end{figure}

In the limit $T \rightarrow 0$, the dominant part in the density profile
is
\begin{eqnarray}
\rho=3\gamma_0(T)(1-4\alpha \gamma_0(T)^2 \; T^2 \; z^2).
\end{eqnarray}
Identifying
\begin{eqnarray}
\beta_0(T) \equiv 3\gamma_0(T)\,, ~~~~
\beta_1(T) \equiv -12 \alpha  T^2 \; \gamma_0(T)^3\,,
\end{eqnarray}
we find that
\begin{eqnarray}
\lim_{T \rightarrow 0} T \; \beta_0(T) = {6 \over \tilde l}, ~~~~
\lim_{T \rightarrow 0} T \beta_1(T) \neq 0\,.
\end{eqnarray}
Therefore, even in the thin-shell limit, a straight interpolation
in the geometry does not correspond to a constant density profile (see fig.
(\ref{F:fig2})).
In the presence of a Gauss-Bonnet term it is not compatible to
impose a simple description for the
interior density profile and for the geometric warp factor
at the same time.
In the limit of strictly zero thickness (distributional limit), one will
unavoidably lose some information on the combined matter-geometry system.
%

To finish this section let us make an additional observation.
From expressions (\ref{straight-interpolation2}) and
(\ref{straight-interpolation2-cond}),
we can see that
\begin{eqnarray}
A''= \frac{2}{T} A'\Big|^{}_{T/2} + \nu(T,z) ~~~~\mbox{with}~~~~~
\lim_{T\rightarrow 0} T\nu(T,z) = 0   \,. \label{propsi}
\end{eqnarray}
Using this property in~(\ref{EGB-equations-rho},\ref{EGB-equations-pt}),
\begin{eqnarray}
\rho +p_T=3A''(1 - 4 \alpha A'^2)={6 \over T} A'(T/2)(1 - 4 \alpha A'^2)\ ,
\end{eqnarray}
and evaluating at $y=T/2$, we find
\begin{eqnarray}
T\rho\Big|_{T/2}=6A'(1 - 4 \alpha A'^2)\Big|_{T/2}.
\label{boundary-value}
\end{eqnarray}
We can see that contrary to what happens in the Einstein case, this condition
is different from the averaged condition (\ref{gb-average-condition}),
\begin{eqnarray}
T\langle \rho \rangle=\lim_{T \rightarrow 0} \int_{T/2}^{-T/2}  \rho\; dy=
(3A'-4 \alpha A'^3)\Big|_{-T/2}^{T/2}=
6A'(1-{4 \over 3} \alpha A'^2)\bigg|_{T/2}.
\label{average3}
\end{eqnarray}
Therefore, the averaged density and the boundary density are
different, and this is independent of the brane thickness.  For this simple
model, in thin-shell limit one can define two different internal
density parameters characterizing the thin brane.
One represents
the total averaged internal density, and can be defined as
\begin{eqnarray}
\rho_{\rm av} \equiv \lim_{T \rightarrow 0} T \langle \rho \rangle.
\end{eqnarray}
The other represents an internal density parameter calculated by
extrapolating to the interior the value of the density on the
boundary. This density can be defined as
\begin{eqnarray}
\rho_{\rm bv} \equiv \lim_{T \rightarrow 0} T \rho\Big|_{T/2}.
\end{eqnarray}
The junction conditions for a thin shell
\be
\rho=\frac{6}{\tilde l}\left(1-\frac{4}{3}\frac{\alpha}{\tilde l^2}\right)
\ee
given in \cite{Low:2000pq},
corresponds to the averaged condition (\ref{gb-average-condition}), or
(\ref{average3})
and therefore relates the total bending of the geometry
in passing through the brane to its total averaged density.
Instead, the particular condition analyzed for the boundary value
of $\rho\Big|_{T/2}$ in eq. (\ref{boundary-value}) yields in the thin-shell
limit the junction condition \cite{Germani}
\be
\rho=\frac{6}{\tilde l}\left(1-4\frac{\alpha}{\tilde l^2}\right)\ .
\ee
This condition is only considering information about the boundary value
of the density and not about its average.

In summary, what this analysis suggests is that in the presence
of the Gauss-Bonnet term, we can not ignore the interior
structure of the brane, by modelling it by a simple model, even
in the thin-shell limit. This point was first made in general terms by \cite{DeruDole}, and we have
made it explicit. We will see again this feature
in the next section on the cosmological dynamics of thick shells.

\section{Dynamical thick shells in Einstein and Lanczos gravity}
\label{S:dynamic}

We use the class of spacetime metrics given in (\ref{themetric}),
which contain a FRW metric in
every hypersurface $\{y={\rm const}.\}$, with a matter content described by
an energy-momentum tensor of the form (\ref{tAB}).  We consider
the additional assumption of a {\em static} fifth dimension: $\dot{b} = 0$.
We can rescale the coordinate $y$ in such a way that $b=1$.
Then the line element (\ref{themetric}) becomes
\begin{equation}
ds^2=-n^2(t,y)dt^2+a^2(t,y)h_{ij}dx^idx^j+dy^2\, .
\end{equation}
In Appendix~\ref{appa} we show that the $\{ty\}-$component of the
Lanczos field equations, for the case with a well-defined
limit in Einstein gravity, leads to the equation~(\ref{keyrel}).
In our case it implies the following relation:
\begin{equation}
n(t,y)=\xi(t)\dot a(t,y)\,. \label{0y}
\end{equation}
The remaining field equations can be written in the form
given in~(\ref{ttcom},\ref{ijcom},\ref{yycom}).  In our case they
become\footnote{The coupling constant $\alpha$ used here is one half the
one used in \cite{GS1}.}
\begin{eqnarray}
\left[a^4\left(\Phi+2\alpha\Phi^2+\frac{1}{l^2}\right)\right]' =
\frac{1}{6}(a^4)'\rho  \,, \label{ttc}
\end{eqnarray}
\begin{eqnarray}
\frac{n'}{n}\frac{\dot{a}}{a'}
\left[a^4\left(\Phi+2\alpha\Phi^2+\frac{1}{l^2}\right)\right]'
- {\left[a^4\left(\Phi+2\alpha\Phi^2+\frac{1}{l^2}\right)\right]^{\displaystyle{\cdot}}\,}' =
2\dot{a}a'a^2p_L\,, \label{ijc}
\end{eqnarray}
\begin{eqnarray}
\left[a^4\left(\Phi+2\alpha\Phi^2+\frac{1}{l^2}\right)\right]^\cdot =
-\frac{1}{6}(a^4)^{\displaystyle{\cdot}} p_T \,, \label{yyc}
\end{eqnarray}
where now $\Phi$ is given by
\begin{eqnarray}
\Phi=\frac{\dot a^2}{n^2 a^2}+\frac{k}{a^2}-\frac{a'^2}{a^2}
= H^2 +\frac{k}{a^2} -\frac{a'^2}{a^2} \,,
\end{eqnarray}
and we define the Hubble function
associated with each $y={\rm const}.$ slice as
\begin{eqnarray}
H(t,y)\equiv \frac{\dot a}{n a}\,.
\end{eqnarray}
With the assumption $\dot{b}=0$, the field equation (\ref{ijc}) leads to a
conservation equation for matter of the standard form
[see eq.(\ref{5ceq})]:
\begin{eqnarray}
\dot{\rho} = -3\frac{\dot{a}}{a}(\rho+p_L) \,. \label{mconeq}
\end{eqnarray}

Following the approach in the static scenario, we consider here the situation
in which there is a $Z_2$ symmetry and a fixed proper thickness $T$
for the brane.  Then one has to solve separately the equations for
the {\it bulk}
($|y|>T/2$) and the equations for the thick {\it brane} ($|y|<T/2$).  The first
step has already been done, and the result is~\cite{GS1}:
\begin{eqnarray}
\Phi+2\alpha\Phi^2+\frac{1}{l^2} = \frac{M}{a^4}\,,~~~~
\mbox{for $|y|>\textstyle{T\over2}$}\,,
\end{eqnarray}
where, as we show in Appendix \ref{B2}, $M$ is a constant that can be identified
with the mass of a black
hole present in the bulk.  Once the solution inside the thick brane has been
found, one has to impose the junction conditions~(\ref{cindm},\ref{cincm}) at
$y=\pm T/2$.

The first thing we can deduce from the junction conditions is that the
quantity $\Phi$ is continuous across the two boundaries $y=\pm T/2\,.$
But in general, its transversal derivative, $\Phi'$, will be discontinuous.
Then, using equation (\ref{yyc}) it follows that the transversal pressure
has to be zero on the boundary, $p_T(t,\pm T/2)=0$.  At the same time, from
(\ref{yyc}) we deduce that we must always have
\begin{eqnarray}
a^4\left(\Phi+2\alpha\Phi^2+\frac{1}{l^2}\right)\bigg|_{y=\pm T/2}=M\,.
\label{intm}
\end{eqnarray}
On the other hand, using again the relation~(\ref{keyrel}), we find that
\begin{eqnarray}
H' = -\frac{a'}{a}H~~~~\Rightarrow~~~~\Phi' = -2\frac{a'}{a}
\left(\Phi + \frac{a''}{a}\right)\,,
\end{eqnarray}
and then, expanding (\ref{ttc}), we obtain
\begin{eqnarray}
(1+4\alpha\Phi)\frac{a''}{a} = \Phi -\frac{2}{l^2} -\frac{1}{3}\rho
\,.
\label{mult}
\end{eqnarray}

In the limit $T \rightarrow 0$, the profiles of the density $\rho$ and of
$a''$ diverge, so that these dominant terms in expression~(\ref{mult})
have to be equated. This results in
\begin{eqnarray}
(1+4\alpha\Phi)\left({a' \over a}\right)'= -\frac{1}{3}\rho \,.
\label{00-thin-limit}
\end{eqnarray}
In what follows we consider the analysis of the Einstein and
Lanczos theories separately.

\subsection{Einstein gravity}

In Einstein gravity it is not difficult to write down an equation describing
the dynamics of every layer in the interior of a thick shell.
To that end we take $\alpha=0$ in the equations above.
By integrating (\ref{ttc}) over the interval $(-T/2,y_\ast)$,
and using (\ref{intm},\ref{00-thin-limit}), we arrive at
\begin{eqnarray}
\left(H^2 +{k \over a^2} + {1 \over l^2}  \right)&=&
\left({a' \over a}\right)^2+
{M \over a^4}+{1 \over 6 a^4}\int_{-T/2}^{y_\ast} (a^4)'\rho\, dy\cr&=&
{1 \over 36} \left(\int_{-y_\ast}^{y_\ast} \rho\, dy \right)^{2}+
{M \over a^4}+{1 \over 6a^4}\int_{-T/2}^{y_\ast} (a^4)'\rho\, dy\,.
\label{layer-Einstein}
\end{eqnarray}
A particular layer of matter inside the shell, located at $y=y_\ast$,
can be seen as separating an internal spacetime from a piece of
external spacetime.  From the previous equation, we can see that the
cosmological evolution of each layer $y=y_\ast$ in the thick shell
depends on the balance between the integrated density beyond the layer
(external spacetime), and a weighted contribution of the integrated
density in the internal spacetime. Therefore, the dynamics of each
shell layer will be influenced by the particular characteristics of
the internal density profile inside the shell. However, by looking at
this same equation, we can see that the dynamics of the boundary layer
$y_\ast=-T/2$ are only influenced by the total integrated density
throughout the shell:
\begin{eqnarray}
\left(H^2 +{k \over a^2} + {1 \over l^2}-
{M \over a^4}  \right)\bigg|_{-T/2}=
{1 \over 36}\left(\int_{-T/2}^{T/2} \rho dy\right)^{2}=
{1 \over 36}\left(T\langle \rho \rangle \right)^{2}={1 \over 36}\rho_{\rm av}^2 .
\end{eqnarray}
This is the modified Friedmann equation for the cosmological evolution
of the brane \cite{bdl}.


In the same manner as with static shells, we analyze
the case in which the density profile tends to a time-dependent value
in the thin-shell limit:
\begin{eqnarray}
\rho =\beta_0(T,t)+ \omega_3(T,t,z), ~~~~
\lim_{T \rightarrow 0} T \beta_0(T,t)=\rho_{\rm av}(t)\,, ~~~~
\lim_{T \rightarrow 0} T\omega_3(T,t,z) =0\,.
\label{constant-density}
\end{eqnarray}
When $\alpha=0$, eq. (\ref{00-thin-limit}) tells us that if the
density profile depends only on $t$ in the thin-shell limit, then, in
this same limit, the divergent part of the geometry $(a'/a)'$ is also
constant through the brane interior, describing what we called before a
straight interpolation. A simple density profile amounts to a simple
and equivalent geometric profile, and vice-versa. In this same case, but
including the Gauss-Bonnet term, $\alpha \neq 0$, the geometrical factor $(a'/a)'$
will exhibit a non-trivial profile in $y$, even in the thin-shell limit, as discussed
in more detail in the next subsection.

In the case in which $\rho$ depends only on time,
eq. (\ref{layer-Einstein}) reads
\begin{eqnarray}
\left.\left(H^2 +{k \over a^2} + {1 \over l^2}  \right)\right|_{y_\ast}=
\left.\left({a' \over a}\right)^2\right|_{y_\ast}+ {M \over a^4(y_\ast)}+
\frac{1}{6}\rho(t)\left.\left(1+ {a^4_{T/2} \over a^4}\right)\right|_{y_\ast}\,.
\end{eqnarray}
From Equation (\ref{00-thin-limit}) we deduce that
\begin{eqnarray}
\left.{a' \over a}\right|_{y_\ast}={1 \over 6}\rho_{\rm av} -
{1 \over 3}\int_{-T/2}^{y_\ast} \rho \, dy
=- {1 \over 3}\rho_{\rm av} z_\ast\,,
\end{eqnarray}
and integrating we obtain
\begin{eqnarray}
a(t,y)=a_0(t)\exp\left(-{1 \over 6}\rho_{\rm av}(t)Tz^{2}  \right)\ .
\end{eqnarray}
(Remember that $z\equiv y/T$.)
Therefore, at the lowest order in $T$ we have an equation for the
internal geometry of the form
\begin{eqnarray}
H_0^2+{k \over a_0^2}+{1 \over l^2}={1 \over 9}\rho_{\rm av}^2 z^2 +
{1 \over 36}\rho_{\rm av}^2(1-4z^2)+\frac{M}{a_0^4}
={1 \over 36}\rho_{\rm av}^2+\frac{M}{a_0^4},
\end{eqnarray}
which is exactly the standard braneworld generalized Friedmann equation~\cite{bdl}.

\subsection{Lanczos gravity}

In the general Lanczos case, eq. (\ref{00-thin-limit}) can be
written as
\begin{eqnarray}
\left[1+4\alpha\left(H^2+{k \over a^2}\right)\right]\left({a' \over a}\right)'
-4\alpha \left({a' \over a}\right)^2 \left({a' \over a}\right)'
=-{1 \over 3}\rho.
\label{gb-expanded}
\end{eqnarray}
Integrating between $-T/2$ and $T/2$ yields
\begin{eqnarray}
2\left[1+4\alpha\left(H^2+{k \over a^2}\right)\right]
\left({a' \over a}\right)\bigg|_{T/2}
-{8\alpha \over 3}\left({a' \over a}\right)^3 \bigg|_{T/2}
=-{1 \over 3}\langle \rho \rangle T=-{1 \over 3}\rho_{\rm av}\,.
\label{gb-blowup}
\end{eqnarray}
The boundary equation (\ref{intm}) can be written as
\begin{eqnarray}
\left.\left(H^2+{k \over a^2}\right)\right|_{T/2}-
\left.\left({a' \over a}\right)^2\right|_{T/2}
+2\alpha\left[\left(H^2+{k \over a^2}\right)-\left({a' \over a}\right)^2
\right]^2_{T/2}-\cr-{M \over a^4(T/2)}+{1 \over l^2}=0\,. \label{bcegb}
\end{eqnarray}
This is a quadratic equation for $(a'/a)^2\Big|_{T/2}$, with solutions
\begin{eqnarray}
\left({a' \over a}\right)^2\bigg|_{T/2}={1 \over 4 \alpha}
\left.\left[1+4\alpha\left(H^2+{k \over a^2}\right)\right|_{T/2}
\pm
\sqrt{1+{8\alpha \over l^2}-{8\alpha M \over a^4}} \right]\,.
\end{eqnarray}
From these two roots we take only the minus sign, as it is the only
one with a well-defined limit when $\alpha$ tends to zero.
By squaring (\ref{gb-blowup}) and substituting the above solution
we arrive at a cubic equation for $H^2+k/a^2$, first found
in~\cite{Charmousis}. This cubic equation has a real root
that can be expressed as~\cite{Gregory-Padilla}
\begin{eqnarray}
H^2+{k \over a^2}={1 \over 8 \alpha}
\left[
(\sqrt{\lambda^2+\zeta^3}+\lambda)^{2/3}
+
(\sqrt{\lambda^2+\zeta^3}-\lambda)^{2/3}
-2
\right]\,, \label{feqchar}
\end{eqnarray}
where
\begin{eqnarray}
\lambda \equiv \sqrt{\alpha \over 2} \rho_{\rm av}\,,
~~~~
\zeta \equiv  \sqrt{1+8\alpha V(a) } \equiv
\sqrt{1+{8\alpha \over l^2}-{8\alpha M \over a^4}}\,.
\end{eqnarray}
In addition, we need the conservation equation
\begin{eqnarray}
\dot \rho = 3 H (\rho+p_L)\,,
\end{eqnarray}
which is valid for each section $y=y_*$, and in particular,
for the boundary, $y=T/2$.
This equation can be averaged to give
\begin{eqnarray}
T\langle \dot \rho \rangle = 3 \langle H T (\rho+p_L) \rangle = 3 H\Big|_{T/2}
(T \langle \rho \rangle+T \langle p_L \rangle)+{\cal O}(T),
\end{eqnarray}
or written in another way,
\begin{eqnarray}
\dot \rho_{\rm av}= 3 H\Big|_{T/2}(\rho_{\rm av}+p_{{\rm av}L})\ .
\end{eqnarray}
This happens because
\begin{eqnarray}
H(t,y)\rightarrow H_0(t,y_0)+{\cal O}(T)
\end{eqnarray}
for any $y_0 \in[-T/2,T/2]$, which we have taken as $y_0=T/2$ for convenience.

We analyze now the simple case of a constant density profile
(\ref{constant-density}). For consistency with the $T \rightarrow 0$
case, we know that
\begin{eqnarray}
a(t,y)=a_0(t)[1+T{\tilde a}(t,z)]+ {\cal O}(T^2)\,,
\end{eqnarray}
and therefore, from~(\ref{0y}),
\begin{eqnarray}
n(t,y)= \xi(t)\left[a_0\left(1+T\tilde{a}(t,z)\right)\right]^{\displaystyle{\cdot}}\,.
\end{eqnarray}
In the same limit, eq. (\ref{00-thin-limit}) becomes
\begin{eqnarray}
{\tilde a}_{,zz}= -{1\over 3}
{\beta_0(T,t)T  \over \left[1+4\alpha
\left(H_0^2+{k \over a_0^2}-{\tilde a}_{,z}^2 \right)\right]}.
\end{eqnarray}
(Here the subscript $,z$ denotes differentiation with respect to $z$.)
A necessary condition to have a straight interpolation for the geometry
is that ${\tilde a}(t,z)=b(t)Z(z)$. To check whether or not a simple density
profile corresponds to a simple geometrical profile, we can therefore try
to solve this equation by separation of variables.
It is not difficult to see that in order to find a solution
with a well defined Einstein limit, we need
\begin{eqnarray}
b(t) = \mu \,,
~~~~ \mu^{-1}\beta_0(T,t)T = \mu^{-1}\beta_0(T)T = \rho_{\rm av}= {\rm constant},
~~~~ H_0^2+{k \over a_0^2}=\Lambda_4\,,
\end{eqnarray}
where $\mu$ is a constant that can be absorbed into the function $Z(z)$,
and we can take it to be $\mu=1$.
In this way we recover the AdS and dS solutions
for the brane (depending on the sign of the effective four-dimensional
cosmological constant).  To find the specific $y$ profile, we have to solve
\begin{eqnarray}
Z_{,zz}= -{1\over 3}
{\rho_{\rm av}  \over \left[1+4\alpha
\left(\Lambda_4-Z_{,z}^2\right)\right]}.
\end{eqnarray}
This equation can be integrated to give
\begin{eqnarray}
(1+4\alpha\Lambda_4)Z_{,z}-{4\alpha \over 3}Z_{,z}^3=-{1\over 3}\rho_{\rm av}\, z\,.
\end{eqnarray}
For our purposes the specific solution of this cubic equation is
not important.  What we want to point out, is that the solution does not
correspond to a straight interpolation as happened in the Einstein case.
So, in general, simple solutions for the matter profile lead to non-trivial
profiles for the scalar curvature even in the thin-shell limit.

Taking a simple model for the
geometry, the straight interpolation model,
\begin{equation}
a(t,z)=a_0(t)-\frac{1}{2}b(t)z^2T\ ,
\label{req}
\end{equation}
we deduce the density profile,
using eq. (\ref{gb-expanded}),
\begin{eqnarray}
\lim_{T \rightarrow 0}T \rho=
3\left[1+4\alpha\left(H_0^2+{k \over a_0^2}\right)\right]
\left({b \over a_0}\right)-12\alpha\left({b \over a_0}\right)^3 z^2.
\end{eqnarray}
As in the static case, even for very small thickness the density
profile has now a non-trivial structure.
We observe that
\begin{eqnarray}
{a' \over a}\bigg|_{T/2}= -{1 \over 2}{b \over a_0} + {\cal O}(T) \,,
~~~~
\left({a' \over a}\right)'= {2 \over T}\left[ {a' \over a}\bigg|_{T/2}
+{\cal O}(T)\right]\,. \label{extraas}
\end{eqnarray}
The second relation and eq.~(\ref{propsi}) coincide in the thin-shell
limit. Therefore, evaluating~(\ref{gb-expanded}) on $y=T/2$, we obtain
\begin{eqnarray}
\left[1+4\alpha\left(H^2+{k \over a^2}\right)\right]\left.\left({a' \over a}\right)
\right|_{T/2}
-4\alpha \left.\left({a' \over a}\right)^3\right|_{T/2}
=-{1 \over 6}T \rho\Big|_{T/2}=-{1 \over 6}\rho_{\rm bv}\,. \label{cubic2}
\end{eqnarray}
Following the same steps as before, but using this condition
instead of eq. (\ref{gb-blowup}), we arrive at a cosmological generalized
Friedmann equation \cite{GS1} different from that in~\cite{Charmousis} in its form
and in the fact that it depends on the quantity associated with the boundary value of
the energy density, $\rho_{\rm bv}$, instead of the value associated with the
average of the energy density, $\rho_{\rm av}$.  Remarkably, the cubic equation
that results from combining the last equation~(\ref{cubic2}) with the boundary
condition~(\ref{bcegb}), becomes in this case linear.  That is, the coefficients of the
terms quadratic and cubic in $H^2+k/a^2$ vanish~\cite{GS1}.  The modified
Friedmann equation found in this case is:
\begin{eqnarray}\label{nonconv}
H^2+{k \over a^2}=
{1 \over \left(1+{8\alpha \over l^2}-{8\alpha M \over a^4}\right)}
{1 \over 36} \rho_{\rm bv}^2 +{1 \over 4 \alpha}\left(
\sqrt{1+{8\alpha \over l^2}-{8\alpha M \over a^4}}-1 \right)\,.
\end{eqnarray}
In contrast with the modified Friedmann equation~(\ref{feqchar}), which
was obtained by using a completely general procedure, in order to obtain
this equation we had to use a procedure which required an
extra assumption, namely equation~(\ref{extraas}). Hence it will
not work  for profiles of the metric function $a(t,z)$ that do not
satisfy these requirements or equivalent ones.   On the other hand,
by looking at the developments here presented, we can conclude that the
different results found in the literature
for the dynamics of a distributional shell have their origin in the
additional internal richness introduced in the brane by the presence
of the GB term.

We have analyzed and compared how the thin-shell limit of static and
cosmological braneworld models is attained in Einstein and
Lanczos gravitational theories. We have seen that the
generalized Friedmann equation proposed in~\cite{Charmousis} is always
valid and relates the dynamical behaviour of the shell's boundary with
its total internal density (obtained by integrating transversally the
density profile).  Instead, the generalized Friedmann equation
proposed in~\cite{GS1} relates the dynamical behaviour of the
shell's boundary with the boundary value of the density within the
brane. This equation is not always valid, only for specific
geometrical configurations.

Einstein's equations in these models transfer the divergent
contributions of the thin-shell internal density profile to the
structure of the internal geometry in a faithful way. If we do not know
the internal structure of the shell, we can always model it in simple
terms by assuming an (almost) constant density profile and an (almost)
constant internal curvature. However, the GB term makes it incompatible
to have both magnitudes (almost) constant.  If the density is (almost)
constant, then the curvature is not, and vice-versa. Therefore
the particular structure of the Lanczos theory
introduces important microphysical features into the matter-geometry
configurations, beyond those in Einstein gravity, that are hidden in the
distributional limit. We see this more clearly in the next section.

\section{Smooth flat brane model}

In this section, using an adequate definition of the brane stress-energy tensor,
we confirm the results obtained previously for a Gauss-Bonnet brane model, extending
the straight interpolation case.
To do so, we use an approach
directly based on the field
equations for a smooth flat brane.

Suppose we can solve the problem of a five-dimensional scalar field with the metric
\be
ds^2=-e^{-2A_B(y)}d\eta^2+dy^2\ ,
\ee
where $d\eta^2$ is the four-dimensional Minkowski metric. Now the field equations read (see
par. (\ref{EinsteinGB})):
\ba
3A''_B\left(1-4\alpha A'^2_B\right)-p_B^t=\rho_B\ ,
6A'^2_B(1-2\alpha A'^2_B)-\frac{6}{l^2}=p^t_B\ ,\nonumber
\ea
where $A'_B(y)$ is a family of solutions parameterized by $B$, subject to the boundary
condition
\be
\lim_{y\rightarrow\infty}A_B'(y)=\frac{1}{\tilde l}\ ,
\ee
($\tilde l$ is defined in (\ref{modlength})), $\rho_B$ is the energy-density of the matter and
$p_B^t$ is the transversal pressure in the $y$ direction. Moreover we have the ``total bending"
condition
\be
\int^\infty_{-\infty}A''_B(y)dy=\frac{2}{\tilde l}\ .
\ee
Now the domain wall limit implies that
\be
\lim_{B\rightarrow\infty}A'_B(y)=\frac{1}{\tilde l}\ \mbox{for}\ \mid y\mid\geq T/2\ ,
\ee
where $T$ defines a ``proper thickness"\footnote{This is not uniquely defined, and can be
for example interpreted as the proper variance of the distribution $A''_B(y)$,
\be
\sigma_B=\left[\int^\infty_{-\infty} A''_B(y)y^2dy/\int^\infty_{-\infty} A''_B(y)dy\right]^{1/2}
\ .\nonumber
\ee}
of the smooth model and is eventually taken to zero.
From it we have
\be\label{totalbending}
\lim_{B\rightarrow\infty}\int^{T/2}_{-T/2}A''_B(y)dy=\frac{2}{\tilde l}\ ,
\ee
and
\be
\lim_{B\rightarrow \infty}p_B^t=0\ \mbox{for}\ \mid y\mid\geq T/2\ .
\ee
The ``total bending" junction condition follows as
\be
\rho_{\rm av}=\lim_{T\rightarrow 0}T\langle \rho\rangle=\lim_{T\rightarrow 0,B\rightarrow
\infty}\int^{T/2}_{-T/2}\rho_B(y)dy=\frac{6}{\tilde l}\left(1-\frac{4\alpha}{3\tilde l^2}
\right)\ .
\ee
We now define another possibility for the junction conditions that we call ``holographic"
junction
conditions. From the integral (\ref{totalbending}) we have, using the average theorem for
integrals,
\be
\lim_{B\rightarrow\infty}\int^{T/2}_{-T/2} A''_B(y)dy=\lim_{B\rightarrow\infty}A''_B(y_s)
T=\frac{2}{\tilde l}\ ,
\ee
where $\mid y_s\mid\leq T/2$ and we call the hypersurface $y=y_s$ the ``screen".
Therefore we have
\be
\lim_{B\rightarrow\infty}TA''_B(y_s)=\frac{2}{\tilde l}=2\lim_{B\rightarrow\infty}
A'_B(T/2)\ .
\ee
Now considering the straight interpolation case we have that $y_s\sim T/2$ for $T\rightarrow 0$, then
\be
\rho_s=\lim_{T\rightarrow 0, B\rightarrow\infty} T\rho_B(y_s)=\frac{6}{\tilde l^2}\left(
1-\frac{4\alpha}{\tilde l^2}\right)\ .
\ee
This confirms that, even in the smooth model,
\be
\rho_{\rm av}\neq \rho_s\ .
\ee
\subsection{A simple explicit example}

Consider the following family of solutions
\be
A'_B(y)=\frac{1}{\tilde l}\tanh(By)\ ,
\ee
for which
\be
\rho_{\rm av}=\frac{6}{\tilde l^2}\left(1-\frac{4\alpha}{3\tilde l^2}\right)\ .
\ee
To find the screen, we can try to solve the equation
\be
\frac{2}{\tilde l}=TA''_B(y_s)\ ,
\ee
for $y_s$. This has a solution
\be
y_s=\frac{1}{B}\tanh^{-1}\sqrt{1-\frac{2}{BT}}\ .
\ee
Then we have
\be
A'_B(y_s)=\frac{1}{\tilde l}\sqrt{1-\frac{2}{TB}}\ ,
\ee
so that
\be
\lim_{B\rightarrow\infty}A'_B(y_s)=\lim_{B\rightarrow\infty}A'_B(T/2)=\frac{1}{\tilde l}\ .
\ee
We can therefore conclude
\be
\rho_s=\frac{6}{\tilde l}\left(1-\frac{4\alpha}{\tilde l^2}\right)\ .
\ee

\section{Holographic description of the dark radiation term}

The holographic principle has been applied extensively to cosmological cases in the original
Randall-Sundrum model (see \cite{padilla} for a review). At the time of writing this
thesis, this concept
has also been applied to the Gauss-Bonnet case (see e.g. \cite{Padilla,Gregory-Padilla}).
Since this field is
relatively new, in this section we concentrate on the more simple Randall-Sundrum scenario.

Here we show an example, without going too much into details, of how one can interpret,
holographically,
the Weyl contribution to the non-conventional Friedmann equation
(\ref{nonconv}), in the case
$\alpha=0$, when the Gauss-Bonnet term is switched off\footnote{See also \cite{SV} for a different perspective. Here
the quantum mechanical properties of the 5D Schwarzschild-AdS black hole have been used.}. Splitting the energy density into the
matter energy density ($\rho$)
and brane tension ($\lambda$),
$\rho_{\rm bv}=\rho+\lambda$, the modified Friedmann equation, replacing all the constants,
reads (see also sec. \ref{star})
\begin{equation}
H^2=\frac{8\pi G_N}{3}\rho\left(1+\frac{\rho}{2\lambda}\right)+\frac{M}{a^4}-\frac{k}{a^2}
+\frac{1}{3}\Lambda\ .
\end{equation}
In order to apply the AdS/CFT description, we will set $k=\Lambda=0$. We follow again
\cite{bida}.

We consider a four-dimensional cosmological model with a Friedmann geometry,
\begin{equation}
ds^2=a(\eta)^2\left[-d\eta^2+d\vec{x}\cdot d\vec{x}\right]\ .
\end{equation}
Since the Universe is accelerating (or decelerating) it produces particles by quantum
vacuum fluctuations. These particles will
back-react on the metric, producing a non-isotropic perturbation of the type
\be
ds^2=a(\eta)^2\left[-d\eta^2+\sum_{i=1}^3\left[1+h_i(\eta)\right] (dx^i)^2\right]\ ,
\ee
where $\mbox{max}\mid h_i(\eta)\mid\ll 1$. For simplicity we will use the constraint
$\sum_{i=1}^3 h_i(\eta)=0$. Considering
the vacuum as a massless scalar field $\phi(x)$, we can decompose it in Fourier space as
\be
\phi(x)=\int d^3k \left[a_k u_k(x)+a_k^\dag u^*_k(x)\right]\ ,
\ee
where $\dag$ means the Hermitian conjugate and $*$ the complex conjugate.

Given the symmetries of the problem, we can use the following separation of variables,
\be
u_k=(2\pi)^{-3/2}e^{ik\cdot x}\frac{\chi_k(\eta)}{a(\eta)}\ .
\ee
Now the evolution equation (\ref{evolscalar}) in the conformal case and in conformal
time reduces to
\be
\frac{d^2\chi_k}{d\eta^2}-\sum^3_{i=1}k_i^2\chi_k=0\ .
\ee
If we also impose the orthonormality of the $u_k$,$u^*_k$,
we have the conditions
\be
\chi_k\partial_\eta\chi^*_k-\chi^*_k\partial_\eta\chi_k=i\ .
\ee
Since we would like an asymptotically flat spacetime, we also impose
\be
\lim_{\eta\rightarrow\infty}h_i(\eta)=0\ .
\ee
The normalized positive frequency solution for $\eta\rightarrow -\infty$ is
\be
\chi_k^{\rm in}(\eta)=(2k)^{-1/2}e^{-ik\eta}\ .
\ee
Then we can write the integral equation
\be\label{chi}
\chi_k(\eta)=\chi_k^{\rm in}(\eta)+k^{-1}\int_{-\infty}^\eta V_k (\eta')\sin\left[k(\eta-\eta')
\right]
\chi_k(\eta')d\eta'\ ,
\ee
with $V_k=\sum_{i=1}^3 k_i^2 h_i(\eta)$. The production of particles will be in the
late region ($\eta\rightarrow\infty$),
and
\be
\chi_k^{\rm out}(\eta)=\alpha_k\chi^{\rm in}_k(\eta)+\beta_k\chi^{\rm in\ *}_k(\eta)\ ,
\ee
where the Bogolubov coefficients to first order in $V_k$ are
\ba
\alpha_k &=& 1+i\int^\infty_{-\infty}\chi^{\rm in\ *}_k(\eta)V_k(\eta)\chi_k(\eta)d\eta\cr
\beta_k &=& -i\int^\infty_{-\infty}\chi^{\rm in}_k(\eta)V_k(\eta)\chi_k(\eta)d\eta\
.\nonumber
\ea
The number of particles created per proper volume ($n$) will be linked to the probability
of non-zero values for $\chi_k$ in the
output region. Then
\be
n=(2\pi a)^{-3}\int\mid\beta_k\mid^2d^3k\ ,
\ee
with an associated energy density
\be
\rho=(2\pi)^{-3}a^{-4}\int \mid \beta_k\mid^2 k d^3 k\ .
\ee
The second-order approximation in $V_k$ gives
\be
\rho=-(3840\pi^2 a^4)^{-1}\int^\infty_{-\infty}d\eta_1\int^{\infty}_{-\infty}
d\eta_2\ln\left[2i(\eta_1-\eta_2)\right]
\sum_i(\partial^3_\eta h_i(\eta_1))(\partial^3_\eta h_i(\eta_2))\ .
\ee
Physically the perturbation will be a composition of damped oscillating functions of the form
\be
h_i(\eta)=e^{-\alpha\eta^2}\cos(\beta\eta^2+\delta_i)\ ,
\ee
where $\delta_i-\delta_{i+1}=2/3\pi$. Then we finally have
\be
\rho=\frac{1}{2880\pi}\frac{(\alpha^2+\beta^2)^{1/2}}{\alpha^3 a^4}\ .
\ee
If we now associate
\be
\frac{1}{2880\pi}\frac{(\alpha^2+\beta^2)^{1/2}}{\alpha^3}=M\ ,
\ee
we have an holographic description of the projected Weyl tensor on the brane. Since this
quantum
effect is due only to the geometry, independently of the matter content, it is physically
reasonable
to connect it with the projected Weyl tensor, which is a purely geometrical quantity
dependent only on the bulk
curvature.

\newpage
\chapter{Conclusions}

In this thesis I studied the Randall-Sundrum mechanism of localization of gravity in the presence
of an infinitely large extra dimension, from the astrophysical and cosmological point of view.
I focused on the exact basic models. As a
motivation for studying these scenarios in light of the holographic principle,
I showed their relations with known quantum
solutions in four-dimensional gravity.

\subsection*{Astrophysics}
I investigated how 5-dimensional gravity can affect static
stellar solutions on the brane. I found exact braneworld
generalizations of the uniform-density stellar solution, and used
this to estimate the local (high-energy) effects of bulk gravity.
I derived astrophysical lower limits on the brane tension.
I also found that the star is less
compact than in general relativity. The smallness of high-energy corrections to
stellar solutions flows from the fact that $\lambda$ is well above
the energy density $\rho$ of stable stars. However nonlocal
corrections from the bulk Weyl curvature (5-dimensional graviton
effects) have qualitative implications that are very different
from general relativity.

The Schwarzschild solution is no longer the unique asymptotically
flat vacuum exterior; in general, the exterior carries an imprint
of nonlocal bulk graviton stresses. The exterior is not uniquely
determined by matching conditions on the brane, since the
5-dimensional metric is involved via the nonlocal Weyl stresses.
I demonstrated this explicitly by giving two exact exterior
solutions, both asymptotically Schwarzschild. Each exterior which
satisfies the matching conditions leads to a bulk metric, which
could in principle be determined locally by numerical integration.
Without any exact or
approximate 5-dimensional solutions to guide us, we do not know
how the properties of the bulk metric, and in particular its
global properties, will influence the exterior solution on the
brane.

Guided by perturbative analysis of the static weak field
limit, I made the following conjecture:
{\em if the bulk for a static stellar solution on the brane is
asymptotically AdS and has regular Cauchy horizon, then the
exterior vacuum which satisfies the matching conditions on the
brane is uniquely determined, and agrees with the perturbative
weak-field results at lowest order.} An immediate implication of
this conjecture is that the exterior is not Schwarzschild, since
perturbative analysis shows that there are nonzero Weyl stresses
in the exterior (these stresses are the manifestation
on the brane of the massive Kaluza-Klein bulk graviton modes). In
addition, the two exterior solutions that I present would be
ruled out by the conjecture, since both of them violate the
perturbative result for the weak-field potential.

The static problem is already complicated, so that analysis of
dynamical collapse on the brane can be very difficult. However,
the dynamical problem gives rise to more striking features.
Energy densities well above the brane tension could be reached
before horizon formation, so that high-energy corrections could be
significant.
In this direction I explored the consequences for
gravitational collapse of braneworld gravity effects, using the
simplest possible model, i.e.\ an Oppenheimer-Snyder-like collapse on a
generalized Randall-Sundrum type brane. Even in this simplest
case, extra-dimensional gravity introduces new features. Indeed using
only the projected 4D equations, I have shown, independent of the
nature of the bulk, that the exterior vacuum on the brane is
necessarily {\em non-static}. This contrasts strongly with GR,
where the exterior is a static Schwarzschild spacetime. Although
I have not found the exterior metric, I know that its non-static
nature arises from (a)~5D bulk graviton stresses, which transmit
effects nonlocally from the interior to the exterior, and (b)~the
non-vanishing of the effective pressure at the boundary, which
means that dynamical information on the interior side can be
conveyed outside. My results suggest that gravitational collapse
on the brane may leave a signature in the exterior, dependent upon
the dynamics of collapse, so that astrophysical black holes on the
brane may in principle have KK hair.

I expect that the non-static property of the exterior will be transient and {\em
non-radiative}, as follows from a perturbative study of non-static
compact objects, showing that the Weyl term ${\cal E}_{\mu\nu}$ in
the far-field region falls off much more rapidly than a radiative
term. It is reasonable to assume that the exterior
metric will tend to be static at late times and tend to Schwarzschild,
at least  at large distances.

Moreover I showed that this non-static behaviour is due to an anomaly of the Ricci scalar.
Indeed assuming a static exterior,
I found that in the absence of the cosmological constant, it does not vanish as one must expect
from
a vacuum solution. This means that there is a ``potential energy" stored on the boundary of
the star. This energy must
be released in some way producing a time signature in the exterior. Since this anomaly resembles
very much the Weyl anomaly due to the Hawking process, it is reasonable to conjecture
that the time signal can be
holographically reproduced by an ``evaporation process". Using this holographic point of view
I derived an indirect measure for the brane vacuum energy, $\lambda\sim 10 M_p^4 u^{-1}$,
where $u$ depends on the number of fields involved in the Hawking process.

\subsection*{Cosmology}

I analyzed and compared how the thin-shell limit of static and
cosmological braneworld models is attained in Einstein and
Lanczos gravitational theories. I showed that the
generalized Friedmann equation proposed in~\cite{Charmousis} is always
valid and relates the dynamical behaviour of the shell's boundary with
its total internal density (obtained by integrating transversally the
density profile).  By contrast, the generalized Friedmann equation
proposed in~\cite{Germani} relates the dynamical behaviour of the
shell's boundary with the boundary value of the density within the
brane. This equation is not always valid, but only for specific
geometrical configurations.

Einstein equations in these models transfer the divergent
contributions of the thin-shell internal density profile to the
structure of the internal geometry in a faithful way. If one does not know
the internal structure of the shell, one can always model it in simple
terms by assuming an (almost) constant density profile and an (almost)
constant internal curvature. However, the Gauss-Bonnet term makes it incompatible
to have both magnitudes (almost) constant.  If the density is (almost)
constant, then the curvature is not, and vice-versa. Therefore, one
can say that the particular structure of the Lanczos theory
introduces important microphysical features to the matter-geometry
configurations beyond those in Einstein gravity, that are hidden in the
distributional limit.

Studying a smooth flat brane model I generalized these conclusions for a more physical model.
In the particular example I gave, one can introduce two types of junction condition
which relate the
``total bending" of the brane with the matter content. The one I called the
``total bending"
junction condition that relates the bending to the total matter inside the brane. The second
junction condition relates instead the matter content in a particular
hypersurface called the ``screen",
with the total bending. In this screen all the information of the
total bending is stored. I call this a ``holographic" junction condition.

Independently of the gravitational
theory used, I showed how to derive the modified Friedmann equation and how it is related
to the black hole solution of the theory.

In particular for the simplest case of the
Randall-Sundrum scenario I showed how to interpret holographically the black hole mass using quantum particle
production with an FRW geometry.


\appendix
\newpage
\chapter{Junction conditions}\label{Appjc}

In this appendix we describe the junction conditions of a non-null hypersurface $\Sigma$,
following
\cite{Synge}.
We use the principle of the continuity of geodesics across any
hypersurface. Since the geodesic equation involves at most first-order derivatives of the metric,
we require the continuity of the metric together with its first derivative.

If $n$ is the unit normal
vector of the hypersurface $\Sigma$, it is always possible, at least locally, to define Gaussian
normal coordinates such that $x^n$ is the adapted coordinate to $n$ and
$a,b,c...$ are the adapted coordinates to $\Sigma$. Then the following are continuous quantities:
\be
g_{\alpha\beta},\ g^{\alpha\beta}\ ,\ \partial_\gamma g_{\alpha\beta}\ ,\ \partial_{\gamma}
g^{\alpha\beta}\ ,
\ \partial_{an}g^{\alpha\beta}\ ,\ \partial_{ab}
g_{\alpha\beta}\ ,
\ \partial_{ab}g^{\alpha\beta}{}\ .
\ee
It follows that the $G^n_\alpha$ component must be continuous.
In a covariant form,
\be\label{Appjc2}
[G^\alpha{}_\beta n^\beta]\Big|_{\Sigma}=0\ ,
\ee
where $[f]\Big|_\Sigma\equiv f\Big|_{\Sigma^+}-f\Big|_{\Sigma^-}$,
and $\Sigma^\pm$ are respectively
the outer and inner face of $\Sigma$. Equivalently
\ba
[g_{\alpha\beta}]\Big|_\Sigma = 0\ ,\
[K_{\alpha\beta}]\Big|_\Sigma = 0\ ,\label{jcK}
\ea
where $K_{\alpha\beta}=\pounds_n g_{\alpha\beta}/2$ is the extrinsic curvature of $\Sigma$.

We now work out the so-called Israel junction conditions for thin shells in this case.
Since we are interested in $Z_2$-symmetric branes, this considerably simplifies the calculations.

Suppose we have two parallel hypersurfaces $\Sigma_1$ and $\Sigma_2$. We again
introduce Gaussian normal coordinates, where the coordinate $x^n$ defines the orthogonal
direction of both $\Sigma_1$ and $\Sigma_2$ . Moreover we fix the point $x^n=0$ as the centre
point, and we call $T$ the proper distance between the hypersurfaces.
In these coordinates, the $Z_2$ symmetry means
\ba
g_{\alpha\beta}(x^n)&=&g_{\alpha\beta}(-x^n)\ ,\cr
K_{\alpha\beta}(x^n)&=&-K_{\alpha\beta}(-x^n)\ . \nonumber
\ea
Therefore from (\ref{jcK}) we have
\ba
K_{\alpha\beta}\Big|_{\Sigma_1}=-K_{\alpha\beta}\Big|_{\Sigma_2}\ ,
\ea
or equivalently, making explicit only the $x^n$ dependence of $K_{\alpha\beta}$,
\be
K_{\alpha\beta}(T/2)=-K_{\alpha\beta}(-T/2)\ .
\ee
We consider the incremental ratio in the thin shell limit ($T\rightarrow 0$),
\be
\lim_{T\rightarrow 0}\frac{K_{\alpha\beta}(T/2)-K(-T/2)}{T}=\partial_{x^n}K_{\alpha\beta}(0)\ .
\ee
This behaves like a Dirac delta function. It is a straightforward exercise to show
that if $\Phi(x^n)$ is a smooth test function, then
\be
\int^{T/2}_{-T/2}\partial_{x^n}K_{\alpha\beta} \Phi(x)dx=2K_{\alpha\beta}(0) \Phi(0)\ .
\ee
Therefore in the thin limit, $K_{\alpha\beta}$ ``jumps" or more rigorously\footnote{Since
in this limit $\Sigma_1\rightarrow \Sigma_2$, $\Sigma$ denotes one of the two equivalent
hypersurfaces.}
\be
[K_{\alpha\beta}]=2K_{\alpha\beta}\Big|_{\Sigma}\ ,
\ee
where here we define $[f]=f(0^+)-f(0^-)$.

\newpage
\chapter{5D geometry}
\section{Lanczos gravity}\label{appa}
In this appendix we present the main geometrical quantities and field
equations associated with the 5D metric
\begin{eqnarray}
ds^2 = \tilde g_{AB}dx^Adx^B = -n^2(t,y)dt^2 + a^2(t,y)h_{ij}(x^k)dx^idx^j + b^2(t,y)dy^2\,,
\label{themetric}
\end{eqnarray}
where $h_{ij}$ is the metric of the three-dimensional maximally symmetric
surfaces $\{y=\mbox{const.\}}$, whose spatial curvature is parametrized
by $k=-1,0,1$.  A particular representation of $h_{ij}$ is
\begin{eqnarray}
h_{ij}dx^idx^j = \frac{1}{\left(1+\frac{k}{4}r^2\right)^2}\left(dr^2 +
r^2d\Omega^2_2\right)\,,
\end{eqnarray}
where $d\Omega^2_2$ is the metric of the 2-sphere.  The metric~(\ref{themetric})
contains as particular cases the metrics used in this thesis.

The non-zero components of the Einstein tensor $\tilde G_{AB}$ corresponding to this line
element are given by ($\dot{Q}=\partial_tQ$, $Q'=\partial_yQ$):
\begin{eqnarray}\label{second}
\tilde G_{tt} & = & 3\left\{ n^2\Phi + \frac{\dot{a}}{a}\frac{\dot{b}}{b}-
\frac{n^2}{b^2}\left[\frac{a''}{a}-\frac{a'}{a}\frac{b'}{b}\right]\right\}
\,, \cr
\tilde G_{ty} & = & 3\left( \frac{\dot{a}}{a}\frac{n'}{n}+\frac{a'}{a}\frac{\dot{b}}{b}
-\frac{\dot{a}'}{a}\right) \,, \cr
\tilde G_{ij} & = & \frac{a^2}{b^2}h_{ij}\left\{\frac{a'}{a}\left(\frac{a'}{a}
+2\frac{n'}{n}\right)-\frac{b'}{b}\left(\frac{n'}{n}+2\frac{a'}{a}\right)
+2\frac{a''}{a} + \frac{n''}{n}\right\}
\cr
& - & \frac{a^2}{n^2}h_{ij}\left\{\frac{\dot{a}}{a}\left(\frac{\dot{a}}{a}-
2\frac{\dot{n}}{n}\right) - \frac{\dot{b}}{b}\left(\frac{\dot{n}}{n}
-2\frac{\dot{a}}{a}\right) + 2\frac{\ddot{a}}{a}+\frac{\ddot{b}}{b} \right\}
- kh_{ij}\,, \cr
\tilde G_{yy} & = & 3\left\{-b^2\Phi + \frac{a'}{a}\frac{n'}{n} - \frac{b^2}{n^2}
\left[\frac{\ddot{a}}{a}-\frac{\dot{a}}{a}\frac{\dot{n}}{n}\right]\right\} \,,
\end{eqnarray}
where
\begin{equation}
\Phi(t,y) = \frac{1}{n^2}\frac{\dot{a}^2}{a^2}-\frac{1}{b^2}\frac{a'^2}{a^2}
+\frac{k}{a^2}\,. \label{defphi}
\end{equation}
Apart from the metric and the Einstein tensor, the field equations in
Lanczos gravity~(\ref{fieldeq}) contain a term quadratic in the
curvature, namely $\tilde H_{AB}$ [see eq.~(\ref{habterm})].
The non-zero components of this tensor can be written as
\begin{eqnarray}
\tilde H_{tt} & = & 6\Phi\left[
\frac{\dot{a}}{a}\frac{\dot{b}}{b}+\frac{n^2}{b^2}\left(
\frac{a'}{a}\frac{b'}{b}-\frac{a''}{a} \right) \right]\,, \nonumber \\
\tilde H_{ty} & = &
6\Phi\left(\frac{\dot{a}}{a}\frac{n'}{n}+\frac{a'}{a}\frac{\dot{b}}{b}
-\frac{\dot{a}'}{a}\right)\,, \nonumber \\
\tilde H_{ij} & = & 2a^2h_{ij}\left\{ \Phi\left[\frac{1}{n^2}\left(\frac{\dot{n}}{n}
\frac{\dot{b}}{b} - \frac{\ddot{b}}{b}\right) - \frac{1}{b^2}\left(
\frac{n'}{n}\frac{b'}{b}-\frac{n''}{n}\right)\right] \nonumber \right. \\
& + & \frac{2}{a^2bn}\left[ \frac{\dot{a}^2\dot{b}\dot{n}}{n^4}
+\frac{a'^2b'n'}{b^4} + \frac{\dot{a}a'}{b^2n^2}\left(
b'\dot{n}-\dot{b}n'\right) \right] \nonumber \\
& - &  2\left[ \frac{1}{n^2}\frac{\ddot{a}}{a}\left(
\frac{1}{n^2}\frac{\dot{a}}{a}
\frac{\dot{b}}{b} + \frac{1}{b^2}\frac{a'}{a}\frac{b'}{b}\right)-\frac{1}{b^2}
\frac{a''}{a}\left(\frac{1}{n^2}\frac{\dot{a}}{a}\frac{\dot{n}}{n}
+\frac{1}{b^2}\frac{a'}{a}\frac{n'}{n}\right)\right] \nonumber \\
& + & \left. \frac{2}{b^2n^2}\left[\frac{\ddot{a}}{a}\frac{a''}{a}-
\frac{\dot{a}^2}{a^2}\frac{n'^2}{n^2}-\frac{a'^2}{a^2}\frac{\dot{b}^2}{b^2}
-\frac{\dot{a}'}{a}\left(\frac{\dot{a}'}{a}-2\frac{\dot{a}}{a}\frac{n'}{n}
-2\frac{a'}{a}\frac{\dot{b}}{b}\right) \right] \right\}\,, \nonumber \\
\tilde H_{yy} & = & 6\Phi\left[\frac{a'}{a}\frac{n'}{n}+\frac{b^2}{n^2}\left(
\frac{\dot{a}}{a}\frac{\dot{n}}{n}-\frac{\ddot{a}}{a}\right)\right]\,.
\end{eqnarray}

In this thesis we consider the situation in which a thick brane is embedded
in the five-dimensional spacetime described by (\ref{themetric}), whose boundaries
are located at $y=\mbox{const.}$ hypersurfaces.  Consider the usual junction
conditions at a hypersurface $\Sigma_{y_c}\equiv\{y=y_c\}$,
that is, the continuity of the induced metric, $g_{AB}=\tilde g_{AB}-n_An_B$ and the
extrinsic curvature, $K_{AB} = -g^{C}_{(A}g^D_{B)}\nabla^{}_C n^{}_D$,
of $\Sigma_{y_c}$:
\begin{eqnarray}
n(t,y^+_c) = n(t,y^-_c) \,, ~~~~ a(t,y^+_c) = a(t,y^-_c) \,, \label{cindm}
\end{eqnarray}
\begin{eqnarray}
\frac{n'(t,y^+_c)}{b(t,y^+_c)} = \frac{n'(t,y^-_c)}{b(t,y^-_c)} \,, ~~~~
\frac{a'(t,y^+_c)}{b(t,y^+_c)} = \frac{a'(t,y^-_c)}{b(t,y^-_c)} \,. \label{cincm}
\end{eqnarray}

We assume a matter content described by an energy-momentum tensor
of the form
\begin{eqnarray}
\tilde \kappa^2\, \tilde T_{AB} = \rho u_{A}u_{B} + p_Lh_{AB} + p_Tn_{A}n_{B}\,,
\end{eqnarray}
where
\begin{eqnarray}
u_A=(-n(t,y),\mb{0},0)\,,~~~~
h_{AB} = \tilde g_{AB} + u_Au_B - n_An_B \,,~~~~
n_A=(0,\mb{0},b(t,y))\,,
\end{eqnarray}
where $\rho\,,$ $p_L\,,$ and $p_T$ denote, respectively, the energy density and
the longitudinal and transverse pressures with respect to the
observers $u^A$.  They are functions of $t$ and $y$.  The energy-momentum
conservation equations, $\nabla_A \tilde T^{AB}=0$, reduce to:
\begin{eqnarray}
\dot{\rho} = -\frac{\dot{b}}{b}(\rho+p_T) -3\frac{\dot{a}}{a}(\rho+p_L)\,,
\label{5ceq}
\end{eqnarray}
\begin{eqnarray}
p_T' = -3\frac{a'}{a}(p_T-p_L)-\frac{n'}{n}(\rho+p_T) \,.
\end{eqnarray}

The $\{ty\}$-component of the field equations for the metric
(\ref{themetric}) in Lanczos gravity [eq.~(\ref{fieldeq})] has the
form
\begin{eqnarray}
\left(1+4\alpha\Phi\right)\left(\frac{\dot{a}}{a}\frac{n'}{n}+
\frac{a'}{a}\frac{\dot{b}}{b}-\frac{\dot{a}'}{a}\right) = 0\,. \label{tycom}
\end{eqnarray}
If we discard the possibility $1+4\alpha\Phi=0$ by restricting ourselves
to models with a well-defined limit in Einstein gravity ($\alpha\rightarrow
0$), we see that the metric functions must satisfy
\begin{eqnarray}
\dot{a}' = \frac{n'}{n}\dot{a} + \frac{\dot{b}}{b}a' \,. \label{keyrel}
\end{eqnarray}
Using this consequence of the $\{ty\}$-component, we can rewrite the
remains components of $\tilde G_{AB}$ and $\tilde H_{AB}$ as
\begin{eqnarray}
\tilde G_{tt} = \frac{3n^2}{2a^3a'}\left(a^4\Phi\right)'\,,~~~~
\tilde G_{yy} = - \frac{3b^2}{2a^3\dot{a}}\left(a^4\Phi\right)^{\displaystyle{\cdot}}   \,,
\end{eqnarray}
\begin{eqnarray}
\tilde G_{ij} = \frac{1}{2\dot{a}a'}h_{ij}\left\{\frac{\dot{b}}{b}\frac{a'}{\dot{a}}
\left(a^4\Phi\right)^{\displaystyle{\cdot}} + \frac{n'}{n}\frac{\dot{a}}{a'}
\left(a^4\Phi\right)' - \left(a^4\Phi\right)^{\displaystyle{\cdot}}{}'  \right\}   \,,
\end{eqnarray}
\begin{eqnarray}
\tilde H_{tt} = \frac{3n^2}{2a^3a'}\left(a^4\Phi^2\right)'\,,~~~~
\tilde H_{yy} = - \frac{3b^2}{2a^3\dot{a}}\left(a^4\Phi^2\right)^{\displaystyle{\cdot}}   \,,
\end{eqnarray}
\begin{eqnarray}
\tilde H_{ij} = \frac{1}{2\dot{a}a'}h_{ij}\left\{\frac{\dot{b}}{b}\frac{a'}{\dot{a}}
\left(a^4\Phi^2\right)^{\displaystyle{\cdot}} + \frac{n'}{n}\frac{\dot{a}}{a'}
\left(a^4\Phi^2\right)' - \left(a^4\Phi^2\right)^{\displaystyle{\cdot}}{}'  \right\}   \,.
\end{eqnarray}

Then the field equations~(\ref{fieldeq}) for the metric~(\ref{themetric})
are equivalent to eq. (\ref{keyrel}) and
\begin{eqnarray}
\left[a^4\left(\Phi+2\alpha\Phi^2+\frac{1}{l^2}\right)\right]' =
\frac{1}{6}(a^4)'\rho  \,, \label{ttcom}
\end{eqnarray}
\begin{eqnarray}
\frac{\dot{b}}{b}\frac{a'}{\dot{a}}
\left[a^4\left(\Phi+2\alpha\Phi^2+\frac{1}{l^2}\right)\right]^{\displaystyle{\cdot}}
+ \frac{n'}{n}\frac{\dot{a}}{a'}
\left[a^4\left(\Phi+2\alpha\Phi^2+\frac{1}{l^2}\right)\right]'-\cr
- {\left[a^4\left(\Phi+2\alpha\Phi^2+\frac{1}{l^2}\right)\right]^{\displaystyle{\cdot}}{}}' =
2\dot{a}a'a^2p_L\,, \label{ijcom}
\end{eqnarray}
\begin{eqnarray}
\left[a^4\left(\Phi+2\alpha\Phi^2+\frac{1}{l^2}\right)\right]^{\displaystyle{\cdot}} =
-\frac{1}{6}(a^4)^{\displaystyle{\cdot}} p_T \,. \label{yycom}
\end{eqnarray}
Introducing (\ref{ttcom}) and (\ref{yycom}) into (\ref{ijcom}), we obtain
the conservation equation~(\ref{5ceq}).

\section{Bulk geometry in static coordinates} \label{B2}

In this section we show how to derive the modified Friedmann equation for any gravitational theory\footnote{See also \cite{Barcelo-edge} for
the closed Friedmann  brane case ($k=+1$).}.
A Friedmann brane at $y=0$ in a bulk metric
\be \label{matric1}
ds^2=-n^2(\tau,y)d\tau^2+a^2(\tau,y)d\vec x\cdot d\vec x+ b(\tau,y) dy^2
\ee
is locally equivalent to a Friedmann brane moving geodesically in a black-hole-type metric
\be\label{metricBH}
ds^2=-f(R)dT^2+R^2d\vec x\cdot d\vec x + \frac{dR^2}{f(R)}\ .
\ee
The form of $f(R)$ is determined by solving the field equations of the theory.

In this picture we imagine the brane as a hypersurface that is moving towards or away
from a black-hole with an expansion factor $a(\tau,y(\tau))=R(T(\tau))$. At
each fixed radial distance from the black-hole, $dR=0=da$, one has
\be
da=\dot ad\tau+a'dy=0\ \Rightarrow\ dy^2=\left(\frac{\dot a}{a'}\right)^2 d\tau^2\ .
\ee
Substituting into (\ref{matric1}), we obtain
\be
ds^2=-\left(n^2-\frac{\dot a^2}{a'^2}b^2\right)d\tau^2\ .
\ee
We suppose that the brane is a hypersurface in the black-hole background in geodesic
motion. Therefore its four-velocity in static coordinates is
\be
u_Adx^A=-\sqrt{f(R(T))+\dot R(T)^2}dT+\frac{\dot R(T)}{f(R(T))}dR=-d\tau\ ,
\ee
where $\tau$ is the proper time on the brane. For $dR=0$ we have
\be\label{tau}
-\sqrt{f(R)+\dot R^2}dT=d\tau\ .
\ee
Substituting into (\ref{metricBH}), we have
\be\label{BH}
ds^2=-\frac{f(R(T))}{f(R(T))+\dot R(T)^2}d\tau^2\ .
\ee
Equating now eq. (\ref{tau}) and (\ref{BH}), using $R(T(\tau))=a(\tau,y(\tau))$, we obtain
\be
n^2-\left(\frac{\dot a b}{a'}\right)^2=\frac{f(a)}{f(a)+\dot a^2}\ .
\ee
Since $\tau$ is the proper time, $n(\tau,0)=1$, and
\be
f(a)=\left(\frac{a'}{b}\right)^2-\dot a^2\ .
\ee
Therefore defining the Hubble rate $H=\dot a/a$ and the function $H_y=a'/ba$, we obtain
\be
\frac{f(a)}{a^2}=H_y^2-H^2\ ,
\ee
where $H_y$ must be found by junction conditions that are dependent on the gravity theory
chosen. In particular, for the Randall-Sundrum-type model, we have
\be
f(a)=k-\frac{M}{a^2}\ ,
\ee
and we find \cite{Kr:99,Id:00}
\be
H^2+\frac{k}{a^2}=H_y^2+\frac{M}{a^4}\ ,
\ee
where $M/a^4$ is the dark radiation term. Instead for Lanczos gravity, we have
\be
f(a) = k +
\frac{a^2}{2\alpha}\left(1-\sqrt{1+\frac{4\alpha M}
{3a^4}+\frac{2}{3}\alpha\Lambda}\right)\ ,
\ee
leading to \cite{GS1}
\be
H^2+\frac{k}{a^2}=H_y^2+\frac{1}{2\alpha}\left(1-\sqrt{1+\frac{4\alpha M}
{3a^4}+\frac{2}{3}\alpha\Lambda}\right)\ .
\ee
Local equivalence of the metrics (\ref{matric1}) and (\ref{metricBH}) has been proved in the
Randall-Sundrum model \cite{BH} and in the Lanczos model \cite{Charmousis}.


\end{document}

%% file: intestazione.tex
\title{ \bf {\Large Astrophysical and Cosmological Consequences of the Dynamical Localization of Gravity}}
\author{{\sc Cristiano Germani} \\\\ {\it
Institute of Cosmology and Gravitation} \\  {\it University of
Portsmouth}\\ \\ \\ \\ \\
Thesis for the award of the degree of \\ \\ {\Large Doctor of Philosophy} \\ \\ \\ \\ \\ \\ {\bf Supervisor:}\\ \\ Prof. Roy Maartens\\
}
\date{24 November 2003}
\maketitle
\begin{center} {\bf {\large To Emanuela, the Sun, and my Family, the Life.}}
\end{center}

\newpage
\begin{center} \begin{quote} {\it A good result cannot be discarded just because the world insists in looking different from our
theories (Carlo Rovelli).}
\end{quote}
\end{center}
\vspace{5cm}
\begin{center} \begin{quote} {\it Conjecture:
It is impossible to change the conception of Nature with white-boards
(myself).}
\end{quote}
\end{center}
\newpage
\begin{center}{\Large
Declaration}
\end{center}
The work presented in this thesis is partly based on
collaborations with C. Barcel\'o (Institute of Cosmology and
Gravitation, Portsmouth University), M. Bruni (Institute of
Cosmology and Gravitation, Portsmouth University), N. Deruelle
(Institute d' Astrophisique de Paris), R. Maartens (Institute of
Cosmology and Gravitation, Portsmouth University) and C. F.
Sopuerta (Institute of Cosmology and Gravitation, Portsmouth
University).

The list below identifies sections or paragraphs which are partially based on the listed
publications:
\begin{itemize}
\item Section 4.1: C. Germani and R. Maartens, Phys. Rev. {\bf
D64}, 124010 (2001). \item Sections 4.2.1-4.2.2: M. Bruni, C.
Germani and R. Maartens, Phys. Rev. Lett. {\bf 87}, 231302 (2001).
\item Sections 5.1-5.2 and Appendix B.1: C. Barcelo, C. Germani
and C. F. Sopuerta, Phys. Rev. {\bf D68}, 104007 (2003).
\item Sections 5.1-5.2: C.
Germani and C. F. Sopuerta, Astrophys. Space Sci. {\bf 283}, 487
(2003). \item Section 5.2 and Appendix B.2: C. Germani and C. F.
Sopuerta, Phys. Rev. Lett. {\bf 88}, 231101 (2002). \item Section
5.3: N. Deruelle and C. Germani, arXiv: gr-qc/0306116 (2003).
\end{itemize}
I hereby declare that this thesis has not been submitted, either in the same or different form,
to this or any other university for a degree and that it represents my own work.
\vspace{3cm}\\
Cristiano Germani.
\newpage
\thispagestyle{empty}
\begin{center}{\Large
Acknowledgments}
\end{center}
In these three years I met so many people who changed my life and
my way of thinking that it is almost impossible to remember everybody
here. Anyway first of all I wish to give a global thanks to any man
or woman who had any interaction with me. I indeed think that any
experience, good or bad, renews our life, making it more exciting
\footnote{No-one knows if fate (or decisions?) make us better or
worse people, since the data we have are not enough to make a
statistic.}. I will try to thank here all the people who really made
this thesis possible.

First of all I thank my family. I remember the first week I
arrived in this island, my father was desperate due to a
fantastically bad English weather, saying `` I will never allow to my
son to live without the light". But then, as often magically
happens in England, soon the light appeared and the sentence
changed to `` I want to pass my retirement in Portsmouth!".

A second special thanks is for Emanuela. She and the Sun made
something amazing. The light gives happiness and makes a man
(especially a Mediterranean one) alive in the fight against
loneliness. Emanuela had the power to do the same without using
nuclear reactions! Thanks to her warm energy and her courage I
could arrive at the end of this experience.

Roy, thanks. You really gave me the opportunity to grow up
scientifically and make me conscious wether I was really able to do
this job. You believed in me, accepting all the risks a supervisor
can have believing in his student. I learned how to be more modest,
even if my Roman descent was resisting it.

Bruce, what can I say. A ``thanks" is not enough for a friend and a
guider in the same time. You are one of the few guys who really
inspired me and renewed my love for the understanding of Nature.
I will never forget our walks on the sea-front, when the
rest of the world was too drunk to understand the beauty of the
Universe and the enigmas it is using to play with us.

MarcoB, thanks for giving me the opportunity of understanding what
I really wanted from my life.

Carlo and MarcoC, I really appreciated your help in donating to me
your experience. You contributed very much to my successes. In particular a special thanks
is for MarcoC who gave me useful suggestions on the ``dark side" \footnote{$\mbox{Dark side}\sim\mbox{Quantum side}$.}
of my thesis.

CarlosB and CarlosS, the tough Spanish guys. Thanks for accepting
my stresses without kicking my back!

Christine, thanks for your fantastic availability and sorry for
the frequent FIN1A (or whatever it is) I asked from you.
You know I love travelling!

Ismael, I do not know if you will read this thesis but I really
wish to thank you. You were so important in my life as one of
the best friends I ever had. Without you I couldn't survive alone
in this strange country. You made me feel at home.

Christopher, your friendship was really unique and I still think of
the sadness I had when you had to leave Portsmouth.

I really wish to thank the ICG as the warmest place to work in the
world!

I don't want to forget Nathalie. I was so amazed by her
attitude to attack and solve problems, her genuine
passion about physics and her way to collaborate. I was really
feeling to navigate in the same sea. Thanks Nathalie.

A very special thanks goes to my best friends of the last part
of my Portsmouth experience, Andrea, Caterina, Frances and Marta.
They brought me back to the real world, often so far from our physics
nut-shell. Their friendship was so important and strong that it
will never be attacked from any distance or time. Thank you guys.

Lastly I wish to thank PPARC for funding me for the whole period of my PhD and the
University of Rome "La Sapienza", Department of Aeronautics and Space Engineering, for hosting me during the writing of
this thesis.
\vspace{3cm}\\
Wait a second.... ciao a tutti gli amichetti !!!!! (compresi Davide e Luchino)

\newpage

\begin{center}{ {\Large Abstract}}\end{center}

In this thesis I review cosmological and astrophysical
exact models for Randall-Sundrum-type braneworlds and their
physical implications. I present new insights and show their
analogies with quantum theories via the holographic idea. In
astrophysics I study the two fundamental models of a
spherically symmetric static star and spherically symmetric
collapsing objects. I show how matching for the pressure of
a static star encodes braneworld effects. In addition
I study the problem of the vacuum exterior conjecturing a
uniqueness theorem. Furthermore I show that a collapsing dust
cloud in the braneworld has a non-static exterior, in contrast to the
General Relativistic case. This non-static behaviour is linked to
the presence of a ``surplus potential energy" that must be released,
producing a non-zero flux of energy. Via holography this can be
connected with the Hawking process, giving an indirect measure of
the brane tension. In cosmology I investigate the generalization of
the Randall-Sundrum-type model obtained by introducing the Gauss-Bonnet
combination into the action. I elucidate the junction conditions
necessary to study the brane model and obtain the cosmological
dynamics, showing that, even in the thin shell limit for the brane,
the Gauss-Bonnet term implies a non-trivial internal structure for
the matter and geometry distributions.
Independently of the gravitational
theory used, I show how to derive the modified Friedman equation
and how it is related
to the black hole solution of the theory.
Via
holography I also show how to interpret quantum mechanically the
mass of this black hole from a four-dimensional perspective in the
simplest Randall-Sundrum-type scenario.
\newpage